\documentclass[12pt,draftcls,onecolumn]{IEEEtran}

\usepackage{amsmath,amssymb,amsthm}
\usepackage{mathrsfs}
\usepackage{verbatim}
\usepackage[english]{babel}
\usepackage{setspace}
\usepackage{graphicx}				
\usepackage{epstopdf}
\usepackage{amsmath}				
\usepackage{amssymb}
\usepackage{cite}
\usepackage{amsfonts}
\usepackage{enumerate}
\usepackage{amsthm}
\usepackage{array}
\usepackage{graphicx}
\usepackage{resizegather}
\usepackage{float}
\usepackage[section]{placeins}
\usepackage{fixltx2e}
\usepackage[justification=centering]{caption}
\usepackage{color}
\usepackage{lipsum} 
\usepackage{fancyhdr}
\usepackage{url}

\renewenvironment{proof}{{\bfseries Proof. }}{\textbf{Proof. }}
\newcommand{\RNum}[1]{\uppercase\expandafter{\romannumeral #1\relax}}
\usepackage{tikz}

\usetikzlibrary{arrows,positioning} 
\usepackage{verbatim}

\author{Zahra Gallehdari$^{1}$, Nader Meskin$^{2}$ and Khashayar Khorasani$^{1}$
\thanks{This publication was made possible by NPRP grant No.  5-045-2-017
from the Qatar National Research Fund (a member of Qatar Foundation).
The statements made herein are solely the responsibility of the authors.}
\thanks{$^{1}$Zahra Gallehdari and Khashayar Khorasani are with the Department
of Electrical and Computer Engineering, Concordia University,
Quebec, Canada
         {\tt\small z\_galle@encs.concordia.ca} and  {\tt\small kash@ece.concordia.ca.}}
\thanks{$^{2}$Nader Meskin is with the Department of Electrical Engineering, Qatar
University, Doha, Qatar
        {\tt\small nader.meskin@qu.edu.qa.}}
}
\date{}

\renewcommand{\baselinestretch}{1.25}

\begin{document}

\title {\LARGE \bf 
An $H_\infty$ Cooperative Fault Recovery Control   of Multi-Agent Systems}
\maketitle

\nocite{Basil:2010:MISC}

\newtheorem {lemmas}{\textbf{Lemma}} 
\newtheorem {theorems}{\textbf{Theorem}} 
\newtheorem {definitions}{\textbf{Definition}} 
\newtheorem {assumptions}{\textbf{Assumption}} 
\newtheorem {corollaries}{\textbf{Corollary}} 
\newtheorem{remarks}{\textbf{Remark}} 
\newtheorem{notations}{\textbf{Notation}} 
\newtheorem{Propositions}{\textbf{Proposition}} 
\newtheorem{facts}{\textbf{Fact}}
\newtheorem{algorithm}{\textbf{Algorithm}}

 \tikzset{
 	head/.style = {fill = orange!50!blue,
 		label = center:\textsf{\Large H}},
 	tail/.style = {fill = blue!70!yellow, text = black,
 		label = center:\textsf{\Large T}}
 }
\begin{abstract}
In this work, an $H_{\infty}$ performance fault recovery control problem for a team of multi-agent systems that is subject to actuator faults is studied. Our main objective is to design a distributed control reconfiguration strategy such that \textbf{a)} in absence of disturbances the  state consensus errors either remain bounded or converge to zero asymptotically, \textbf{b)} in presence of actuator fault the output of the faulty system behaves exactly  the same as that of the healthy system, and \textbf{c)}  the specified  $H_{\infty}$ performance bound is guaranteed to be minimized in presence of bounded energy disturbances. 
The gains of the reconfigured  control laws are selected first by employing a geometric approach  where a set of controllers guarantees that the output of the faulty agent imitates that of the healthy agent and the consensus achievement objectives are satisfied. Next, the remaining degrees of freedom in the selection of the control law gains are used to minimize the bound on a specified $H_{\infty}$ performance index. 
The effects of  uncertainties and imperfections in the FDI module decision in correctly estimating the fault severity as well as delays in invoking the reconfigured control laws are investigated and a bound on the maximum tolerable estimation uncertainties and time delays are obtained.
Our proposed distributed and cooperative control recovery approach is applied to a team of five autonomous underwater vehicles  to demonstrate its capabilities and effectiveness in accomplishing the overall team requirements subject to various actuator faults, delays in invoking the recovery control, fault estimation and isolation imperfections and unreliabilities under different  control recovery scenarios. 
\end{abstract}
\section{Introduction}
Utilization of unmanned vehicles (agents)  in operations where human involvement is dangerous, or impossible as in deploying mobile robots for planetary surface exploration, autonomous underwater vehicles for surveying  deep sea,  among others,  has  recently received extensive interest by the research community. In addition, deployment of multiple vehicles such as spacecraft, mobile robots,  or unmanned underwater vehicles instead of using  a single vehicle increases the  system performance and reliability, while it will ultimately reduce the cost of the overall mission. \par

 In safety critical missions, the agents should have the capability to  cope with unexpected external influences such as  environmental changes  or internal events such as actuator and sensor faults. If these unexpected events are not managed successfully, they can lead to the team  instability or cause sever overall team  performance degradations. For example, the crash of the NASA's DART spacecraft in 2006 was due to a fault in its position sensors  \cite{croomes2006}.\par 
 
The development of control reconfiguration for multi-agent systems is distinct from the control design problem of healthy multi-agent systems \cite{breger2008,Semsar2009,Ren2010a,Movric2014}. This is so in the sense that the former should be ideally solved on-line and use only local information given that faults  occur at unknown times, have unknown patterns,  and the existing fault detection and isolation (FDI) module in the team information may  be available only locally, while the latter problem can be solved  off-line and by potentially using the entire system  information. Moreover,  due to the  information sharing structure of multi-agent systems,  the fault tolerant control approaches that have extensively been studied in the literature for single agent systems  \cite{Yang2010,Wang2010,Liu2012,Xavier2012,Khosrowjerdi2004} \underline{will not be} directly applicable to multi-agent systems.\par

Recently, the control reconfiguration problem of multi-agent systems has been studied in  \cite{Tousi2012,Azizi11,ACC2014,ECC2014,Zhou2014,Zhao2014,Chen2014,Wang2015-2,Wang2015-3,Mehrabian2011,Mehrabian2011-2,Li2012,Xiao2013}. In \cite{Tousi2012,Azizi11}, formation flight problem in a network subject to loss of effectiveness (LOE) faults is considered and in  \cite{ACC2014,ECC2014,Zhou2014} the consensus achievement problem in faulty multi-agent systems is studied.  In  \cite{Tousi2012}, a discrete-event supervisory module is designed to recover the faults that cannot be recovered by the agents using only local recovery solutions. In \cite{Azizi11}, a  high-level performance monitoring  module is designed  that monitors all the agents and detects deviations of the error signals from their acceptable ranges. This module would then activate a high-level supervisor to compensate for the deviations in the performance specifications due to limitations of the low-level recovery strategy. In \cite{Zhao2014,Chen2014,Wang2015-2,Wang2015-3} adaptive control approaches are employed to compensate for actuator faults and in \cite{Mehrabian2011,Mehrabian2011-2} control reconfiguration problem in a team of Euler Lagrange systems subject to actuator faults and environmental disturbances is studied. Finally in \cite{Li2012,Xiao2013},  attitude synchronization problem for a team of satellites in presence of actuator faults is studied.  \par

In this work,  $H_{\infty}$ performance  control reconfiguration problem in multi-agent systems subject to occurrence of \underline{three} types of faults, namely, the loss of effectiveness (LOE), stuck and outage faults is studied. 
The proposed $H_\infty$-based control reconfiguration  strategy  guarantees that the faulty agent outputs imitate those of the healthy system while  the state consensus errors are either ensured to be asymptotically stable or  remain bounded 
 in  absence of disturbances and  the disturbance attenuation bound is minimized when the disturbances exist. {Furthermore, this approach can compensate for the outage and stuck faults which cause rank deficiency and change the agent structure, whereas in the adaptive approaches it is assumed that the fault does not cause rank deficiency.  }
 
 Our proposed approach is similar to the works in \cite{Zhou2014, ECC2014}, but it has the following  distinctions, namely: (i) in \cite{Zhou2014} it is assumed that all the followers have access to the leader input signal while in this work {we do not} require this assumption,  (ii) in \cite{Zhou2014} environmental disturbances have not been considered whereas in this work {we do include} disturbances in our analysis and design, (iii) in this work agents could be subject to simultaneous LOE, outage and stuck faults, however in \cite{ECC2014} only a single LOE fault has been studied and in \cite{Zhou2014} only LOE and outage faults have been considered, (iv) in  both \cite{Zhou2014,ECC2014} the network topology is assumed to be indirected  whereas in this work we have considered a {directed network topology}, 
 and (v)  in this work we ensure that the outputs of the faulty agent are exactly  forced to follows those of the healthy agent and the state consensus errors remain bounded, whereas in \cite{Zhou2014,ECC2014} the consensus problem is considered. {The main motivation for enforcing outputs of the agents outputs to follow that of the leader is that in some applications like small light weight under vehicles, a small deviation in the speed can cause a big deviation in the agent position which may cause the network become disconnected or the agent becomes lost. In order to reach this objective, we formulated the problem as disturbance decoupling problem with stability and we use the Geometric approach \cite{Basile92} and controlled invariant subspaces to solve the problem along with linear algebra and matrix theory to address exact output following and state consensus error stability in the team as well as disturbance attenuation. To the best of our knowledge this problem has not been considered in the current literature in multi-agent systems.  }
 \par
 In view of the above discussion, the \underline{main contributions} of this work can be summarized as follows:
 \begin{enumerate}
 	\item [1)] A distributed control reconfiguration strategies for multi-agent systems subject to  LOE, outage and stuck faults are proposed and developed. Towards this end, associated with each agent a novel ``virtual auxiliary system'' is constructed for the first time in the literature. Each agent will receive information from {only} the states of its associated auxiliary agent and the nearest neighboring auxiliary agents. This is in contrast with conventional cooperative schemes where each agent will be receiving the actual state information from its nearest neighboring agents. The proposed strategy guarantee an $H_\infty$ performance control reconfiguration with stability.
 	\item [2)] The proposed  reconfiguration control  laws  guarantee that 
	the output of the faulty agent behaves the same  as that of the healthy system, and  moreover a specified $H_{\infty}$ performance index  is minimized  in presence of environmental disturbances.
 	\item [3)] The effects of  uncertainties and imperfections in the FDI module decision in correctly estimating the fault severity as well as delays in invoking the reconfigured control laws are investigated and a bound on the maximum tolerable estimation uncertainties and time delays are obtained.
 	\item [4)] The proposed distributed  reconfiguration control laws are capable of and designed specifically for accommodating single, concurrent and simultaneous actuator faults in multi-agent systems.
 \end{enumerate}
The remainder of this work is as follows. In Section \ref{section2}, the required background information are provided and the problem is formally defined. In Section \ref{proposed methodology}, the proposed reconfigured control law and the effects of uncertainties on the proposed solution are investigated. In Section \ref{simulation results}, the proposed control laws are applied to a network of Autonomous  Underwater Vehicles (AUV)s and extensive simulation results and various case studies are studied and presented. Finally, Section  \ref{conclusion} concludes the paper.
\section{ Background and Problem Definition}\label{section2}
\subsection{Graph Theory}\label{subsection 1.1}
 The communication network among $N+1$ agents can be represented by a graph. A directed graph $\mathcal{G} = (\mathcal{V}, \mathcal{E})$ consists of a nonempty finite set of vertices $\mathcal{V} =\{ v_0, v_1, . . . , v_N \}$ and a finite set of arcs $\mathcal{E} \subset \mathcal{V} \times \mathcal{V}$. The $i$-{th} vertex represents the $i$-{th} agent and the directed edge from $i$ to $j$ is denoted as the ordered pair $(i,j) \in \mathcal{E} $, which implies that agent $j$ receives information from agent $i$. The neighbor set of the $i$-th agent  in the network is denoted by ${\mathcal{N}}_i=\{ j|(j,i) \in \mathcal{E}\}$. 
The adjacency matrix of the graph $\mathcal{G}$ is given by $G = [g_{ij} ] \in \mathbb{R}^{(N+1)\times (N+1)}$, where $g_{ij} =1$ if $j \in {\mathcal{N}}_i$, otherwise $g_{ij} =0$. The Laplacian matrix for the graph $\mathcal{G}$ is defined as $L=D-G$, where $D=\text{diag} \{d_i\}$ and $d_i=\sum_{j=0}^N{g_{ij}}$. 
\subsection{Leader-Follower Consensus Problem in a Network of Multi-Agent Systems}\label{subsection 1.2}
The main objective of the consensus problem in a leader-follower (LF) network architecture  is to ensure all the team members  follow the leader's specified  trajectory/states.
Consider a network with $N$ follower agents that are governed by
\begin{eqnarray}
&&\dot x_i(t)=A x_i(t)+B u_i(t)+B_\omega \omega_i(t), \ x_i(t_0)=x_{i0}, \ i=1,..,N, \label{LL or F}\\
&&y_i(t)=Cx_i(t),\nonumber
\end{eqnarray}
and a leader  agent that has the dynamics given by
\begin{eqnarray}
&&\dot x_0(t)=Ax_0(t)+Bu_0(t)+B_\omega \omega_0(t),\label{L1}\\
&&y_0(t)=Cx_0(t),\nonumber
\end{eqnarray}
where {matrices $A$, $B$, $C$, $B_\omega$ represent the agents dynamics matrices and  are known,}  $x_i(t) \in \mathbb{R}^n$, $y_i(t)\in \mathbb{R}^q$, $u_i(t) \in \mathbb{R}^{m}$, and $\omega_i(t)\in \mathbb{R}^{p}$, $i=0,\dots,N$ are the agents states, outputs, control signals, and  exogenous disturbance inputs. In this work,  bounded energy disturbances are considered, i.e. $\omega_i(t)\in\mathcal{L}_2$, $i=0,\dots,N$ ($\omega_i(t)$ belongs to $\mathcal{L}_2$ if   $\int_{t_0}^{\infty}\omega_i^{\text{T}}(t)\omega_i(t)\text{d}t\leq\infty$).  

In the architecture considered in this paper, certain and \underline{a very few} of the followers, that are designated as \underline{pinned agents}, are communicating with the leader and receive data from it directly. The other \underline{followers are not in communication} with the leader and exchange information \underline{only} with their own nearest neighbor follower agents. {On the other word, each agent only communicate with its neighbors and at least one agent is a neighbour of the leader.}
The consensus error signal for the $i$-th follower is now defined by
\begin{IEEEeqnarray}{rCl}
e_i(t)&=&g_{i0}(x_i(t)-x_0(t))+\sum_{j \in {\mathcal{N}_i}} (x_i(t)-x_j(t)),\label{agents consensus error}
\end{IEEEeqnarray}
where $g_{i0}=1$ if agent $i$ is a pinned agent or is directly communicating with the leader and is zero otherwise. When there are no environmental disturbances, i.e. $\omega_i(t)\equiv 0$, $t>0$, $i=0,\dots,N$ the team reaches a consensus if $e_i(t)$ converges to origin asymptotically as $t\to \infty$. However, when there exist environmental disturbances, $e_i(t)$  cannot converge to origin, although it should remain in a bounded region around the origin. We refer and designate both of these cases as achieving consensus through out this paper. 

{Based on the above representation for the network, the aim is that all follower agents follow the leader agent trajectory. Accordingly, we  partition the network Laplacian matrix defined in Subsection \ref{subsection 1.1}, 
as $L=\begin{bmatrix}L_{11}&L_{12}\\L_{21}&L_{22} \end{bmatrix}$, $L_{11}=0$, $L_{21}=0$, where $L_{21}$ is a $N\times 1$ vector and represents the  leader's links to the followers and $L_{22}$ is an $N\times N$ matrix and specifies the followers' connections. This will help us  to discuss the effects of the  leader agent and follower agents to reach the entire team objectives.}
\subsection{The Types and Description of the Actuator  Faults} 
Before formally defining the \underline{three} fault types that are considered in this work, we let $B=\begin{bmatrix}b^1,b^2,\dots,b^{m}\end{bmatrix}$ denote the matrix of  input channels of the healthy agent, where $b^k$ denotes the $k$-th column of the matrix $B$, $B^{fk}$  denote the matrix of the  faulty agent with a fault in only the $k$-th input channel,  and $B^f$ denote the matrix of the  faulty agent subject to several concurrent faulty channels. \\
\underline{\emph{Loss of Effectiveness (LOE) Fault}}: 
For the LOE fault, only a percentage of the generated control effort is available to the agent for actuation, therefore the dynamics of the $i$-th faulty agent after the occurrence of a fault at $t=t_f$ is modelled according to 
\begin{eqnarray}
&&\dot x_i^f(t)=A x_i^f(t)+B^f u_i(t)+B_\omega \omega_i(t),\ t\geq t_f,\ i=1,\dots,N, \label{faulty system}\\
&&y_i^f(t)=Cx_i^f(t),\nonumber
\end{eqnarray}
where $x_i^f(t)\in\mathbb{R}^n$ denotes the state of the faulty agent, $B^f=B \Gamma_i$, $ \Gamma_i=\operatorname{diag} \{ \Gamma_i^k \}$, for $k=1,\dots,m$, $\Gamma_i^k$ represents the fault effectiveness  of the $k$-th channel of the $i$-th agent, $0<\Gamma_i^k<1$ if the $k$-th actuator is faulty, and $\Gamma_i^k=1$ if it is healthy. \par
\noindent \underline{\emph{Outage Fault}}: If the $k$-th actuator of the $i$-th agent is completely  lost at the time $t=t_f$, then we have $u_i^k(t)=0$ for $t\geq t_f$, where $u_i(t)=\begin{bmatrix}u_i^1(t),\dots,u_i^m(t)\end{bmatrix}^\text{T}$. The dynamics of the $i$-th agent with an outage fault in its $k$-th actuator can be represented by 
\begin{eqnarray}
&&\dot x_i^f(t)=A x_i^f(t)+B^{fk} u_i(t)+B_\omega \omega_i(t),\ t\geq t_f,\ i=1,\dots,N,\label{otage}\\
&&y_i^f(t)=Cx_i^f(t)\nonumber,
\end{eqnarray} 
where  $B^{fk}=\begin{bmatrix}b^1,b^2,\dots,b^{k-1},0,b^{k+1},\dots,b^{m}\end{bmatrix}$.\par
\noindent \underline{\emph{Stuck Fault}}: If at the time $t=t_f$ the $k$-th actuator of the $i$-th agent  freezes at a certain value and does not respond to subsequent commands, the fault is then designated as the stuck fault.  The dynamics of the $i$-th faulty agent under this  fault type can be modelled as
\begin{eqnarray}
&&\dot x_i^f(t)=A x_i^f(t)+B u_i^f (t)+B_\omega \omega_i(t),\ t\geq t_f,\ i=1,\dots,N, \label{LIP}\\
&&y_i^f(t)=Cx_i^f(t),\nonumber
\end{eqnarray} 
where $u_i^f(t)=\begin{bmatrix}u_i^1(t),\dots,u_i^{k-1}(t),\underline{u}_i^k,u_i^{k+1}(t),\dots,u_i^{m}(t)\end{bmatrix}^{\text{T}}$, and $\underline{u}_i^k=u_i^k(t_f)$ for all $t\geq t_f$ denotes the value of the stuck command. \par
We are now in a position to state the following assumptions. 
\begin{assumptions}\label{network structure}
(a) The network graph is directed and  has a spanning tree, and 
(b) The leader control input is bounded and the upper bound is known.
 \end{assumptions}
 \begin{assumptions}\label{assumption 2}
(a) The agents are stabilizable and remain stabilizable even after the fault occurrence.\\
(b) Each agent is equipped with a local FDI module which detects with possible delays and correctly isolates the fault in the agent and also estimates the severity of the fault with possible errors in the case of the LOE or stuck faults. 
\end{assumptions}
Regarding the above assumptions the following clarifications are in order. First, the Assumptions \ref{network structure}-(a) and  \ref{assumption 2}-(a)  are  quite common   for consensus achievement and fault recovery control design problems, respectively.   
Second, it is quite necessary that in most
practical applications one considers a leader whose states
are ensured to be bounded. Moreover, in practical scenarios
the actuators are quite well understood and described
and their maximum deliverable control effort and bound they can tolerate are readily available and known. Therefore Assumptions \ref{network structure}-(b) is also not  restrictive. Furthermore, in Subsection \ref{subsection 3} we analyze the system behavior for situations where either Assumption \ref{assumption 2}-(c) does not hold or 
  the estimated fault severities by the FDI module are not accurate. We obtain the maximum uncertainty bound that our proposed approaches can tolerate.  However, as stated in Assumption \ref{assumption 2}-(b),  we require the correct actuator location  as well as the type of the fault for guaranteeing that our proposed reconfigured control laws will yield the desired design specifications and requirements. The Scenario $4$ in Section \ref{simulation results} does demonstrate the consequences of violating this assumption.
 
 As far as Assumption 2-(c) is concerned, it should be
noted that this assumption is indeed quite realistic for
the following observations and justications. The transient
time that any cooperative or consensus-based controller
takes to settle down and the overall team objectives
are satisfied is among one of the design consideration
and specification for the controller selection. In most
practical consensus achievement scenarios dealing with a healthy team, the transient time associated with the
agent response is ensured to be settled down in a very
small fraction of the entire mission time, and in most
cases the healthy transient time takes a few seconds to
minutes to die out. Therefore, it is quite realistic and
indeed practical that during this very short and initial
operation of the system, the agents are assumed to be
fault free. In other words, we will not initiate the mission
with agents that are faulty from the outset. It is highly
unlikely that during the very first few moments after the
initiation of the mission a fault occurs in the agents. For
all the above explanations and observations we believe
that Assumption 2-(c) is meaningful and quite realistic.
\subsection{Notations and Preliminaries}
For a vector $x=\left[x_1,\dots,x_n\right]^{\text{T}}$ we define  $L^1$, $L^2$ (Euclidean norm ) and $L^\infty$ norm as $\|x\|_1=\sum_{i=1}^n|x_i|$, $\|x\|_2=\sqrt {x_1^2+\dots+x_n^2}$, $\|x\|_{\infty}=\max (|x_1|,\dots,|x_n|)$. The signal $x(t)$ is also represented as $x(t)=\text{col}\{x_i(t)\}$. The function $\text{sgn}\{x(t)\}$ is defined as
\begin{eqnarray}
\text{sgn}\{x(t)\}=\begin{bmatrix}\text{sgn}\{x_1(t)\},\dots,\text{sgn}\{x_n(t)\}\end{bmatrix}^{\text{T}},\ \text{sgn}\{x_i(t)\}=\left \{ \begin{array}{rcl}0&x_i(t)=0\\ \frac{x_i(t)}{|x_i(t)|}& x_i(t)\ne 0 \end{array}\right.\label{sgn}.
\end{eqnarray}
 For the vector $x$ the notation $\text{diag}\{x\}$ denotes a diagonal matrix that has diagonal entries $x_i$'s. The notations $I_n$, $1_n$ and $0_{n\times m}$ denote  an identity matrix of dimension $n\times n$, a unity $n\times 1$ vector with all its entries as one, and a zero matrix of dimension $n \times m$, respectively. For a matrix $X\in \mathbb{R}^{n\times n}$, the notation $X>0$ ($X\leq 0$) or $X<0$ ($X\leq 0$) implies that $X$ is a positive definite (positive semi-definite) or a negative definite (negative semi-definite) matrix. For a matrix $A\in\mathbb{R}^{m\times n}$, its $2$-norm is defined by 
 $$\|A\|_2=\left\{\sup\frac{\|Ax\|_2}{\|x\|_2}:x\in\mathbb{R}^n,\ x\ne0\right\}.$$
  The term $X^{-L}$ ($X^{-R}$) denotes the generalized left (right) inverse of the matrix $X$. The terms $\lambda_{i}(X)$, $\lambda_{\text{min}}(X)$ and $\lambda_{\text{max}}(X)$ denote the $i$-th  eigenvalue, the smallest, and the largest eigenvalues of the matrix $X$, respectively. For the matrix $X$, $\sigma_i(X)$, $\sigma_{\min}(X)$, $\sigma_{\max}(X)$,  denote the $i$-th  singular value, the minimum singular value, and the largest singular value of $X$. 
The notations $\text{Im}\{X\}$ and $\text{Ker}\{X\}$  denote the image and the kernel of $X$. 
\begin{theorems}\label{theorem 0.3}\cite{Yedavalli14}
Consider the system 
\begin{equation}
\dot x(t)=Ax(t)+f(x(t),t),\label{eq 0.10}
\end{equation}
where $A$ is Hurwitz stable and $x(t)\in\mathbb{R}^n$ is the state vector. The system (\ref{eq 0.10}) is stable if 
$$\frac{\|f(x(t),t)\|_2}{\|x(t)\|_2}<\frac{1}{\sigma_{\text{max}}(P)},$$
for all $x(t)\in\mathbb{R}^n$ and $t>0$, where $P$ is the solution to 
$$PA+A^{\text{T}}P+2I=0.$$
\end{theorems}
\begin{facts}\label{Fact1}
For any two matrices $X$ and $Y$ and a positive scaler $\alpha$ we have 
$$X^{\text{T}}Y+Y^{\text{T}}X\leq \alpha X^{\text{T}}X+\alpha^{-1}Y^{\text{T}}Y.$$
\end{facts}
 \subsection{Problem Definition}\label{subsection 2-E}
 In this work, our main goal and objective is to design a state feedback reconfigurable  or recovery control strategy in a directed network of multi-agent systems that seek consensus in presence of three types of actuator faults and environmental disturbances. 
 Suppose the $i$-th agent becomes faulty and its first $m_{o}$ actuators are subject to the outage fault, $m_{o}+1$ to $m_{s}$ actuators are subject to the stuck fault, while the remaining $m-m_{s}$ actuators are either subject to the LOE fault or are healthy. Using equations (\ref{faulty system})-(\ref{LIP}) the model of $i$-th faulty agent that is subject to three types of actuator faults can be expressed as
\begin{eqnarray}
&&\dot x_i^f(t)=A x_i^f(t)+B_i^{f} u_i^f(t)+B_{\omega}\omega_i(t),\ x_i^f(t_f)=x_i(t_f),\  t\geq t_f,\label{faulty system pre.}\\ 
&&y_i^f(t)=Cx_i^f(t),\nonumber
\end{eqnarray} 
where  
$B_i^{f}=\begin{bmatrix}B_i^o&B_{i}^s&B_i^r\end{bmatrix}$, $B_{i}^o=\begin{bmatrix}b^{1},\dots,b^{m_{o}}\end{bmatrix}$, $B_{i}^s=\begin{bmatrix}b^{m_{o}+1},\dots,b^{m_{s}}\end{bmatrix}$, 
$B_{i}^r=\begin{bmatrix}b^{m_{s}+1},\dots,b^{m}\end{bmatrix}\Gamma_i$,  $\Gamma_i=\text{diag}\{\Gamma_i^k\}$, $k=m_{s}+1,\dots,{m}$, $\Gamma_i^{k}$ denotes the $k$-th actuator effectiveness and fault severity factor, $u_i^f(t)=\begin{bmatrix}0_{1\times m_o}&(\underline u_i^{s})^\text{T}&(u_i^r(t))^\text{T}\end{bmatrix}^\text{T}$, $\underline u_i^s=\begin{bmatrix}u_i^{m_{o}+1}(t_f),\dots,u_i^{m_{s}}(t_f)\end{bmatrix}^{\text{T}}$, $u_i^r(t)=\begin{bmatrix} u_i^{m_{s}+1}(t),\dots,u_i^{m}(t)\end{bmatrix}^{\text{T}}$. \par
Considering the structure of the control law $u_i^f(t)$ and the matrix $B_i^{f}$, it follows that only  the actuators $m_s+1$ to $m$ are available to be reconfigured. Therefore, to proceed with our proposed control recovery strategy the model (\ref{faulty system pre.}) is rewritten as follows 
\begin{eqnarray}
&&\dot x_i^f(t)=A x_i^f(t)+B_i^r u_i^r(t)+ B_i^s \underline u_i^{s}+ B_{\omega}\omega_i(t),\ x_i^f(t_f)=x_i(t_f), \  t\geq t_f, \label{faulty system 2}\\ 
&&y_i^f(t)=Cx_i^f(t).\nonumber
\end{eqnarray} \par
The main objective of the control reconfiguration or control recovery is to design and select   $u_i^r(t)$  such that the state consensus errors either remain bounded 
and $y_i^f(t)=y_i^h(t)$,  for $t\geq t_f$, when $\omega_i(t)\equiv 0$, $i=0,\dots,N$, and the environmental disturbances are  attenuated for $\omega_i(t)\ne 0$, where $y_i^h(t)=y_i(t)$, $i=1,\dots,N$, and $y_i(t)$ is defined as in equation (\ref{LL or F}).\par 
 To develop our proposed reconfiguration control laws, a \textbf{\underline{virtual auxiliary system}} associated with each agent is now introduced as follows
 \begin{eqnarray}
 &&\dot x_i^a(t)=A x_i^a(t)+B u_i^a(t), \ x_i^a(t_0)=x_{i0}^a, \ i=1,\dots,N,\label{aux LF dynamics}\\
 &&y_i^a(t)=Cx_i^a(t),\nonumber
 \end{eqnarray}
where $x_i^a(t)\in\mathbb{R}^n$, $u_i^a(t)\in\mathbb{R}^m$ and $y_i^a(t)\in\mathbb{R}^q$ denote the state of the auxiliary system corresponding to the $i$-th agent,  its control and output signals, respectively.  
Furthermore, the disagreement error for each auxiliary system is also defined as
 \begin{equation}
 e_i^a(t)=\sum_{j\in\mathcal{N}_i}(x_i^a(t)-x_j^a(t))+g_{i0}(x_i^a(t)-x_0(t)).\label{auxiliary consensus}
 \end{equation}
 The auxiliary system that is defined in (\ref{aux LF dynamics}) is ``virtual" and is not subject to actuator faults or disturbances, and hence it can be used as the {reference model} for designing the reconfigured  control laws of the actual  system (\ref{LL or F}) once it is subjected to actuator faults.\par 
 The  $H_{\infty}$ performance index corresponding to the $i$-th healthy agent (\ref{LL or F}) and the $i$-th faulty agent (\ref{faulty system 2}) is now defined according to 
 \begin{eqnarray}
 J_i&=&\int_{t_0}^{\infty}\big((x_i(t)-x_0(t))^{\text{T}}(x_i(t)-x_0(t))-\gamma^2(\omega_i^{\text{T}}(t)\omega_i(t)+\omega_0^{\text{T}}(t)\omega_0(t))\big)\text{d}t,\label{agent cost index}\\
 J_i^f&=&\int_{t_f}^{\infty}(\xi_i^{f^\text{T}}(t)\xi_i^f(t)-\gamma_f^2\omega_i^{\text{T}}(t)\omega_i(t))\text{d}t,\label{auxiliary cost index}
 \end{eqnarray}
 where 
$  \xi_i^f(t)=x_i^f(t)-x_i^a(t)$, 
 and $\gamma$ and $\gamma_f$ represent the disturbance attenuation bounds. Based on the above definitions, the team performance index is now defined by $J=\sum_{i=1}^NJ_i$. Under the control laws $u_i(t)$, $i=1,\dots,N$, the $H_{\infty}$ performance index bound for the healthy team is attenuated if $J=\sum_{i=1}^NJ_i\leq 0$, $\forall\ \omega_i\in \mathcal{L}_2$.  Furthermore, the $H_{\infty}$ performance index for the $i$-th faulty agent is attenuated if $J_i^f\leq 0$, $\forall\ \omega_i\in \mathcal{L}_2$, $i=0,\dots,N$. {It should be noted that
 the performance indices (\ref{agent cost index}) and (\ref{auxiliary cost index}) are not and cannot be calculated directly as the disturbance is unknown and the aim of the proposed approach is to minimize the performance indices without directly calculating them.}
 
We are now in a position to formally state  the problem that we consider in this work.
 \begin{definitions}\label{def. 1}
 (a) The   state consensus $H_{\infty}$ performance control problem for the  healthy  team is solved if in  absence of  disturbances, the agents   follow the leader states and consensus errors converge to zero asymptotically, and in  presence of disturbances, the prescribed  $H_{\infty}$ performance bound for the healthy team is attenuated, i.e. $J=\sum_{i=1}^NJ_i\leq 0$.\\ 
 (b) Under Assumptions \ref{network structure} and \ref{assumption 2}, the $H_{\infty}$ performance  control reconfiguration problem with stability is solved if in absence of  disturbances the state consensus errors remain bounded  
 while the output of the faulty agent behaves the same as those of the healthy system outputs, and in presence of disturbances the disturbance attenuation bound is minimized and $J_i^f\leq 0$.
 \end{definitions}
\section{$H_{\infty}$ Performance Cooperative and Distributed  Control Reconfiguration Strategy}\label{proposed methodology}
In this section, our proposed  reconfigurable control law is introduced and developed.  Since each agent only shares its information with its nearest neighbors,  the reconfiguration control strategy also employs the same information as well as the agent's FDI module information. \par
Consider the dynamics of the $i$-th faulty agent is given by (\ref{faulty system 2}). As defined above  $\xi_i^f(t)=x_i^f(t)-x_i^a(t)$,  with $x_i^f(t)$ denoting the $i$-th faulty agent  state and $x_i^a(t)$  defined in (\ref{aux LF dynamics}),  we let $z_i(t)=C\xi_i^f(t)$ to  denote the deviation of the output of the faulty agent from its associated auxiliary agent output. Then, the dynamics associated with $\xi_i^f(t)$ can be obtained as
\begin{eqnarray}
&&\dot \xi_i^f(t)=A \xi_i^f(t)+B_i^r u_i^r(t)+B_i^s\underline u_i^s-Bu_i^a(t)+B_{\omega}\omega_i(t),\ t\geq t_f, \nonumber \\ 
&&z_i(t)=C\xi_i^f(t).\label{eq. 1.11}
\end{eqnarray}
Moreover, the faulty agent consensus error is defined as 
\begin{equation}
e_i^f(t)=\sum_{j\in\mathcal{N}_i}(x_i^f(t)-x_j(t))+g_{i0}(x_i^f(t)-x_0(t)).\label{cons. er}
\end{equation}
\begin{lemmas}\label{Lemma 1}
The faulty agent consensus error (\ref{cons. er}) is stable 
if $e_i^a(t)$ and $\xi_i(t)=x_i(t)-x_i^a(t)$ are asymptotically stable and $\xi_i^f(t)$ is stabilized.
\end{lemmas}
\proof
From the auxiliary error dynamics  (\ref{eq. 1.11}), one can express the state consensus error dynamics for the $i$-th faulty agent that is denoted by $e_i^f(t)$ according to
\begin{eqnarray*}
e_i^f(t)&=&\sum_{j\in \mathcal{N}_i}(x_i^f(t)-x_j(t))+g_{i0}(x_i^f(t)-x_0(t))\\
&=&e_i^a(t)+(d_i+g_{i0})\xi_i^f(t)-\sum_{j\in \mathcal{N}_i}\xi_j(t),
\end{eqnarray*}
Therefore if the control law $u_i^r(t)$ can be reconfigured such that $\xi_i^f(t)$ is stabilized then it follows that $e_i^f(t)$ will be stable. This completes the proof of the lemma.   \qed
\par
The above lemma shows that stability of the faulty agent's consensus error can be guaranteed by  reconfiguring the control law $u_i^r(t)$ such that  $\xi_i^f(t)$ is stable. This implies that  one can transform the control reconfiguration problem to that of the stabilization problem. Consequently, in the next two subsections we consider the problem of stabilizing $\xi_i^f(t)$. However, as  seen from (\ref{eq. 1.11}), the dynamics of $\xi_i^f$ depends on the control of the healthy agents. Hence, before presenting our proposed control reconfiguration strategy, the control law for   the healthy team (where it is assumed without loss of any generality that all the agents are healthy) is presented below.
 \par
In this work, the following general  control law structure is utilized,
\begin{eqnarray}
u_i(t)=K_{1i}\xi_i(t)+K_{2i}e_i^a(t)+c_{i0}\text{sgn}(Ke_i^a(t)),\label{proposed healthy control law}
\end{eqnarray}
which is the generalization of the one  developed in \cite{Li13} and is given by 
\begin{eqnarray}
u_i(t)=c_1Ke_i(t)+c_2\text{sgn}(Ke_i(t)),\label{Li control law}
\end{eqnarray}
where $\xi_i(t)=x_i(t)-x_i^a(t)$, and  $e_i^a(t)$ and $e_i(t)$ are given by (\ref{auxiliary consensus}) and (\ref{agents consensus error}), respectively.

\begin{remarks}
The main challenge in developing the reconfigurable control law in multi-agent system as compared to that in single agent is that  in single agent control recovery the agent is redesigned its control law to maintain its stability. However, in multi-agent system the agent should redesign its control law such that the entire team remains stable and loosing one agent can cause a disconnected network and failing the entire mission. The main difficulty  in the design which is not the case in single agent is that   each agent only share information with its nearest neighbours and communication channels are limited, so that the design should be performed using only local information.  
\end{remarks}

The followings comments summarize the main characteristics of  the control law (\ref{proposed healthy control law}) : 

\textbf{(1)} In the control law (\ref{proposed healthy control law}) an agent employs and communicates \underline{only} the auxiliary states $x_i^a(t)$ that are   unaffected by  both disturbances and  faults. 
In contrast  in standard consensus control schemes such as  (\ref{Li control law})  the actual states $x_i(t)$ are employed and communicated from the nearest neighbor agents.  
 Hence, the utilization of  (\ref{proposed healthy control law})   avoids the propagation of the adverse effects of the disturbances and faults  through out the team of multi-agent systems. This along with the degrees of freedom in designing the control recovery laws  allow us to manage the $i$-th faulty agent by {only} reconfiguring the control law of the faulty agent, and moreover it also provides us with the capability to recover simultaneous faults in multiple agents. 
 
  \textbf{(2)} The gain $K_{1i}$ is designed such that the states of the $i$-th agent follow the states of its associated auxiliary agent,  while the gain $K_{2i}$ is designed such that the states of the auxiliary agents reach a consensus and follow the leader state. 
\footnote{The states $x_i^a(t),\ i=1,\dots,N$ are virtual; however, since $u_i^a(t)$ depends on the leader state, $x_i^a(t)$ also depends on the leader state  (which is  available to only a very few follower agents in the network). Therefore, $x_i^a(t)$ should be communicated between the neighboring agents.}  
\par
\textbf{(3)} Each agent receives only the auxiliary agents states in its nearest neighbor set as opposed to  their actual states 
 that is conventionally required in standard multi-agent consensus approaches. 
 
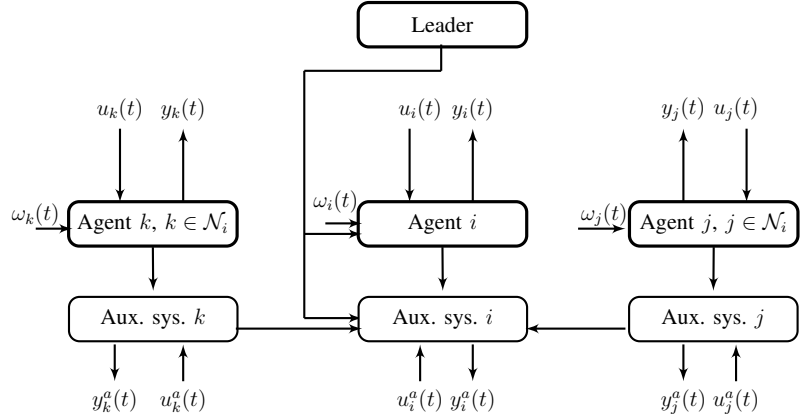
\begin{figure}
	\centering
	{\begin{tikzpicture}[thick,scale=0.7, every node/.style={transform shape}]
		[node distance=1cm, auto]  
		\tikzset{
			mynode/.style={rectangle,rounded corners,draw=black, top color=white, bottom color=white,very thick, minimum height=2em,  minimum width=3.2cm, text centered},
			mynode1/.style={rectangle,rounded corners,draw=black, top color=white, bottom color=white, minimum height=2em,  minimum width=3.2cm, text centered},		
			myarrow/.style={->, >=latex', shorten >=2pt, thick},	
			myarrow2/.style={<-, >=latex', shorten >=2pt, thick},
			myarrow3/.style={->, >=latex', shorten >=2pt, dashed},		
			myarrow4/.style={<-, >=latex', shorten >=2pt, dashed},
			myarrow5/.style={<-, >=latex', shorten >=2pt, dashed},			
			mylabel/.style={text width=7em, text centered} 
		}  
		\node at (0,-0.2) (leader) {};  
		\node[ below=2.5cm of leader] (cen1) {}; 
		\node[mynode, left=6.3cm of cen1] (agentk) {Agent $k$, $k\in\mathcal{N}_i$ }; 
		\node[mynode, left=0.8cm of cen1] (agent2) {Agent $i$};
		\node[mynode, right=0.8cm of cen1]  (agent3) {Agent $j$, $j\in\mathcal{N}_i$};
		\node[ mynode,  above=2.9cm of agent2] (Leader) {Leader}; 
            \draw  (-5.2,-0.1)  --  (-2.6,-0.1);
            \draw  (-2.6,-0.1)  --  (-2.6,0.4);            
            \draw  (-5.2,-0.1)  --  (-5.2,-4.8);
            \draw[myarrow]  (-5.2,-3.2)  --  (-4,-3.2);
            \draw[myarrow]  (-5.2,-4.8)  --  (-4,-4.8);
		\draw[myarrow]  (-8.7,-1.2)  --  (-8.7,-2.7);
		\node[above] at (-8.7,-1.2)	{$u_k(t)$ };	
		\draw[myarrow2]  (-7.5,-1.2)  --  (-7.5,-2.7);
		\node[above] at (-7.5,-1.2)	{$y_k(t)$ };	
		\draw[myarrow]  (-10.3,-3.1)  --  (-9.5,-3.1);
		\node[above] at (-10.3,-3.2)	{$\omega_k(t)$ };	
		
		\draw[myarrow]  (-3.2,-1.2)  --  (-3.2,-2.7);
		\node[above] at (-3,-1.2)	{$u_i(t)$ };	
		\draw[myarrow2]  (-2,-1.2)  --  (-2,-2.7);
		\node[above] at (-2,-1.2)	{$y_i(t)$ };	
		\draw[myarrow]  (-4.8,-3.0)  --  (-4.0,-3.0);
		\node[above] at (-4.6,-2.95)	{$\omega_i(t)$ };

		\draw[myarrow]  (3.2,-1.2)  --  (3.2,-2.7);
		\node[above] at (3,-1.2)	{$u_j(t)$ };	
		\draw[myarrow2]  (2,-1.2)  --  (2,-2.7);
		\node[above] at (2,-1.2)	{$y_j(t)$ };	
		\draw[myarrow]  (0,-3.1)  --  (1,-3.1);
		\node[above] at (0.5,-3.2)	{$\omega_j(t)$ };

		\node[ below=1.5cm of leader] (cen2) {};				
		\node[ below=1.5cm of cen1] (cen3) {}; 
		\node[mynode1, left=6.3cm of cen3] (auxk) {Aux. sys. $k$}; 
		\node[mynode1, left=0.8cm of cen3] (aux2) {Aux. sys. $i$};
		\node[mynode1, right=0.8cm of cen3]  (aux3) {Aux. sys. $j$};
		\draw[myarrow] (agentk.south)  --  (auxk.north);
		\draw[myarrow] (agent2.south)  --  (aux2.north);
		\draw[myarrow] (agent3.south)  --  (aux3.north);

		\draw [myarrow2] (-1,-5.0) -- (+1,-5.0);
		\node[above] at (0,-4.9)	{};
		\draw [myarrow] (-6.5,-5.0) -- (-4,-5.0);
		\node[above] at (-4.6,-4.95)	{ };		
		
		\draw[myarrow2] (-8.8,-6.0)  --  (-8.8,-5.2);
		\node[below] at (-8.8,-6.0)	{$y_k^a(t)$ };
		\draw[myarrow]  (-7.5,-6.0)  --  (-7.5,-5.2);
		\node[below] at (-7.5,-6.0)	{$u_k^a(t)$ };			
		
		\draw[myarrow2] (-2,-6.0)  --  (-2,-5.2);
		\node[below] at (-2,-6.0)	{$y_i^a(t)$ };
		\draw[myarrow]  (-3,-6.0)  --  (-3,-5.2);
		\node[below] at (-3,-6.0)	{$u_i^a(t)$ };
		
		\draw[myarrow2] (2,-6.0)  --  (2,-5.2);
		\node[below] at (2,-6.0)	{$y_j^a(t)$ };
		\draw[myarrow]  (3,-6.0)  --  (3,-5.2);
		\node[below] at (3,-6.0)	{$u_j^a(t)$ };
		
		\end{tikzpicture}} 
	\caption{The schematic of the $i$-th pinned agent and its nearest neighbor agents $j$ and $k$, which are not pinned.
	} 
	\label{network schematic}
\end{figure}

\begin{figure}
	\centering
	{\begin{tikzpicture}[thick,scale=0.6, every node/.style={transform shape}]
		[node distance=1cm, auto]  
		\tikzset{
			mynode/.style={rectangle,rounded corners,draw=black, top color=white, bottom color=white,very thick, minimum height=2em,  minimum width=3.2cm, text centered},
			mynode1/.style={rectangle,rounded corners,draw=black, top color=white, bottom color=white, minimum height=2.5cm,  minimum width=0.5cm, text centered},	
			mynode2/.style={circle,draw=black, very thick, minimum size=0.9cm, text centered},	
			mynode3/.style={rectangle,rounded corners,draw=black, top color=white, bottom color=white, minimum height=1cm,  minimum width=1cm, text centered},						
			myarrow/.style={->, >=latex', shorten >=2pt, thick},	
			myarrow2/.style={<-, >=latex', shorten >=2pt, thick},
			myarrow3/.style={->, >=latex', shorten >=2pt, dashed},		
			myarrow4/.style={<-, >=latex', shorten >=2pt, dashed},
			mylabel/.style={text width=7em, text centered} 
		}  
		
		\draw[myarrow2] (-0.2,0.7)  --  (-0.9,0.7);	
		\draw[myarrow2] (-1.5,0.7)  --  (-3,0.7);	
		\node[left] at (-3.1,0.7)	(label xi) {  $x_i^a(t)$ };
		\node[mynode3, right=1.25cm of label xi] (gain2){$d_i$};				
		\draw[myarrow2] (-0.2,0)  --  (-3,0);
		\node[left] at (-3.1,0)	{$\sum_{j\in\mathcal{N}_i}x_j^a(t)$};			
		\draw[myarrow2] (-0.2,-0.7)  --  (-3,-0.7);
		\node[left] at (-3.1,-0.7)	{$g_{i0}(x_i^a(t)-x_0(t))$ };		
		\node[mynode1] (sum1) {};  
		\node[ left=-0.3cm of sum1] (cen1) {}; 
		\node[mylabel, above= 0.3cm of cen1] (label1) { $+$};
		\node[mylabel, above= -0.5cm of cen1] (label2) { $-$};		 
		\node[mylabel, above= -1.1cm of cen1] (label2) { $+$};	
		
		\draw[myarrow] (sum1)  --  (2.0,0);
		\node[above] at (1.0,0)	{$e_i^a(t)$ };
		
		\node[mynode1, right=1.8cm of cen1] (aux2) (aux cont){$K_{2i}e_i^a(t)+K_{i0}(t)$};	
		\draw[myarrow] (aux cont)  --  (6.4,0);
		\node[above] at (5.7,0)	{$u_i^a(t)$ };
		\node[above] at (7.1,0.5)	{$+$ };	
		\node[above] at (6.0,-0.65)	{$+$ };	
		
		\node[mynode2, right=6.2cm of cen1] (sum2){$\sum$};	
		\draw[myarrow] (sum2)  --  (8.5,0);
		\node[right] at (8.5,0)	{$u_i(t)$ };
		
		\draw[-] (-2.2,0.7)  --  (-2.2,4.0);
		
		\draw[myarrow] (-2.2,4)  --  (6.4,4);		
		\node[mynode2, above=3.15cm of sum2] (sum3){$\sum$};
		\node[mynode3, below=0.9cm of sum3] (con2){$K_{1i}$};	
		\draw[myarrow] (sum3)  --  (con2);
		\draw[myarrow] (con2)  --  (sum2);			
		\draw[myarrow] (6.7,6)  --  (sum3);
		\node[above] at (6.7,6)	{$x_i(t)$ };	
		\node[right] at (6.7,5.1)	{$+$ };	
		\node[right] at (5.5,4.3)	{$-$ };		
		\node[right] at (6.8,3.4)	{$\xi_i(t)$ };							
		\end{tikzpicture}} 
	\caption{The $i$-th agent cooperative control structure and its associated auxiliary system control laws, where  $\xi_i(t)=x_i(t)-x_i^a(t)$ and $e_i^a(t)$ is defined in (\ref{auxiliary consensus}).} 
	\label{ith agent control schematic}
\end{figure}
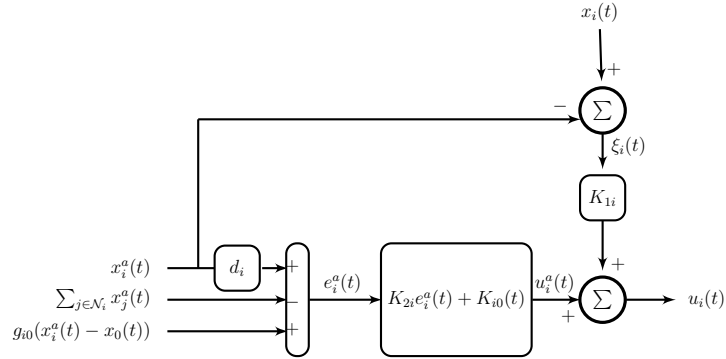

 \textbf{(4)} The control law (\ref{proposed healthy control law}) is shown subsequently to solve the consensus problem in a \underline{directed} network topology that is {subject to environmental disturbances}, whereas the control law (\ref{Li control law}) solves the consensus problem in {disturbance free} environment and where the network topology is assumed to be \underline{undirected}. The procedure for selecting and designing the gains of the control law (\ref{proposed healthy control law}) is provided in Theorem \ref{theorem 1}.  Moreover, 
 the structure of the proposed control law of this agent are provided in Figures \ref{network schematic} and \ref{ith agent control schematic}. 
\begin{theorems}\label{theorem 1}
	The control law 
	$u_i(t)=u_{it}(t)+u_{ic}(t)$
	 solves the $H_{\infty}$ performance state consensus problem in a team of $N$ follower agents whose dynamics are given by (\ref{LL or F}) and the leader dynamics that is given by (\ref{L1}), if $u_{it}(t)$ and $u_{ic}(t)$ are selected as follows:
	\begin{eqnarray*}
	&&u_{it}(t)=K_{1i}\xi_i(t)\\
	&&u_{ic}(t)=u_i^a(t)=K_{2i}e_i^a(t)+K_{i0}(t),
	\end{eqnarray*}
	 where $e_i^a(t)$ is  defined as in (\ref{auxiliary consensus}), 
	$K_{1i}=c_1K$, $K_{2i}=c_{2i}K$, $K_{i0}(t)=c_{i0}\text{sgn}(Ke_i^a(t))$, $\text{sgn}\{.\}$ is  defined as in (\ref{sgn}), $K=-B^{\text{T}}P$,   $c_1=\frac{c_3}{2}$, and finally the positive definite matrix $P$ is the solution to
	\begin{eqnarray*}
		&&A^{\text{T}}P+PA- c_3PBB^{\text{T}}P+2\gamma^{-2}c_4^{-1}PB_{\omega}B_{\omega}^{\text{T}}P+d_0^*I<0, 
	\end{eqnarray*}
	and $c_{2i}$ and $c_3$ are solutions to
	$$C_2L_{22}^{\text{T}}+L_{22}C_2>c_3I,\ c_3>0,\ C_2=\text{diag}\{c_{2i}\}>0,$$
	where $d_0^*$ denotes the number of pinned agents, $\gamma^2$ is the desired disturbance attenuation bound,  $c_4^{-1}=\max\{1,N^{-1}\lambda_{\text{min}}^{-1}(L_{22}^{\text{T}}L_{22})\}$,   
	and $ c_{i0}$'s are the solutions to the inequalities
	$$
	u_{0M}-d_ic_{i0}+\sum_{j\in\mathcal{N}_i}c_{j0} < 0, \ c_{i0}>0,\ i,j=1,\dots,N,
	$$
	where $u_{0M}$ denotes the upper bound of the leader control signal, i.e., $\|u_0(t)\|_{\infty}\leq u_{0M}$ for all $t\geq t_0$.
\end{theorems}
\proof
The team reaches a consensus if $x_i(t)\to x_j(t)\to x_0(t)$. This goal is also achieved if agents' controls are designed such that $x_i(t)\to x_i^a(t)$ ($\xi_i(t)\to 0$) and $x_i^a(t)\to x_0(t)$ ($e_i^a(t)\to 0$) for $i=1,\dots,N$. 
This implies that the consensus achievement problem can be re-stated as  the problem of asymptotically stabilizing  
$\xi_i(t)$ and $e_i^a(t)$ simultaneously. \par
In the following, first we discuss the stability criterion and disturbances attenuation for  $e_i^a(t)$ and $\xi_i(t)$ in Parts A and B, respectively and  then in Part C, we derive the conditions that  satisfy the requirements for both Parts A and B that in fact solve the $H_{\infty}$ performance state consensus.\par
\underline{\textbf{Part A}}: From  (\ref{aux LF dynamics}) and (\ref{auxiliary consensus}), the dynamics of $e^a(t)=\text{col}\{e_i^a(t)\}$ can be obtained as 
\begin{eqnarray}
\dot e^a(t)&=&\mathcal{A}e^a(t)+\mathcal{B}u^a(t)+\mathcal{B}_0u_0(t)+\mathcal{B}_{\omega}{\omega}_0(t),\label{eq. 2.1}
\end{eqnarray}
where  $u^a(t)=\text{col}\{u_i^a(t)\}$, $\mathcal{A}=I_N\otimes A$, $\mathcal{B}=L_{22}\otimes B$, $\mathcal{B}_0={L}_{21}\otimes B$, $\mathcal{B}_{\omega}={L}_{21}\otimes B_{\omega}$. Let us select $u_i^a(t)$  as $u_i^a(t)=K_{2i}e_i^a(t)+c_{i0}\text{sgn}(Ke_i^a(t))$, then the system (\ref{eq. 2.1}) becomes
\begin{eqnarray}
\dot e^a(t)&=&(\mathcal{A}+L_{22}C_2\otimes BK) e^a(t)+(L_{22}C_0\otimes B)\text{sgn}((I\otimes K)e^a(t))+\mathcal{B}_0u_0(t)+\mathcal{B}_{\omega}{\omega}_0(t),\nonumber\\
\label{augmented error auxiliary}
\end{eqnarray}
where $C_2=\text{diag}\{c_{2i}\}$ and $C_0=\text{diag}\{c_{i0}\}$. Since the $\text{sgn}$ function is discontinuous, in order to conduct  the stability analysis of the system  (\ref{augmented error auxiliary}), it is replaced with its differential inclusion (for more details refer to \cite{Shevitz94,Bacciotti99}) representation as follows 
\begin{eqnarray}
\dot e^a(t)&\in^{a.e.}&\mathcal{K}[(\mathcal{A}+L_{22}C_2\otimes BK) e^a(t)+(L_{22}C_0\otimes B)\text{sgn}((I\otimes K)e^a(t))+\mathcal{B}_0u_0(t)+\mathcal{B}_{\omega}{\omega}_0(t)],\nonumber\\
\label{augmented error auxiliary 1}
\end{eqnarray}
where the operator $\mathcal{K}[.]$ is defined as in \cite{Shevitz94,Bacciotti99} to investigate its Filipov solutions. Now,  we require to define the Lyapunov function candidate $V(e^a(t))$ to study the stability properties of the error dynamics system. 
For this purpose, let us select $V(e^a(t))=e^{a^{\text{T}}}(t)\mathcal{P}e^a(t)$, as a Lyapunov function candidate for the system (\ref{augmented error auxiliary 1}), where $\mathcal{P}=I_N\otimes P$. Also, let $K=-B^{\text{T}}P$, so that 
the set-valued derivative of $V(e^a(t))$ along the trajectories of the system (\ref{augmented error auxiliary 1}) is given by 
\begin{eqnarray}
\dot {\bar{V}}(e^a(t))&=&\mathcal{K}[e^{a^{\text{T}}}(t)\big(I_N\otimes (A^{\text{T}}P+PA)-(C_2L_{22}^{\text{T}}+L_{22}C_2)\otimes PBB^{\text{T}}P\big)e^a(t)+2e^{a^{\text{T}}}(t)(I\otimes PB)\nonumber
\\
&&(L_{21}\otimes I)u_0(t)-2e^{a^{\text{T}}}(t)(I\otimes PB)(L_{22}C_0\otimes I)\text{sgn}((I\otimes B^{\text{T}}P)e^a(t))+2\omega_0^{\text{T}}(t)\mathcal{B}_{\omega}^{\text{T}}\mathcal{P}e^a(t)].\nonumber\\
\label{eq. 6.5}
\end{eqnarray}\par
Let $T_1(t)=e^{a^{\text{T}}}(t)(I\otimes PB)(L_{21}\otimes I)u_0(t)$, $T_2(t)=e^{a^{\text{T}}}(t)(I\otimes PB)(L_{22}C_0\otimes I)\text{sgn}((I\otimes B^{\text{T}}P)e^a(t))$, $\bar e_i(t)=B^{\text{T}}Pe_i^a(t)$ and $\bar e(t)=\text{col}\{\bar e_i(t)\}$. Since $T_1(t)$ is a scaler, $T_1(t)\leq \|T_1(t)\|_1$, and one has 
\begin{eqnarray}
T_1(t)&\leq&\|{T_1}(t)\|_1\leq \|(L_{21}^\text{T}\otimes I)(I\otimes B^\text{T}P)e^{a}(t) \|_1\|u_0(t)\|_{\infty}.
\end{eqnarray}
Then by using the Holder's inequality 
\begin{eqnarray}
T_1(t)&\leq& \|(L_{21}^\text{T}\otimes I)\|_\infty\|(I\otimes B^\text{T}P)e^{a}(t) \|_1\|u_0(t)\|_{\infty}\nonumber\\
&\leq&
\|(I\otimes B^\text{T}P)e^{a}(t)\|_1u_{0M}=u_{0M}\sum_{i=1}^N\sum_{k=1}^{m}|\bar e_i^k(t)|,\label{eq 15.1}
\end{eqnarray}
where $\bar e_i^k(t)$ is the $k$-th element of $\bar e_i(t)=\begin{bmatrix}\bar e_i^1(t),\dots,\bar e_i^{m}(t)\end{bmatrix}^{\text{T}}$ and we use the fact that $\|L_{21}\|_\infty=1$.  
On the other hand,  $T_2(t)$ can be written as
\begin{eqnarray}
T_2(t)=\bar e^{\text{T}}(t)(L_{22}C_0\otimes I)\text{sgn}(\bar e(t))=\sum_{i=1}^NT_{2i}(t),
\end{eqnarray}
where 
\begin{eqnarray*}
T_{2i}(t)&=&\bar e_i^{\text{T}}(t)(d_ic_{i0}\text{sgn}\{\bar e_i(t)\}-\sum_{j\in\mathcal{N}_i}c_{j0}\text{sgn}\{\bar e_j(t)\})\nonumber\\
&=&\sum_{k=1}^{m}\bar e_i^k(t) \big(d_ic_{i0}\text{sgn}\{ \bar e_i^k(t)\}-\sum_{j\in\mathcal{N}_i}c_{j0}\text{sgn}\{\bar e_j^k(t)\}\big).\label{eq 2.12}
\end{eqnarray*}\par
Let $T_{2i}^k(t)=\bar e_i^k(t) \big(d_ic_{i0}\text{sgn}\{ \bar e_i^k(t)\}-\sum_{j\in\mathcal{N}_i}c_{j0}\text{sgn}\{\bar e_j^k(t)\}\big)$, then three cases can be considered depending on the value of $\bar e_i^k(t)$ as follows:\\
i) $\bar e_i^k(t)=0$, then $T_{2i}^k(t)=0$.\\
ii) $\bar e_i^k(t)>0$, then $\text{sgn}\{ \bar e_i^k(t)\}=1$. Since $c_{j0}>0$ and $\text{sgn}\{\bar e_j^k(t)\}\in\{-1,0,1\}$, it follows that
$$d_ic_{i0}-\sum_{j\in\mathcal{N}_i} c_{j0}\leq d_ic_{i0}\text{sgn}\{ \bar e_i^k(t)\}-\sum_{j\in\mathcal{N}_i}c_{j0}\text{sgn}\{\bar e_j^k(t)\}\leq d_ic_{i0}+\sum_{j\in\mathcal{N}_i}c_{j0},$$
and if $c_{i0}$, $i=1,\dots,N$ are designed such that $d_ic_{i0}-\sum_{j\in\mathcal{N}_i} c_{j0}>0$, then 
\begin{equation}
|\bar e_i^k(t)|( d_ic_{i0}-\sum_{j\in\mathcal{N}_i}c_{j0}) \leq T_{2i}^k(t)\leq |\bar e_i^k(t)|(d_ic_{i0}+\sum_{j\in\mathcal{N}_i}c_{j0}).\label{eq 2.13}
\end{equation}
iii) $\bar e_i^k(t)<0$, then $\text{sgn}\{ \bar e_i^k(t)\}=-1$ and $e_i^k(t)=-|e_i^k(t)|$. Therefore,
$$-d_ic_{i0}-\sum_{j\in\mathcal{N}_i} c_{j0}\leq d_ic_{i0}\text{sgn}\{ \bar e_i^k(t)\}-\sum_{j\in\mathcal{N}_i}c_{j0}\text{sgn}\{\bar e_j^k(t)\}\leq -d_ic_{i0}+\sum_{j\in\mathcal{N}_i}c_{j0}.$$
Again if $c_{i0}$, $i=1,\dots,N$ are designed such that, $d_ic_{i0}-\sum_{j\in\mathcal{N}_i} c_{j0}>0$, then
\begin{equation}
|\bar e_i^k(t)|( d_ic_{i0}-\sum_{j\in\mathcal{N}_i}c_{j0}) \leq T_{2i}^k(t)\leq |\bar e_i^k(t)|(d_ic_{i0}+\sum_{j\in\mathcal{N}_i}c_{j0}).\label{eq 2.14}
\end{equation}\par
Let $T_3(t)=T_1(t)-T_2(t)$. From the  inequalities  (\ref{eq 15.1})-(\ref{eq 2.14})  it follows that
\begin{eqnarray}
T_3(t)&\leq& u_{0M}\sum_{i=1}^N{\sum_{k=1}^{m}|\bar e_i^k(t)|}-\sum_{i=1}^N{\sum_{k=1}^{m} |\bar e_i^k(t)|( d_ic_{i0}-\sum_{j\in\mathcal{N}_i}c_{j0})}\nonumber\\
&=&\sum_{i=1}^N\sum_{k=1}^{m}|\bar e_i^k(t)|\big(u_{0M}-d_ic_{i0}+\sum_{j\in\mathcal{N}_i}c_{j0}\big).\label{eq 2.17}
\end{eqnarray}
Suppose that $c_{2i}$s and $c_3$ are obtained such that
\begin{eqnarray}
C_2L_{22}^{\text{T}}+L_{22}C_2>c_3I,\ c_3>0,\ C_2=\text{diag}\{c_{2i}\}>0.\label{eq. 1.12}
\end{eqnarray} 
Now by using the Fact \ref{Fact1} for the last term in the right-hand side of  (\ref{eq. 6.5}) with $X=(L_{21}\otimes I_{m})\omega_0(t)$, $Y=(I_N\otimes B_{\omega}^{\text{T}}P)e^a(t)$ and $\alpha=\frac{\gamma^2}{2}c_4$, and also the inequalities (\ref{eq 2.17}) and (\ref{eq. 1.12}), the expression (\ref{eq. 6.5}) can be replaced with the following inequality
\begin{eqnarray*}
	\dot{\bar V}(e^a(t))&\leq&\mathcal{K}[e^{a^{\text{T}}}(t)\big(I_N\otimes ( A^{\text{T}}P+PA-c_3PBB^{\text{T}}P)\big)e^a(t)
	+2\sum_{i=1}^N\sum_{k=1}^{m}|\bar e_i^k|\big(u_{0M}-d_ic_{i0}+\sum_{j\in\mathcal{N}_i}c_{j0}\big)\\
	&&+\frac{\gamma^2}{2}c_4\omega_0^{\text{T}}(t)(L_{21}^{\text{T}}L_{21}\otimes I_m)\omega_0(t)+{2}{\gamma^{-2}}c_4^{-1}e^{a^{\text{T}}}(t)(I\otimes PB_{\omega}B_{\omega}^{\text{T}}P)e^a(t)].
\end{eqnarray*}\par
Since now the right hand side of the above inequality is continuous, the operator $\mathcal{K}[.]$ can be removed. Let $d_0^*=L_{21}^{\text{T}}L_{21}$  and add $d_0^* e^{a^{\text{T}}}(t)e^a(t)$ to both sides of the above inequality then it follows that 
\begin{eqnarray}
\dot {\bar V}(e^a(t))-\frac{\gamma^2}{2}d_0^*c_4\omega_0^{\text{T}}(t)\omega_0(t) +d_0^* {e}^{a^\text{T}}(t)e^a(t)\leq g(e^a(t)),\label{eq 2.19}
\end{eqnarray}
where
\begin{eqnarray*}
g(e^a(t))&=&e^{a^{\text{T}}}(t) \big(I_N\otimes (A^{\text{T}}P+PA-c_3PBB^{\text{T}}P+d_0^* I+{2}{\gamma^{-2}}c_4^{-1}PB_{\omega}B_{\omega}^{\text{T}}P)\big)e^a(t)\nonumber\\
&&+2 \sum_{i=1}^N\sum_{k=1}^{m}|\bar e_i^k(t)|\big(u_{0M}-d_ic_{i0}+\sum_{j\in\mathcal{N}_i}c_{j0}\big)   \label{eq. 1.9}.
\end{eqnarray*}
From \cite{Shevitz94}, we require  $g(e^a(t))$ to be negative definite, which will be achieved if $P$ is obtained such that
\begin{equation}
A^{\text{T}}P+PA-c_3PBB^{\text{T}}P+{2}{\gamma^{-2}}c_4^{-1}PB_{\omega}B_{\omega}^{\text{T}}P+d_0^* I<0,  \label{eq. 1.1}
\end{equation}
and $c_{i0}$ are selected such that
\begin{equation}
u_{0M}-d_ic_{i0}+\sum_{j\in\mathcal{N}_i}c_{j0} < 0, i=1,\dots,N.\label{eq 2.18}
\end{equation}
Therefore, if $c_{i0}$, $i=1,\dots,N$ and $P$ are selected as the solutions to (\ref{eq 2.18}) and (\ref{eq. 1.1}),  the function $g(.)$ will be negative definite  and for $\omega_0(t)\equiv 0$, it follows that  $\dot{\bar V}(e^a(t))<0$, or equivalently the consensus errors are asymptotically stable. \par
Now, if the initial conditions are set to zero and the disturbance is the only input to the agents, then by integrating the left-hand side of (\ref{eq 2.19}) one gets 
\begin{equation}
\int_{t_0}^{\infty}( e^{a^{\text{T}}}(t)e^a(t)-\frac{\gamma^2}{2}c_4\omega_0^{\text{T}}(t)\omega_0(t)) \text{d}t<0.\label{eq. 1.2}
\end{equation}
Given that $e^a(t)=(L_{22}\otimes I_n) \xi^a(t)$, 
		 $\xi^a(t)=x^a(t)-1_N\otimes x_0(t)$ and $x^a(t)=\text{col}\{x_i^a(t)\}$,   it follows that  
\begin{equation}
\lambda_{\text{m}}\xi^{a^{\text{T}}}(t)\xi^a(t) \leq e^{a^{\text{T}}}(t)e^a(t)\leq \lambda_{\text{M}}\xi^{a^{\text{T}}}(t)\xi^a(t), \label{eq. 1.3}
\end{equation}
where $\lambda_{\text{m}}=\lambda_{\text{min}}(L_{22}^{\text{T}}L_{22})$ and $\lambda_{\text{M}}=\lambda_{\text{max}}(L_{22}^{\text{T}}L_{22})$.  Hence, from the inequalities (\ref{eq. 1.2}) and (\ref{eq. 1.3}) it follows that  
\begin{eqnarray*}
	&&\int_{t_0}^{\infty}  \lambda_{\text{m}}\xi^{a^{\text{T}}}(t)\xi^a(t)\text{d}t  -\int_{t_0}^{\infty}\frac{\gamma^2}{2}c_4\omega_0^{\text{T}}(t)\omega_0(t) \text{d}t <0,
\end{eqnarray*}
and by selecting $c_4=N\lambda_{\text{m}}$ one gets  
\begin{equation}
\frac{\int_{t_0}^{\infty} \xi^{a^{\text{T}}}(t)\xi^a(t)\text{d}t} {\int_{t_0}^{\infty}\omega_0^{\text{T}}(t)\omega_0(t) \text{d}t} <\frac{N}{2}\gamma^2.\label{eq. 1.6}
\end{equation}\par
\underline{\textbf{Part B}}: Under our proposed  control law the dynamics of the $i$-th auxiliary agent tracking error, $\xi_i(t)$, can be expressed as 
\begin{equation}
\dot \xi_i(t)=(A+c_1 BK) \xi_i(t)+ {B}_{\omega}\omega_i(t).\label{error tracking 1}
\end{equation}
Consider $V_i(\xi_i(t))=\xi_i^{\text{T}}(t) P\xi_i(t)$ as a Lyapunov function candidate for the system (\ref{error tracking 1}) and  select $K=-B^{\text{T}}P$. It then follows that 
\begin{eqnarray*}
	\dot V_i(\xi_i(t))&=& \xi_i^{\text{T}}(t)(A^{\text{T}}P+PA-2c_1 PBB^{\text{T}}P)\xi_i(t)+2 \xi_i^{\text{T}}(t) PB_{\omega}\omega_i(t),
\end{eqnarray*}
and by following along the same steps as in Part A, the above equality can be written as 
\begin{eqnarray*}
	&&\dot V_i(\xi_i(t))-\frac{\gamma^2}{2}\omega_i^{\text{T}}(t)\omega_i(t)+\xi_i^{\text{T}}(t)\xi_i(t)\leq \xi_i^{\text{T}}(t)(A^{\text{T}}P+PA-2c_1 PBB^{\text{T}}P+{2}{\gamma^{-2}}PB_{\omega}B_{\omega}^{\text{T}}P+I)\xi_i(t).
\end{eqnarray*}
Now if $P>0$ is obtained such that 
\begin{equation}
A^{\text{T}}P+PA- 2c_1PBB^{\text{T}}P+{2}{\gamma^{-2}}PB_{\omega}B_{\omega}^{\text{T}}P+I<0,\label{eq 1.5}
\end{equation}
then $\dot V_i(\xi_i(t))-\frac{\gamma^2}{2}\omega_i^{\text{T}}(t)\omega_i(t)+\xi_i^{\text{T}}(t)\xi_i(t)<0$. This implies that for $\omega_i(t)\equiv 0$, we have $\dot V_i(\xi_i(t))<0$, and for $\omega_i(t)\ne 0$, one gets
\begin{equation}
\frac{\int_{t_0}^{\infty}\xi_i^{\text{T}}(t)\xi_i(t)\text{d}t}{\int_{t_0}^{\infty}\omega_i^{\text{T}}(t)\omega_i(t)\text{d}t}\leq \frac{\gamma^2}{2}.\label{eq. 1.13}
\end{equation}
\underline{\textbf{Part C}}: In order to obtain the positive definite matrix $P$ that satisfies the inequalities (\ref{eq. 1.1}) and (\ref{eq 1.5}) and also guarantees the disturbance bound attenuation, let us set  $c_1$ and $c_4$ as $c_1=\frac{c_3}{2}$ and $c_4^{-1}=\max\{1,N^{-1}\lambda_{\text{m}}^{-1}\}$, respectively. Given that $d_0^*\geq 1$, it can be observed that if $P$ satisfies 
\begin{equation}
A^{\text{T}}P+PA- c_3PBB^{\text{T}}P+{2}{\gamma^{-2}}c_4^{-1}PB_{\omega}B_{\omega}^{\text{T}}P+d_0^*I<0,\label{eq 1.10}
\end{equation}
then inequalities (\ref{eq. 1.1}) and (\ref{eq 1.5}) will both hold, where $c_3$ is the solution to (\ref{eq. 1.12}). 
On the other hand
\begin{eqnarray*}
&&\frac{1}{2}\sum_{i=1}^N\int_{t_0}^\infty(x_i(t)-x_0(t))^\text{T}(x_i(t)-x_0(t))\text{d}t=\frac{1}{2}\sum_{i=1}^N\int_{t_0}^\infty(\xi_i(t)+\xi_i^a(t))^\text{T}(\xi_i(t)-\xi_i^a(t))\text{d}t=\\
&&\frac{1}{2}\sum_{i=1}^N\int_{t_0}^\infty(\xi_i^\text{T}(t)\xi_i(t)+{\xi_i^a}^\text{T}(t)\xi_i^a(t)+2\xi_i^\text{T}(t)\xi_i^a(t))\text{d}t\leq \\
&&\frac{1}{2}\sum_{i=1}^N\int_{t_0}^\infty(\xi_i^\text{T}(t)\xi_i(t)+{\xi_i^a}^\text{T}(t)\xi_i^a(t)+\xi_i^\text{T}(t)\xi_i(t)+\xi_i^a(t){\xi_i^a}^\text{T}(t))\text{d}t=\\
&&\sum_{i=1}^N\int_{t_0}^\infty(\xi_i^\text{T}(t)\xi_i(t)+{\xi_i^a}^\text{T}(t)\xi_i^a(t))\text{d}t
\end{eqnarray*}
Now from equations (\ref{eq. 1.6}) one has
\begin{eqnarray*}
\int_{t_0}^\infty{\xi^a}^\text{T}(t)\xi^a(t)\text{d}t=\sum_{i=1}^N\int_{t_0}^\infty{\xi^a}_i^\text{T}(t)\xi_i^a(t)\text{d}t\leq\frac{N}{2}\gamma^2\int_{t_0}^\infty\omega_0^\text{T}(t)\omega_0(t)\text{d}t=\frac{\gamma^2}{2}\sum_{i=1}^N\int_{t_0}^\infty\omega_0^\text{T}(t)\omega_0(t)\text{d}t
\end{eqnarray*}
and by using  (\ref{eq. 1.13})
\begin{eqnarray*}
\sum_{i=1}^N\int_{t_0}^\infty\xi_i^\text{T}(t)\xi_i(t)\text{d}t\leq\frac{\gamma^2}{2}\sum_{i=1}^N\int_{t_0}^\infty\omega_i^\text{T}(t)\omega_i(t)\text{d}t.
\end{eqnarray*}
then it follows that 
\begin{eqnarray*}
	&&\frac{1}{2}\sum_{i=1}^N\int_{t_0}^{\infty}(x_i(t)-x_0(t))^{\text{T}}(x_i(t)-x_0(t))\text{d}t\leq
	\frac{\gamma^2}{2}{\sum_{i=1}^N\int_{t_0}^{\infty}(\omega_i^{\text{T}}(t)\omega_i(t)+\omega_0^{\text{T}}(t)\omega_0(t))\text{d}t}.
\end{eqnarray*}
Therefore, the team $H_{\infty}$ performance upper bound can be expressed as
\begin{equation*}
\frac{\sum_{i=1}^N\int_{t_0}^{\infty}(x_i(t)-x_0(t))^{\text{T}}(x_i(t)-x_0(t))\text{d}t}{\sum_{i=1}^N\int_{t_0}^{\infty}(\omega_i^{\text{T}}(t)\omega_i(t)+\omega_0^{\text{T}}(t)\omega_0(t))\text{d}t}\leq \gamma^2.\label{eq. 1.14}
\end{equation*}
The above inequality implies that $J\leq 0$, or equivalently the healthy team $H_{\infty}$ performance criterion holds. This along with the  properties of the stability of $e_i^a(t)$ and $\xi_i(t)$, as stated in Parts A and B, imply that our proposed control law solves the $H_{\infty}$ performance state consensus problem for the healthy team. \qed

\subsection{$H_{\infty}$ Performance Control Reconfiguration}\label{subsection 1}
Consider the representation of an agent subject to presence of faults be  specified as in Subsection \ref{subsection 2-E}, and given by the equation (\ref{faulty system 2}) or equivalently by the transformed model (\ref{eq. 1.11}). Our proposed reconfigured control law for the $i$-th faulty agent is now given by
\begin{equation}
u_i^r(t)=K_{1i}^{r}\xi_i^f(t)+K_{2i}^{r}u_i^a(t)+\underline u_i^C,\label{reconfigured control}
\end{equation}
{where  $K_{1i}^{r}$, $K_{2i}^{r}$ are control gains and $\underline u_i^C$ is the control command to be designed later}. Therefore the dynamics of the closed-loop faulty agent (\ref{eq. 1.11})  becomes 
\begin{eqnarray}
\dot \xi_i^f(t)&=&(A+B_i^rK_{1i}^{r})\xi_i^f(t)+(B_i^rK_{2i}^{r}-B)u_i^a(t)+B_i^s\underline u_i^s+B_i^r\underline u_i^C+B_{\omega}\omega_i(t),\label{faulty agent tracking}\\
z_i(t)&=&C\xi_i^f(t).\nonumber
\end{eqnarray}\par
 As per Definition \ref{def. 1}, the $H_\infty$ control reconfiguration objectives can now be stated as that of selecting the gains $K_{1i}^{r}$ and $K_{2i}^{r}$ and the control command $\underline  u_i^r$ such that (a) $\xi_i^f(t)$ is stable, (b) $z_i(t)\equiv0$ (that is, $y_i^f(t)=y_i^a(t)$) for $\omega_i(t)\equiv0$, $t\geq t_f$, and  (c)  $\frac{\int_{t_f}^{\infty}\xi_i^{f^\text{T}}(t)\xi_i^f(t)\text{d}t}{\int_{t_f}^{\infty}\omega_i^{\text{T}}(t)\omega_i(t)\text{d}t}\leq \gamma_f^2$ for $\omega_i(t)\ne 0$.  In order to pursue the reconfiguration strategy we required the following assumption, we later discuss how deviation of this assumption affect the results.
{ \begin{assumptions}\label{stuck condition1}
 Under the fault scenario, there still enough actuator redundancy to compensate for the fault, i.e.
\begin{equation}
B_i^s\underline u_i^s\subset \text{Im}\{B_i^r\}.\label{stuck condition}
\end{equation}
\end{assumptions}
 then there exists a control signal  $\underline u_i^C$ such that }
 \begin{equation}
 B_i^s\underline u_i^s+B_i^r\underline u_i^C=0.\label{stuck term solution}
 \end{equation} 
 Subject to the above condition, equation (\ref{faulty agent tracking}) now becomes
\begin{eqnarray}
\dot {\xi}_i^f(t)&=&(A+B_i^rK_{1i}^{r})\xi_i^f(t)+(B_i^rK_{2i}^{r}-B)u_i^a(t)+B_{\omega}\omega_i(t),\label{faulty agent tracking 2}\\
z_i(t)&=&C\xi_i^f(t).\nonumber
\end{eqnarray}\par
 Let us  temporarily assume that  $\omega_i(t)\equiv 0$, then
 \begin{eqnarray}
 z_i(t)&=&Ce^{(A+B_i^rK_{1i}^{r})(t-t_f)}\xi_i^f(t_f)+\int_{t_f}^tCe^{(A+B_i^rK_{1i}^{r})(t-s)}(B_i^rK_{2i}^{r}-B)u_i^a(s)\text{d}s.\label{eq. 2.3}
 \end{eqnarray}
{ From (\ref{eq. 2.3}), to ensure that the outputs of the faulty agent do not deviate after fault, both terms should be zero or negligible. The first term will be negligible if the agents reach a consensus before fault occurrence i.e. $\xi_i^f(t_f)\simeq 0$ or if  $K_{1i}^{r}$ is designed such that  $e^{(A+B_i^rK_{1i}^{r})(t-t_f)}$ damps very fast.  This can be  achieved easily if $\lambda_{max}\{A+B_i^rK_{1i}^{r}\}$ is small enough. }
On the other hand,   according to Theorem \ref{theorem 1}, $u_i^a(t)=u_{ic}(t)=K_{2i}e_i^a(t)+K_{i0}(t)$. Given that the control gains are designed such that $e_i^a(t)$ is asymptotically stable and $K_{i0}(t)$ is bounded,  $u_i^a(t)$ also remains bounded.  Considering that $u_i^a(t)$ does not depend on the dynamics of $\xi_i^f(t)$, it can be treated as a disturbance to the system (\ref{faulty agent tracking 2}).  
Consequently, the problems of \emph{(i)} enforcing $z_i(t)\equiv 0$ (for $t\geq t_f$, $\omega_j(t)\equiv 0$, $j=0,\dots,N$ and any $u_i^a(t)$), and \emph{(ii)} stabilizing  $\xi_i^f(t)$,  is similar to that of the disturbance decoupling problem with stability (DDPS), as studied in \cite{Lunze2006}. \par
 The geometric approach that is based on the  theory of subspaces \cite{Basile92} is the most popular method for solving the DDPS problem. Towards this end, we first  introduce the required subspaces as follows:
 $\mathcal{B}_i^r=\text{Im}\{B_i^r\}$, $\mathcal{C}=\text{Ker}\{C\}$, $\mathcal{V}^*$ and $\mathcal{V}_g^*$ denote the maximal $(A,B_i^r)$ controlled invariant  subspace that is contained in $\mathcal{C}$, and the maximal \underline{ internally stable} $(A,B_i^r)$ controlled invariant  subspace that is contained in $\mathcal{C}$, respectively. \par
 Following the procedure in \cite{Basile92}, if $K_{2i}^{r}$ and $K_{1i}^r$ are selected such that 
 \begin{eqnarray}
\text{Im}\{B_i^rK_{2i}^{r}-B\}\subset \mathcal{V}^*,(A+B_i^rK_{1i}^r)\mathcal{V}^*\subset\mathcal{V}^*\label{DDP}
\end{eqnarray}
 then  the second term in (\ref{eq. 2.3}) will also vanish. On the other hand, if $K_{2i}^{r}$ and $K_{1i}^r$ are selected such that 
  \begin{eqnarray}
\text{Im}\{B_i^rK_{2i}^{r}-B\}\subset \mathcal{V}_g^*,(A+B_i^rK_{1i}^r)\mathcal{V}_g^*\subset\mathcal{V}^*_g\label{DDPS},
\end{eqnarray}
 then the second term in (\ref{eq. 2.3})  will also vanish and $\xi_i^f(t)$ will be stable due to the stability of the subspace $\mathcal{V}_g^*$.
Unfortunately, there is no systematic approach to explicitly obtain $\mathcal{V}_g^*$, implying that $\mathcal{V}_g^*$ cannot be computed and employed directly for obtaining $K_{2i}^r$ that satisfies the condition (\ref{DDPS}). Therefore, we are required to transform the  condition (\ref{DDPS}) into a verifiable one. Once such a controller is obtained, one can then ensure that $z_i(t)\equiv 0$ and $\xi_i^f(t)$ will remain stable. \par
Given that $\mathcal{V}^*$ is $(A,B_i^r)$ controlled invariant, \underline{there exists a matrix $K_{1i}^{r}$, a friend of $\mathcal{V}^*$}, \cite{Basile92} such that $A_c\mathcal{V}^*\subset \mathcal{V}^*$, where $A_c=A+B_i^rK_{1i}^{r}$. Now, by invoking the Theorem 3.2.1 of \cite{Basile92}, for a matrix $A_c$ and its associated $\mathcal{V}^*$, there always exists a nonsingular transformation $T$ such that 
\begin{equation}
\bar A_{c}=T^{-1}A_cT=\begin{bmatrix}\bar{A}_c^{1}&\bar{A}_c^{2}\\0&\bar{A}_c^{3}\end{bmatrix},\label{eq. 3.1}
\end{equation}
where $T=\begin{bmatrix}T_1&T_2\end{bmatrix}$, $\text{Im}\{T_1\}=\mathcal{V}^*$ and $T_2$ is any matrix that renders $T$ nonsingular.
By substituting $A_c=A+B_i^rK_{1i}^{r}$ into (\ref{eq. 3.1}), it follows that 
\begin{eqnarray}
\bar A_{c}=\bar A+\bar B_i^r \bar K_{1i}^{r},\label{eq. 3.2}
\end{eqnarray}
where $\bar A=T^{-1}AT=\begin{bmatrix}\bar A_{11}&\bar A_{12}\\ \bar A_{21}&\bar A_{22}\end{bmatrix}$, $\bar B_i^r=T^{-1}B_i^r=\begin{bmatrix}\bar B_{i1}^r\\ \bar B_{i2}^r\end{bmatrix}$ and $\bar K_{1i}^{r}=K_{1i}^{r}T$. Now, if $\bar K_{1i}^{r}$ is  partitioned  as $\bar K_{1i}^{r}=\begin{bmatrix}\bar K_{1i}^{r1}& \bar K_{1i}^{r2}\end{bmatrix}$, from  (\ref{eq. 3.1}) and (\ref{eq. 3.2}) it can be concluded that there exists $\bar K_{1i}^{r1}$ such that 
\begin{equation*}
\bar A_{21}+\bar B_{i2}^r \bar K_{1i}^{r1}=0. \label{eq 3.2}
\end{equation*}\par
Furthermore, under the transformation $T,$ the system (\ref{faulty agent tracking 2}) can be re-written as
\begin{eqnarray}
\dot{\bar \xi}_i^f(t)&=&\bar A_c\bar\xi_i^f(t)+\bar E_iu_i^a(t)+\bar B_{\omega}\omega_i(t),\label{eq. 2.6}\\
 z_i(t)&=&\bar C\bar \xi_i^f(t),\nonumber
\end{eqnarray}
where $\bar\xi_i^f(t)=T^{-1}\xi_i^f(t)$, $\bar A_c=\begin{bmatrix}\bar A_{c}^1&\bar A_{c}^2\\ 0&\bar A_{c}^3\end{bmatrix}$, $\bar A_{c}^1=\bar A_{11}+\bar B_{i1}^r\bar K_{1i}^{r1}$, $\bar A_{c}^2=\bar A_{12}+\bar B_{i1}^r\bar K_{1i}^{r2}$, $\bar A_{c}^3=\bar A_{22}+\bar B_{i2}^r\bar K_{1i}^{r2}$, $\bar E_i=\bar B_i^r K_{2i}^{r}-\bar B=\begin{bmatrix}\bar B_{i1}^r K_{2i}^{r}-\bar B_1\\ \bar B_{i2}^r K_{2i}^{r}-\bar B_2 \end{bmatrix}$, $\bar B=T^{-1}B=\begin{bmatrix}\bar B_{1}\\ \bar B_{2}\end{bmatrix}$, $\bar C=CT=\begin{bmatrix}0& \bar C_2\end{bmatrix}$, $\bar B_{\omega}=T^{-1}B_{\omega}$. We are now in a position to state the main result of this subsection.  
\begin{theorems}\label{theorem 2}
Consider a team that consists of  a leader that is governed by (\ref{L1}) and $N$ follower agents that are governed by (\ref{LL or F}), and their control laws are designed and specified according to Theorem \ref{theorem 1}. Suppose at time $t=t_f$ the $i$-th agent becomes faulty and its dynamics is now governed by (\ref{faulty system 2}) where Assumption \ref{assumption 2} also hold. The control law (\ref{reconfigured control}) solves the $H_{\infty}$ performance  control reconfiguration problem with stability  where the  $H_{\infty}$ upper bound is given by  $\gamma_f^2=\alpha^{-1}\lambda_{\text{min}}^{-1}\{(TT^{\text{T}})^{-1}\}$ if $\underline u_i^C$ is obtained as a solution to (\ref{stuck term solution}),  $K_{1i}^{r}=\begin{bmatrix}
Y_1X_1^{-1}&Y_2X_2^{-1}
\end{bmatrix}T^{-1}$, and $K_{2i}^{r}$ is the solution to 
\begin{equation}
  \bar B_{i2}^rK_{2i}^{r}-\bar B_{2}=0,\label{matching equation}
\end{equation}
 where $T$ is  defined in  (\ref{eq. 3.1}),  $X_i$ and $Y_i$'s, $i=1,2$ are solutions to 
\begin{eqnarray}
\text{max} \ \alpha \ \text{s.t.}&&  \begin{bmatrix}
\Theta&X\\
X&-I
\end{bmatrix}
<0, \ X=\text{diag}\{X_1,X_2\}>0,\ \bar A_{21}X_1+\bar B_{i2}^r Y_1= 0, \label{OPP}
\end{eqnarray}
where 
$
\Theta=\begin{bmatrix}\Theta_1&\Theta_2\\\Theta_2^{\text{T}}&\Theta_3\end{bmatrix}$, $
\Theta_1=X_1\bar A_{11}^{\text{T}}+Y_1^{\text{T}}\bar B_{i1}^{r^\text{T}}+\bar A_{11}X_1+\bar B_{i1}^r Y_1+\alpha \bar B^1_{\omega}\bar B_{\omega}^{1^{\text{T}}}$, $\Theta_2=\bar A_{12}X_2+\bar B_{i1}^rY_2+\alpha \bar B^1_{\omega}\bar B_{\omega}^{2^{\text{T}}}$, $\Theta_3=X_2\bar A_{22}^{\text{T}}+Y_2^{\text{T}}\bar B_{i2}^{r^{\text{T}}}+\bar A_{22}X_2+\bar B_{2}Y_2+\alpha\bar B^{2}_{\omega}\bar B_{\omega}^{2^\text{T}}$, $\bar A_{11}$, $\bar A_{21}$, $\bar A_{12}$, $\bar B_{i1}$ and $\bar B_{i2}$ are defined as in (\ref{eq. 3.2}) and $\bar B_{\omega}$ and $\bar B_2$ are  defined as in (\ref{eq. 2.6}).
 \end{theorems}
 \proof 
Consider the system (\ref{eq. 2.6}). Given that the two inputs $\omega_i(t)$ and $u_i^a(t)$  are bounded and independent from each other, one can investigate their effects separately. Therefore, the proof is provided in three parts, namely: in Part A we assume that $\omega_i(t)\equiv0$ and the set of all control gains that guarantee $z_i(t)=0$ and stabilize $\xi_i^f(t)$ are obtained. Next, in Part B we assume that the disturbance is the only input to the agent and obtain the gains that minimize the $H_{\infty}$ performance index and guarantee stability as well.  Finally, in Part C, the control gains that satisfy both Parts A and B are obtained.\par
\underline{\textbf{Part A}}: Let  $\omega_i(t)\equiv 0$ so that we have
\begin{eqnarray*}
\dot{\bar \xi}_i^f(t)&=&\bar A_c\bar\xi_i^f(t)+\bar E_iu_i^a(t),\\
 z_i(t)&=&\bar C\bar \xi_i^f(t).\nonumber
\end{eqnarray*}
Since $\bar A_c$ is an upper-triangular matrix, the matrix $e^{\bar A_ct}$ is also upper-triangular and can be written  as 
$e^{\bar A_ct}=\begin{bmatrix}e^{\bar A_c^1t}&F_2(t)\\0&e^{\bar A_c^3t} \end{bmatrix}$, where $F_2(t)=\int_{t_0}^{t}e^{\bar A_c^1(t-s)}\bar A_c^2e^{\bar A_c^3s}\text{d}s$. Under Assumption \ref{assumption 2}-(c), $\xi_i^f(t_f)=0$ and $\bar z_i(t)$ can be written as 
\begin{eqnarray}
 z_i(t)=\int_{t_f}^{t}\bar C_2e^{\bar A_c^3(t-s)}(\bar B_{i2}^rK_{2i}^{r}-\bar B_2)u_i^{a}(s) \text{d}s. \label{eq. 2.9}
\end{eqnarray}\par
If $ K_{2i}^{r}$ is obtained such that 
$$\bar B_{i2}^rK_{2i}^{r}-\bar B_2=0,$$
then  $ z_i(t)\equiv 0$, which implies that the above condition is equivalent to (\ref{DDP}). Moreover, if $\bar K_{1i}^{r1}$ and $\bar K_{1i}^{r2}$ are selected such that $\bar A_{11}+\bar B_{i1}^r \bar K_{1i}^{r1}$ and $ \bar A_{22}+\bar B_{i2 }^r\bar K_{1i}^{r2}$ are Hurwitz, then $\bar A_c$ will also be Hurwitz. Given that $u_i^a(t)$ is bounded and $\bar A_c$ is Hurwitz, then $\bar \xi_i(t)$ will also be bounded. Therefore, condition (\ref{DDPS}) is equivalent  to  obtaining the matrices $\bar K_{1i}^{r1}$, $\bar K_{1i}^{r2}$ and $K_{2i}^{r}$  such that
\begin{eqnarray}
&&\bar A_{21}+\bar B_{i2}^r \bar K_{1i}^{r1}=0, \label{const 1}\\
&&\bar A_{11}+\bar B_{i1}^r \bar K_{1i}^{r1}\  \text{is Hurwitz},\label{const 2} \\
&&  \bar A_{22}+\bar B_{i2 }^r\bar K_{1i}^{r2}\ \text{is Hurwitz},\label{const 3}\\
&& \bar B_{i2}^rK_{2i}^{r}-\bar B_{2}=0. \label{eq 6}
\end{eqnarray}
\underline{\textbf{Part B}}: Let the agents be only affected by the disturbances, then  we obtain
\begin{eqnarray}
\dot{\bar \xi}_i^f(t)&=&\bar A_{c}\bar\xi_i^f(t)+\bar B_{\omega}\omega_i(t),\label{eq. 2.8}\\
 z_i(t)&=&\bar C\bar \xi_i^f(t).\nonumber
\end{eqnarray}
 Consider a Lyapunov function candidate  $V_i^f(\bar{\xi}_i^f(t))=\bar \xi_i^{f^{\text{T}}}(t)P\bar \xi_i^f(t)$, where $P=\text{diag}\{P_1,P_2\}>0$. The time derivative of $V_i^f(t)$ along the trajectories of the system (\ref{eq. 2.8}) is given by
$$
	\dot V_i^f(t)=\bar \xi_i^{f^{\text{T}}}(t)(\bar A_{c}^{\text{T}}P+P\bar A_{c}){\bar \xi_i^f}(t)+2\bar \xi_i^{f^{\text{T}}}(t)P\bar B_{\omega}\omega_i(t).
$$
By applying Fact \ref{Fact1} to the second term in the right hand side of the above equation with $X^{\text{T}}=\bar \xi_i^{f^{\text{T}}}(t)P\bar B_{\omega}$, $Y=\omega_i(t)$ and $\alpha=\gamma^{-2}$, and adding $\bar \xi_i^{\text{T}}(t)\bar \xi_i(t)$ to both sides one gets 
\begin{eqnarray}
&&\dot V_i^f(t)-\gamma^{2}\omega_i^{\text{T}}(t)\omega_i(t)+\bar \xi_i^{f^\text{T}}(t)\bar \xi_i^f(t) \leq \bar \xi_i^{f^{\text{T}}}(t)\Lambda  {\bar \xi_i^f}(t),\label{eq 2.22}
\end{eqnarray}
where
 $$\Lambda=\begin{bmatrix} \bar A_{c}^{1^\text{T}}P_1+P_1\bar A_{c}^1 & P_1\bar A_{c}^2 \\ \bar A_{c}^{2^\text{T}} P_1 & \bar A_{c}^{3^\text{T}}P_2+P_2\bar A_{c}^{3} \end{bmatrix}+ \gamma^{-2} \begin{bmatrix} P_1 \bar B_{\omega}^1 \bar B_{\omega}^{1^{\text{T}}} P_1 &  P_1 \bar B_{\omega}^1 \bar B_{\omega}^{2^{\text{T}}} P_2 \\  P_2 \bar B_{\omega}^2 \bar B_{\omega}^{1^{\text{T}}} P_1 &  P_2 \bar B_{\omega}^2 \bar B_{\omega}^{2^{\text{T}}} P_2 \end{bmatrix}+I,
$$
and $\bar B_{\omega}^1$ and $\bar B_{\omega}^2$ are such that $\bar B_{\omega}=\begin{bmatrix}\bar B_{\omega}^1\\\bar B_{\omega}^2\end{bmatrix}$.
If the matrices $P_1$ and $P_2$ are obtained such that 
\begin{equation}
\Lambda=\begin{bmatrix} \bar A_{c}^{1^\text{T}}P_1+P_1\bar A_{c}^1 & P_1\bar A_{c}^2 \\ \bar A_{c}^{2^\text{T}} P_1 & \bar A_{c}^{3^\text{T}}P_2+P_2\bar A_{c}^{3} \end{bmatrix}+ \gamma^{-2} \begin{bmatrix} P_1 \bar B_{\omega}^1 \bar B_{\omega}^{1^{\text{T}}} P_1 &  P_1 \bar B_{\omega}^1 \bar B_{\omega}^{2^{\text{T}}} P_2 \\  P_2 \bar B_{\omega}^2 \bar B_{\omega}^{1^{\text{T}}} P_1 &  P_2 \bar B_{\omega}^2 \bar B_{\omega}^{2^{\text{T}}} P_2 \end{bmatrix}+I<0,\label{Nolinear MI}
\end{equation}
 then  the right hand side of (\ref{eq 2.22}) will be negative definite and we have 
$$\dot V_i^f(t)-\gamma^{2}\omega_i^{\text{T}}(t)\omega_i(t)+\bar \xi_i^{f^\text{T}}(t)\bar \xi_i^f(t)<0.$$\par
Consequently, by integrating both sides of the above inequality, one gets
$$\frac{\int_{t_f}^{\infty}\bar \xi_i^{f^\text{T}}(t)\bar \xi_i^f(t)\text{d}t}{\int_{t_f}^{\infty} \omega_i^{\text{T}}(t)\omega_i(t)\text{d}t}<\gamma^2.$$
Now, given that $\bar \xi_i^{f^\text{T}}(t)=T^{-1}\xi_i^f(t)$, the $H_{\infty}$ performance  bound for  $\xi_i^f(t)$ can be obtained as 
  \begin{equation*}
 \frac{\int_{t_0}^{\infty} \xi_i^{\text{T}}(t) \xi_i(t)\text{d}t}{\int_{t_0}^{\infty}\omega_i^{\text{T}}(t)\omega_i(t)\text{d}t}\leq \gamma^2\lambda_{\text{min}}^{-1}(T^{\text{-T}}T^{-1})=\gamma^2_f.
 \end{equation*}\par
\underline{\textbf{Part C}}: From Parts A and B, it follows that $\bar K_{1i}^{r1}$ should satisfy (\ref{const 1}) and (\ref{const 2}),  $\bar K_{1i}^{r2}$ should satisfy (\ref{const 3}) and $ K_{2i}^r$ should  satisfy (\ref{eq 6}), while the inequality (\ref{Nolinear MI}) should also hold. Note that if there exist matrices $P_1$ and $P_2$ such that (\ref{Nolinear MI}) holds then $\bar A_c$ will be Hurwitz. This implies that if the inequality (\ref{Nolinear MI}) holds then  (\ref{const 2}) and (\ref{const 3}) will hold. Therefore, the problem is reduced to solving the equality (\ref{eq 6}) for  $ K_{2i}^r$ and solving (\ref{const 1})  and (\ref{Nolinear MI}) simultaneously for  $\bar K_{1i}^{r1}$ and $\bar K_{1i}^{r2}$. Equation (\ref{eq 6}) is linear with respect to $ K_{2i}^r$ and can be solved easily,
whereas considering the structure of $\bar A_{c}^i$ for $i=1,2,3$, the inequality (\ref{Nolinear MI}) is nonlinear with respect to $P_1$, $P_2$ and $\gamma$. However, by multiplying both sides by $P^{-1}$ and using the known change of variables $X=\text{diag}\{X_1,X_2\}$, $X_1=P_1^{-1}$, $X_2=P_2^{-1}$, $Y_1=\bar K_{i1}^{r1}P_1^{-1}$, $Y_2=\bar K_{i1}^{r2}P_2^{-1}$, $\alpha=\gamma^{-2}$ and using the Schur complement, the  inequality (\ref{Nolinear MI}) can be transformed into the following LMI condition:
\begin{equation}
\begin{bmatrix}
\Theta&X\\
X&-I
\end{bmatrix}
<0,\label{faulty agent performance index constraint}
\end{equation}
where 
$\Theta=\begin{bmatrix}\Theta_1&\Theta_2\\\Theta_2^{\text{T}}&\Theta_3\end{bmatrix}$, $
\Theta_1=X_1\bar A_{11}^{\text{T}}+Y_1^{\text{T}}\bar B_{i1}^{r^\text{T}}+\bar A_{11}X_1+\bar B_{i1}^r Y_1+\alpha \bar B^1_{\omega}\bar B_{\omega}^{1^{\text{T}}}$, $\Theta_2=\bar A_{12}X_2+\bar B_{i1}^rY_2+\alpha \bar B^1_{\omega}\bar B_{\omega}^{2^{\text{T}}}$
 $\Theta_3=X_2\bar A_{22}^{\text{T}}+Y_2^{\text{T}}\bar B_{i2}^{r^{\text{T}}}+\bar A_{22}X_2+\bar B_{2}Y_2+\alpha\bar B^{2}_{\omega}\bar B_{\omega}^{2^\text{T}}$. Therefore, the control gains $\bar K_{1i}^{r1}$ and $\bar K_{1i}^{r2}$ satisfy the requirements of Parts A and B if the solutions to the inequality  (\ref{faulty agent performance index constraint}) also satisfy (\ref{const 1}). These requirements can be achieved provided that the gains are obtained as solutions to the following optimization problem, namely
\begin{eqnarray*}
\text{max} \ \alpha \ \text{s.t.}&&  \begin{bmatrix}
\Theta&X\\
X&-I
\end{bmatrix}
<0, \ X>0,\ \bar A_{21}X_1+\bar B_{i2}^r Y_1= 0.
\end{eqnarray*}
Subject to the above conditions the upper bound for the $H_{\infty}$ performance index  and the reconfigured control gain $K_{1i}^{r}$ are now specified according to $\gamma^2=\alpha^{-1}\lambda_{\text{min}}^{-1}\{(TT^{\text{T}})^{-1}\}$ and $K_{1i}^{r}=\begin{bmatrix}
Y_1X_1^{-1}&Y_2X_2^{-1}
\end{bmatrix}T^{-1}$, and this completes the proof of the theorem.\qed
 
The following algorithm summarizes the required steps that one needs to follow for designing the reconfigured control law gains.\par
\underline{\textbf{Algorithm for Design of the Fault Reconfiguration Controller Gains}}:\par
\begin{itemize}
\item [1)] Obtain the maximal $(A,B_i^r)$ controlled invariant subspace, $\mathcal{V}^*$, either by using the iterative algorithm that is proposed in \cite{Basile92} or by using the  Geometric Approach Toolbox\cite{Basil:2010:MISC} (available online). Set $T_1$ such that $\mathcal{V}^*=\text{Im}\{T_1\}$ and select $T_2$ such that $T=\begin{bmatrix}T_1&T_2\end{bmatrix}$ is a nonsingular matrix.
\item [2)] Obtain $\bar A_{11}$, $\bar A_{21}$, $\bar A_{12}$, $\bar B_{i1}$ and $\bar B_{i2}$ as in (\ref{eq. 3.2}) and $\bar B_{\omega}$ and $\bar B_2$   as in (\ref{eq. 2.6}).
\item [3)] Solve the optimization problem (\ref{OPP}) for $X_1,X_2,Y_1$ and $Y_2$.
\item [4)] Set $K_{1i}^{r}$ as $K_{1i}^{r}=\begin{bmatrix}
Y_1X_1^{-1}&Y_2X_2^{-1}
\end{bmatrix}T^{-1}$.
\item [5)] Solve    equation (\ref{matching equation}) for $K_{2i}^{r}$.
\item [6)] Solve equation (\ref{stuck term solution}) for $\underline u_i^C$.
\item [7)] Set $u_i^{r}(t)=K_{1i}^r\xi_i^f(t)+K_{2i}^ru_i^a(t)+\underline{u}_i^C$.
\item [8)] Set $u_i^f(t)=\begin{bmatrix}0_{1\times m_o}&(\underline u_i^s)^\text{T}&(u_i^{r}(t))^\text{T}\end{bmatrix}^\text{T}$.
\end{itemize}\par
In view of  Theorem \ref{theorem 2} and the above Algorithm the following  results can be obtained immediately.
\begin{corollaries}[Presence of only the LOE fault]\label{only loe}
Suppose the actuators are either healthy or subject to the LOE fault. 
In this case, $B_i^r$ in  (\ref{eq. 3.2}) is given by $B_i^r=B\Gamma_i$, where $\Gamma_i=\text{diag}\{\Gamma_i^k\}$, $k=1,\dots,m$. Furthermore,  the faulty control law $u_i^f(t)$, and the reconfigured control law, $u_i^r(t)$,  for the $i$-th faulty agent   are designed according to
 \begin{eqnarray*}
&& u_i^f(t)=u_i^r(t)\\
&&u_i^r(t)=K_{1i}^r\xi_i^f(t)+K_{2i}^ru_i^a(t),
\end{eqnarray*}
 where the control gains $K_{1i}^r$ and $K_{2i}^r$ are designed according to the Steps 4 and 5 of the above algorithm.
\end{corollaries}
\begin{corollaries}[Presence of only the outage fault]\label{outage}
Suppose the actuators $1$ to $m_o$ are subject to the outage fault and the remaining actuators are healthy. In this case, $B_i^r$ in  (\ref{eq. 3.2}) is given by  $B_i^r=\begin{bmatrix}b^{m_o+1}&\dots&b^m\end{bmatrix}$. Furthermore,  the faulty control law $u_i^f(t)$, and the reconfigured control law, $u_i^r(t)$,  for the $i$-th faulty agent  are designed according to
 \begin{eqnarray*}
&&u_i^f(t)=\begin{bmatrix}0_{1\times m_o}&(u_i^r(t))^\text{T}\end{bmatrix}^\text{T},\\
 &&u_i^r(t)=K_{1i}^r\xi_i^f(t)+K_{2i}^ru_i^a(t),
 \end{eqnarray*}
 where the control gains $K_{1i}^r$ and $K_{2i}^r$ are designed according to the Steps 4 and 5 of the above algorithm. 
\end{corollaries}
\begin{corollaries} [Presence of only the stuck fault]\label{only stuck}
Suppose the actuators $1$ to $m_s$ are subject to the stuck and the remaining actuators are healthy. In this case, $B_i^r$ in  (\ref{eq. 3.2}) is given by  $B_i^r=\begin{bmatrix}b^{m_s+1}&\dots&b^m\end{bmatrix}$. Furthermore,  the faulty control law $u_i^f(t)$, and the reconfigured control law, $u_i^r(t)$,  for the $i$-th faulty agent  are designed according to  
 \begin{eqnarray*}
&&u_i^f(t)=\begin{bmatrix}(\underline u_i^s)^\text{T}&(u_i^r(t))^\text{T}\end{bmatrix}^\text{T},\\
&&u_i^r(t)=K_{1i}^r\xi_i^f(t)+K_{2i}^ru_i^a(t)+\underline{u}_i^C,
\end{eqnarray*}
 where the control gains $K_{1i}^r$ and $K_{2i}^r$ are designed according to the Steps 4 and 5 and the control command $\underline{u}_i^C$ is obtained according to the Step 6 of the above algorithm. 
\end{corollaries}
Similar results corresponding to the combination of any two of the considered three types of faults can also be developed. These straightforward results that follow from Theorem \ref{theorem 2} and the Corollaries \ref{only loe}-\ref{only stuck} are not included here for brevity.
\subsection{The Existence of Solutions and  Analysis}\label{subsection 3}\par
In the previous two subsections, two cooperative control strategies to ensure consensus achievement and control reconfiguration in  multi-agent systems subject to actuator faults and environmental disturbances are proposed and conditions under which  these objectives are guaranteed are provided. In the following, we discuss the properties of solutions if certain required conditions are not satisfied. We consider five cases that are designated as \RNum{1} to \RNum{4} below. \par
\underline{\textbf{Case \RNum{1}}}: If the Assumption \ref{assumption 2}-(c) does not hold, i.e., the fault occurs during the transient period, then $\xi_i^f(t_f)\ne 0$, and  the first term in  (\ref{eq. 2.3}) will be non-zero. However, since $K_{1i}^r$ is designed such that $A+B_i^rK_{1i}^r$ is Hurwitz this term will  vanish asymptotically. Note that the delay in receiving the information from the FDI module and  activating the control reconfiguration  will also result in $\xi_i^f(t_f)\ne 0$, and causes a similar effect.  \par
\underline{\textbf{Case \RNum{2}}}: If  $\bar B_2\not\subset \text{Im}\{ \bar B_{i2}^r\}$, then   (\ref{matching equation}) does not have a solution. In this case, we may obtain $K_{2i}^{r}$ as a solution to
\begin{equation*}
\min_{K_{2i}^r} \ \text{trace}\{B_{i2}^rK_{2i}^r-\bar B_2\}.
\end{equation*}
Corresponding to this choice of $K_{2i}^r$, the second  term of   (\ref{eq. 2.3}) will remain non-zero and we have $z_i(t)\ne 0$ but bounded. However, if  $\bar K_{1i}^{r}$ is designed according to Part B in the proof of Theorem \ref{theorem 2},  one can still guarantee boundedness of the state consensus errors.

\underline{\textbf{Case \RNum{3}}}: Suppose that equation (\ref{stuck term solution}) does not have a solution, which is the case if \\
the estimated value of the stuck fault command, that is $\underline u_i^s$, is not accurate, or if Assumption  \ref{stuck condition1} does not hold or both. For generality, suppose that  $\underline u_i^s$ is not accurate and condition (\ref{stuck condition}) does not hold.  
In this case $\underline{u}_i^s=\underline{ \hat u}_i^s+\epsilon_i$, where $\underline{ \hat u}_i^s$ and $\epsilon_i$ denote the estimated then equation (\ref{stuck term solution}) can then be expressed as
$$ (B_i^s\underline{ \hat u}_i^s+B_i^r\underline u_i^C)+B_i^s\epsilon_i=0.$$
Since $\epsilon_i$ is unknown, therefore to obtain  $\underline u_i^C$  we instead use the following optimization problem, namely
\begin{equation*}
\min_{\underline u_i^C}\ \text{trace}\{B_i^s\underline{ \hat u}_i^s+B_i^r\underline u_i^C\}.
\end{equation*}
Let $\eta_i=(B_i^s\underline{ \hat u}_i^s+B_i^r\underline u_i^C)+B_i^s\epsilon_i$. 
Consider the control law (\ref{reconfigured control}) as designed in Theorem \ref{theorem 2}. It follows that for $\omega_i(t) \equiv 0$, equation (\ref{faulty agent tracking 2}) becomes 
\begin{equation*}
\dot{\xi}_i^f(t)=A_c\xi_i^f(t)+(B_i^rK_{2i}^r-B)u_i^a(t)+\eta_i,
\end{equation*}
where $A_c=A+B_i^rK_{1i}^r$. For $K_{2i}^r$ as a solution to (\ref{matching equation}), it follows that
\begin{equation*}
z_i(t)=C A_{c}^{-1}(e^{A_{c}(t-t_f)}-I)\eta_i.
\end{equation*}
Given that $A_{c}$ is Hurwitz, the above equation implies that after a transient period the error between the output of the faulty agent and its associated auxiliary system, or equivalently the output tracking error reaches a constant steady state value, i.e. $\lim_{t\to\infty} z_i(t)=-CA_c^{-1}\eta_i$. Consequently, under this scenario one can still observe that the state consensus errors remain bounded. \par

\underline{\textbf{Case \RNum{4}}}: Let the estimated  actuator loss of effectiveness factor or severity be subject to uncertainties, i.e. $\Gamma_i^k=\hat\Gamma_i^k+\epsilon_i^k$, where $\hat\Gamma_i^k$ is the estimate of the fault severity that is provided by the FDI module, and $\epsilon_i^k$ is an unknown estimation error uncertainty. 
Consider equation (\ref{eq. 3.2}). Since $\Gamma_i\ne \hat{\Gamma}_i$ we have $\bar B_i^r=\hat B_i^r+\bar B_i^{r\epsilon}$, where 
$\hat{\bar B}_i^r=T^{-1}B_i^{fr}\hat{\Gamma}_i$,  $\bar{B}_i^{r\epsilon}=T^{-1}B_i^{fr}\Upsilon_i$,  $\Upsilon_i=\text{diag}\{\epsilon_i^k\}$, $k=m_{s}+1,\dots,m$ and $B_i^{fr}=\begin{bmatrix}b^{m_{s}+1}\dots,B^{m}\end{bmatrix}$. 
In order to analyze the impact of these uncertainties on our previous results, we need to investigate both the matching condition, namely equation (\ref{matching equation}), and the stability of the tracking error $\xi_i^f(t)$. \par
Since $\Upsilon_i$ is unknown, one cannot determine the gain $K_{2i}^r$ such that (\ref{matching equation}) holds. This implies that unlike (\ref{eq. 2.9}) one cannot ensure $z_i(t)\equiv0$.  
On the other hand, $\bar{A}_c$ in  (\ref{eq. 2.6}) should be replaced by
$$
\tilde{A}_c={\bar A}_c+ \bar A_c^{\epsilon}=\bar A+\hat{\bar B}_i^r \bar{K}_{1i}^r+\bar{B}_i^{r\epsilon} \bar K_{1i}^r.
$$
Following along the same steps as those utilized in Subsection \ref{subsection 1} for now  $\bar A_c=\bar A+\hat{\bar B}_i^r \bar{K}_{1i}^r$, one can obtain the control gain $\bar{K}_{1i}^r$ that makes $\bar{A}_c$ Hurwitz. Hence, for $\omega_i(t)\equiv 0$, equation (\ref{eq. 2.6}) can be written as
$$\dot{\bar{\xi}}_i^f(t)=\big(\bar A_c +\bar A_c^{\epsilon} \big)\xi_i^f(t)+\bar E_iu_i^a(t).$$
In order to utilize the results in Theorem \ref{theorem 0.3}, we rewrite the above equation as follows
$$
\dot{\bar{\xi}}_i^f(t)=\bar A_c \xi_i^f(t)+\bar E_iu_i^a(t)+f(\xi_i^f(t)),
$$
where $f(\xi_i^f(t))\triangleq\bar A_c^{\epsilon} \xi_i^f(t)$. Given that $\bar A_c^{\epsilon}=\sum_{l=m_s+1}^m\epsilon_i^lb^lk^{l}$, we get
\begin{equation*}
\| \bar A_c^{\epsilon}\|_2\leq \sum_{l=m_s+1}^{m}|\epsilon_i^l|  \|b^lk^{l}\|_2,
\end{equation*}
where $b^l$ and $k^l$ denote the   $l-m_s$ column of $B_i^{fr}$ and the $l-m_s$  row  of $\bar K_{1i}^r$, respectively. Now, by using Theorem \ref{theorem 0.3}, if there exist $ \bar{\epsilon}_{\text{imax}}^l>0$, $l=1,\dots,m-m_s$ such that
 \begin{equation}
  \sum_{l=m_s+1}^{m} \bar{\epsilon}_{\text{imax}}^l \|b^lk^{l} \|_2 \leq \frac{1}{\sigma_{\min}(P)},\label{uncertainty bound}
  \end{equation}
 and $|\epsilon_i^l |\leq \bar{\epsilon}_{\text{imax}}^l$,  then  the matrix $\tilde{A}_c$ remains Hurwitz, where $P$ is a positive definite matrix solution to
  $$P{\bar A}_c+{\bar A}_c^{\text{T}}P=-2I.$$
This along with the boundedness of $u_i^a(t)$ implies that $\bar\xi_i^f(t)$ will also remain bounded.\par
\underline{\textbf{Case \RNum{5}}}: Suppose that the fault \underline{is recovered after a delay} of $\Delta$ s, i.e. $t_r= t_f+\Delta$, where $t_f$ and $t_r$ denote the time that the fault occurs and the time that the control reconfiguration is invoked. During the time $t_f\leq t\leq t_r$, the tracking dynamics  of the $i$-th agent, i.e., $\xi_i^f(t)$, becomes
$$
\dot \xi_i^f(t)=(A+c_1B_i^fK)\xi_i^f(t)+(B_i^f-B)u_i^a(t),\ t_f\leq t< t_r.
$$
Therefore, one gets
$$x_i^f(t)=\text{exp}((A+c_1B_i^fK)(t-t_f))(x_i(t_f)-x_i^a(t_f))+\int_{t_f}^t\text{exp}((A+c_1B_i^fK)(t-s))u_i^a(s)\text{d}s+x_i^a(t).$$\par
If the fault causes $A+c_1B_i^fK$ to become non-Hurwitz, then $x_i^f(t)$ will grow exponentially. Now, let $x_i^{M}$ denote the maximum allowable upper bound on the agent's state (this can be specified  for example based on the maximum speed of the moving agent or the maximum depth for surveying under the water), then invoking the reconfigured control law cannot be delayed beyond $\Delta$s, where the maximum delay in invoking the reconfigured controller is denoted by $\Delta$ and can be obtained by solving the following equation:
$$x_i^{M}=x_i^a(t_f+\Delta)+\text{exp}((A+c_1B_i^fK)\Delta)(x_i(t_f)-x_i^a(t_f))+\int_{t_f}^{t_f+\Delta}\text{exp}((A+c_1B_i^fK)(t_f+\Delta-s))u_i^a(s)\text{d}s.$$  
This implies that if the fault is not recovered before $t=t_f+\Delta$ s, the faulty agent may no longer be recoverable to satisfy the overall mission requirements and specifications at all times. 
\section{Simulation Results}\label{simulation results}
In this section, our proposed control recovery approach is applied to a network  of  Autonomous Underwater Vehicles (AUVs). The team behavior is studied under several  scenarios, namely when the  agents are healthy and also when the agents are subject to simultaneous LOE,  outage and stuck actuator faults, uncertainties in the FDI module information and delays in invoking the control reconfiguration.  The team is considered to  consist  of  five Sentry Autonomous Underwater Vehicles (AUVs). Sentry, made by the Woods Hole Oceanographic Institution \cite{jakuba2003}, is a fully autonomous underwater vehicle that is capable of surveying to the depth of 6000 m and is efficient for forward motions.   \par
The nonlinear six degrees of freedom equations of motion in the body-fixed frame in the horizontal plane is given by \cite{fossen94}:
\begin{eqnarray*}
&&\mathbb{M}\dot\nu +\mathbb{C}(\nu)\nu +\mathbb{D}(\nu,\phi_f)\nu +g(\eta)=b(\phi,h),\\
&&\dot\eta=\mathbb{J}(\eta)\nu,
\end{eqnarray*}
where $\mathbb{M}$, $\mathbb{C}$, $\mathbb{D}$ and $\mathbb{J}$ denote the inertia matrix, the moment/forces matrix, the damping matrix and the transformational matrix, respectively. The terms $g(\eta)$ and $b(\phi_f,h)$ denote the hydrostatic restoring forces and the truster input, respectively, and are given by 
$$b(\phi_f,h)=\begin{bmatrix}(h_{fp}+h_{fs})\cos \phi_{ff}+(h_{ap}+h_{as})\cos \phi_{af}\\0\\(h_{fp}+h_{fs})\sin \phi_{ff}+(h_{ap}+h_{as})\sin \phi_{af}\\ b_t(h_{fp}-h_{fs})\sin \phi_{ff}+b_t(h_{ap}-h_{as})\sin \phi_{af}\\-a_{ff}(h_{fp}+h_{fs})\sin \phi_{ff}-a_{af}(h_{ap}+h_{as})\sin \phi_{af}\\b_t(h_{fp}-h_{fs})\cos \phi_{ff}+b_t(h_{ap}-h_{as})\cos \phi_{af}\
\end{bmatrix},\ g(\eta)=\begin{bmatrix} 0\\0\\0\\z_{BG}\cos\theta \sin\phi W\\z_{BG}\sin \theta W\\0\end{bmatrix},$$
where 
$\phi_f=[\phi_{ff} \ \phi_{af}]^{\text{T}}$ and 
$h=[h_{fp}\ h_{fs}\ h_{ap}\ h_{as}]^\text{T}$ denote the foil angles and the truster inputs, respectively. 
The term 
$\eta=[\eta_1^\text{T} \ \eta_2^\text{T}]^\text{T}$, where $\eta_1=[x\ y\  z]^\text{T}$ denotes the inertial position and $\eta_2=[\phi \ \theta\ \psi]^\text{T}$ denotes the  inertial orientation. Also, $\nu=[\nu_1^\text{T}\ \nu_2^\text{T}]^\text{T}$, with $\nu_1=[\bar u \ v \ w]^\text{T}$ denotes the body-fixed linear velocity and $\nu_2=[p\ q \ r]^\text{T}$  denotes the angular velocity. Finally, $z_{BG}$ and $W$  denote the vertical distance between the center of the buoyancy and the center of the mass and the vehicle weight, respectively. 
\par
For the Sentry vehicle, the horizontal position is controlled indirectly through the heading subsystem, i.e. $v,r,\psi$, and surge speed subsystem, i.e. $\bar{u}$. Therefore, for control purposes the states $x$ and $y$ are ignored.  Moreover, 
under the assumptions that (a) the truster and foil angles do not affect each other, (b) the pitch and the pitch rate, i.e. $\theta$ and $q$ are sufficiently small, and (c) the foil angles are  sufficiently small, then the states $p,\phi$ are also ignored for control design and are considered passive \cite{jakuba2003}.  Therefore, 
 for the control design in the near horizontal maneuver under the operating point $\nu_1^o=[u^o\ 0\ w^o]^\text{T}$, $\nu_2^o=0_{3\times 1}$ and $\eta^o=0_{6\times 1}$,  the linear model of the Sentry AUV is reduced to the following subsystems: 
\begin{eqnarray*}
&&
\begin{bmatrix}\dot{\bar u}(t)\\ \dot v(t)\\ \dot r(t)\\ \dot \psi(t)\end{bmatrix}=
\begin{bmatrix}a_{11}u^o&0&0&0\\ 0&a_{22}u^o&a_{26}u^o&0\\ 0&a_{62}u^o&a_{66}u^o&0\\0&0&1&0\end{bmatrix}
\begin{bmatrix}\bar u(t)\\ v(t)\\ r(t)\\ \psi(t) \end{bmatrix}
+\begin{bmatrix}
m_{11}&m_{11}&m_{11}&m_{11}\\
m_{26}b_h&-m_{26}b_h&m_{26}b_h&-m_{26}b_h\\
m_{66}b_h&-m_{66}b_h&m_{66}b_h&-m_{66}b_h\\
0&0&0&0
\end{bmatrix}
\begin{bmatrix} h_{fp}(t)\\ h_{fs}(t)\\ h_{ap}(t)\\ h_{as}(t)\end{bmatrix},
\\
&&\begin{bmatrix}\dot w(t)\\ \dot q(t)\\ \dot z(t)\\ \dot\theta(t)\end{bmatrix}
=\begin{bmatrix}
a_{33}u^o&a_{35}u^o&0&-m_{m35}z_{GB}W\\ 
a_{53u^o}&a_{55}u^o&0&-m_{55}z_{GB}W\\
1&0&0&u^o\\
0&1&0&0
\end{bmatrix}
\begin{bmatrix}w(t)\\q(t)\\z(t)\\ \theta(t) \end{bmatrix}+
 \begin{bmatrix}\alpha_h^{11}&\alpha_h^{12} \\ \alpha_h^{21}&\alpha_h^{22} \\0&0\\ 0&0
  \end{bmatrix}\begin{bmatrix}\phi_{ff}(t)\\ \phi_{af}(t)\end{bmatrix},
\end{eqnarray*}
where $\alpha_h^{11}=\frac{\beta_h}{1+\beta_h}h_{31}+(u^o)^2f_{31}$, $\alpha_h^{12}=\frac{1}{1+\beta_h}h_{32}+(u^o)^2f_{32}$, $\alpha_h^{21}=\frac{\beta_h}{1+\beta_h}h_{51}+(u^o)^2f_{51}$, and $\alpha_h^{22}=\frac{1}{1+\beta_h}h_{52}+(u^o)^2f_{52}$. The detail relationships between the above parameters and the system parameters are  provided in \cite{jakuba2003}.
\par
For underwater vehicles, the ocean current is considered as a disturbance to the system, i.e. $\omega(t)=V_c(t)$, where $V_c(t)$ denotes the ocean current. In \cite{fossen2002}, the ocean current   is  modeled  by a first order Gauss-Markov Process as governed by $\dot V_c(t)+\mu V_c(t)=v(t)$, where $\mu\geq 0$ and $v(t)$ is a Gaussian white noise. For $\mu=0$, the model becomes a random walk, i.e. $\dot V_c(t)=v(t)$. Therefore, the disturbance signal that  is applied to the $i$-th agent is expressed as $\omega_i(t)=V_{ci}(t)=\int_{t_0}^tv_i(t)\text{d}t+V_{ci}(t_0)$. \par
 In conducting our simulations 
 we only consider the  speed-heading subsystems, i.e. $\bar u,v,r,\psi$.  
To obtain a linear model, the  forward (surge) speed  $u^o$ is set to $u^o=1$ and  all the parameters are considered to be the same as those in 
\cite{jakuba2003,elec2:2015:MISC}. 
The  numerical values of the triple $(A,B,C)$  for the $i$-th agent is governed  by \\ 
$A=\begin{bmatrix} 
         -0.0401&0 &0&0\\
         0&-0.709&-0.648&0\\
         0 &-1.770&1.414&0\\
         0 &0&1.00&0\end{bmatrix}$,
            $B=1.0e-03\begin{bmatrix}
      0.44&0.44&0.44&0.44\\
       0.06&-0.06&0.06&-0.06\\
       0.49&-0.49&0.49&-0.49\\
        0&0&0&0
         \end{bmatrix}$,\\   
$B_{\omega}=\begin{bmatrix}0.023&0.017&0.03&0\end{bmatrix}$,  $C=\begin{bmatrix}1&0&0&0\end{bmatrix}$, $x_{i}(t)=\begin{bmatrix}\bar u_i(t)& v_i(t)& r_i(t)& \psi_i(t)\end{bmatrix}^\text{T}$,\\ $u_{i}(t)=\begin{bmatrix}h_{ifp}(t)&h_{ifs}(t)&h_{iap}(t)&h_{ias}(t)\end{bmatrix}^\text{T}$.
 \par
\begin{figure}[thpb]
      \centering
        \includegraphics[ scale=0.27]{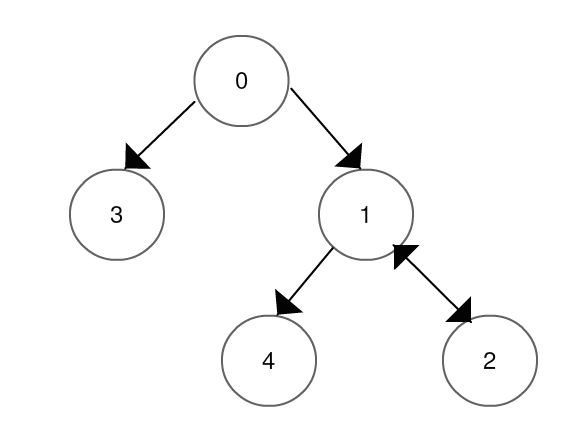}
      \caption{The topology of the leader-follower network of given AUVs.}     
      \label{network}
   \end{figure}
\begin{figure}[thpb]
      \centering
        \includegraphics[ scale=0.23]{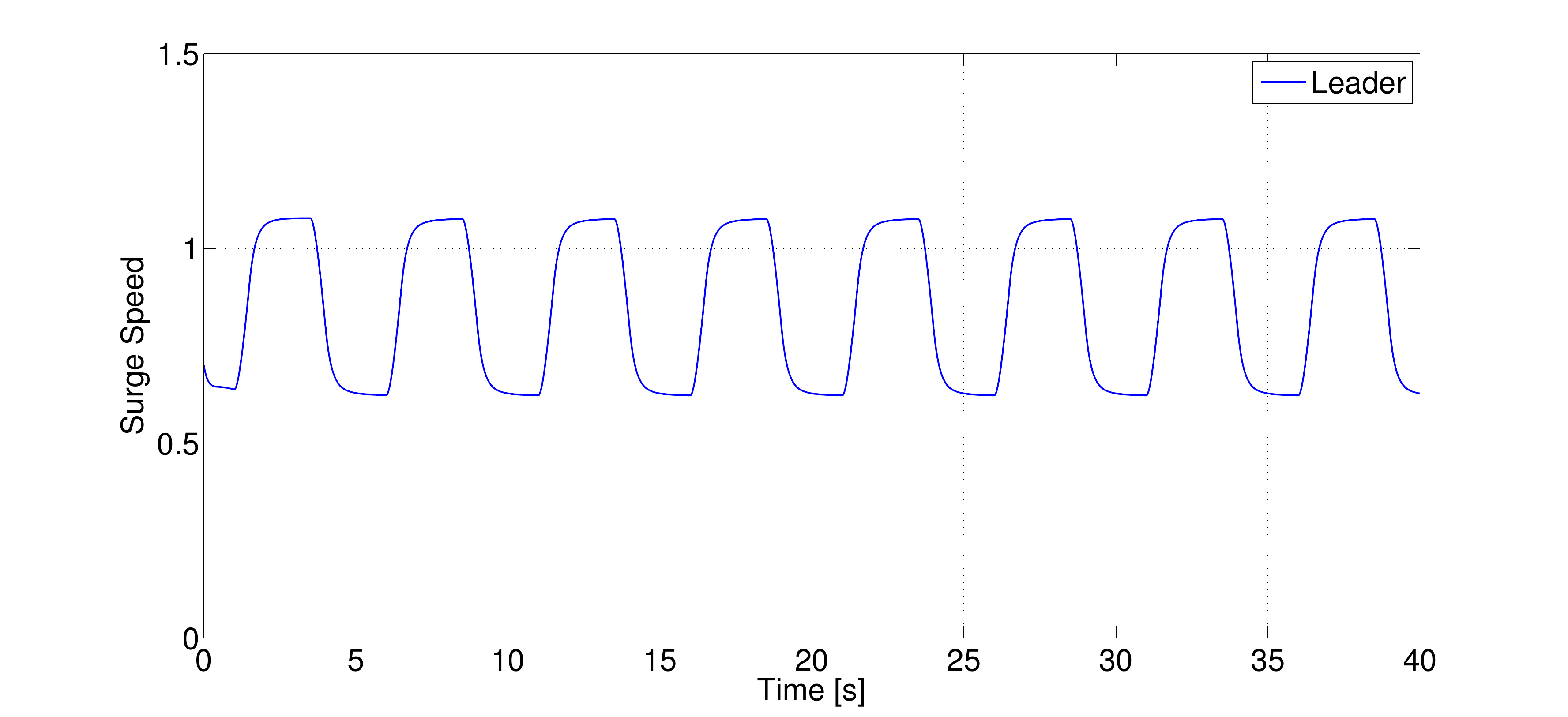}
      \caption{The desired leader surge speed trajectory.}     
      \label{leader trajectory}
   \end{figure}
\begin{figure}[thpb]
      \centering
                  \includegraphics[scale=0.4]{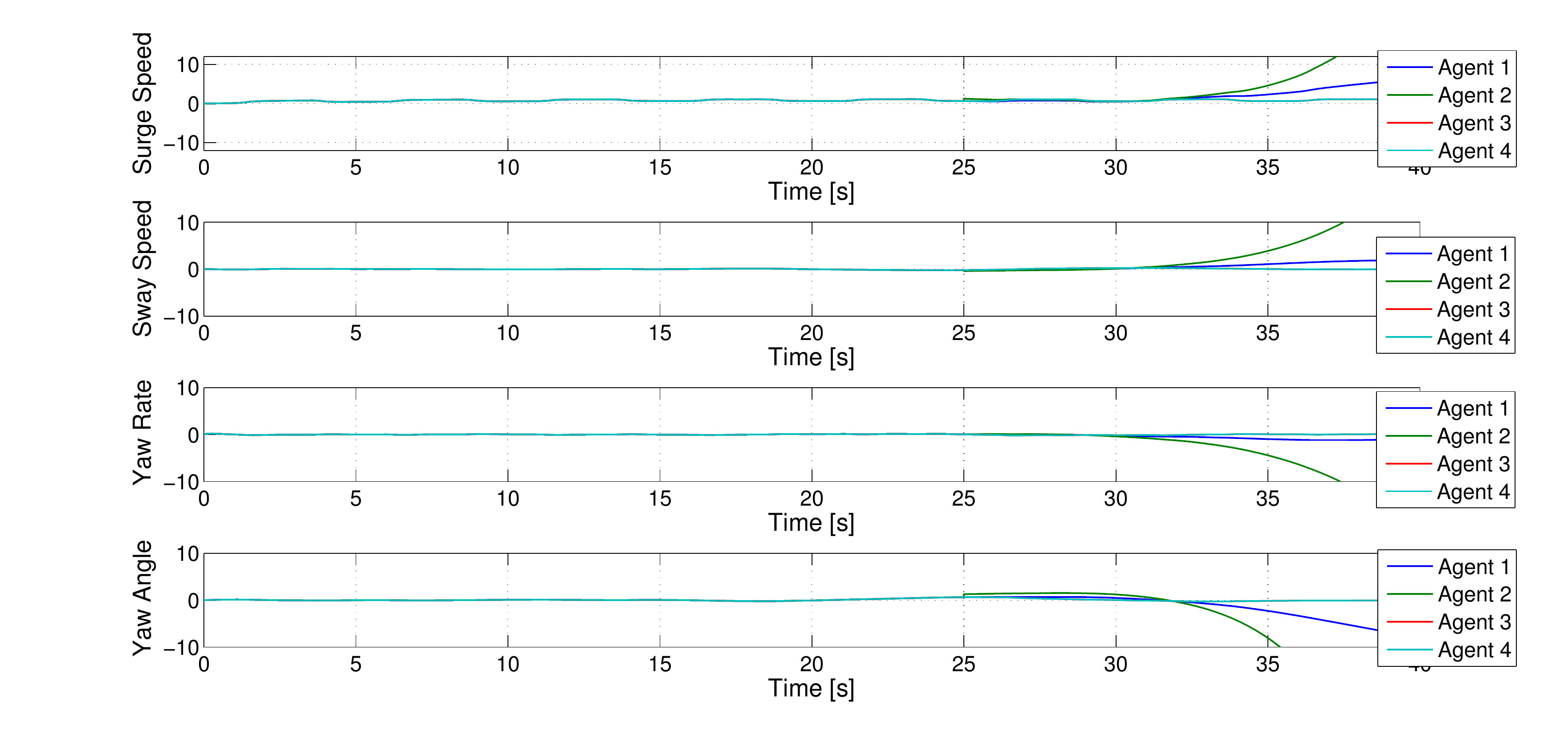}
      \caption{The  followers trajectories corresponding to the Scenario 1.}     
      \label{followers trajectory S1}
   \end{figure}

  The network topology  considered is as shown in Figure \ref{network}.  
   The leader control law is selected as $u_0(t)=K_0x_0(t)+F_0r(t)$, with  $K_0=1.0e+04 [k_{01};k_{02};k_{03};k_{04}]$,  
    $k_{01}=[ -0.52,   -3.5,   -0.65,   -8.0]$,  
   $k_{02}=[-0.22,-0.19,0.39,0.48]$, 
   $k_{03}=[-0.04,3.48 -0.04, 6.47]$, 
   $k_{04}=[-0.25,0.17,0.45,1.19]$, and $F_0= 1.0e+03 [f_{01};f_{02};f_{03};f_{04}]$, 
    $f_{01}=[3.02,-0.14,0.36,6.53]$, 
   $f_{02}=[-0.14,4.02,-1.93,-2.46]$, 
    $f_{03}=[0.36,-1.93,-1.55,3.61]$, 
    $f_{04}=[6.53,-2.46,3.61,-9.55]$, and the desired leader  speed $r(t)=\bar u_{\text{desired}}(t)$ is  defined according to Figure \ref{leader trajectory}. The objective of the team cooperative control is to ensure that all the agents follow the leader output (surge speed) trajectory, while their yaw angle,  sway and yaw rate remain bounded. The acceptable errors  between the desired trajectory and the actual trajectories are considered to be less than $10\%$ in the steady state. The following scenarios are now considered:\\
 \underline{  \textbf{Scenario 1}: Faulty team without control reconfiguration}: In this scenario, it is assumed that no control reconfiguration is invoked after the occurrence of the faults. The specifics for the mission considered are  as follows where the followers state trajectories are depicted in Figure \ref{followers trajectory S1}.\\
\textbf{A}) All the agents are healthy and the agent control law is designed according to Theorem \ref{theorem 1} and using YALMIP toolbox \cite{Yalmip} for MATLAB. The gains are obtained as $c_1=3.926e+06$, 
$K=1.0e03[k_1;k_2;k_3;k_4]$, 
$k_1=[-0.23,0.169 ,$ $-0.275,-0.039]$, 
  $k_2=[ -0.26,-0.175,$ $0.278,0.031]$,  
  $k_3=[ -0.23,0.169,-0.497,-0.039]$, 
   $k_4=[-0.026,-0.175 ,0.278,0.031]$,
     $C_2=1.0e+07\text{diag}\{1.23,1.39,0.39,1.48\}$ and  the $H_{\infty}$  upper bound is computed to be  $\gamma=0.3687$.
 \\
\textbf{B}) At time $t=t_f=25 \ s$, the agents $1$ and $2$ become faulty. Agent $1$ loses its second actuator i.e. $h_{1fs}^f(t)\equiv 0$, $t\geq25$. Agent $2$ loses $30\%$ of its first actuator and its second actuator gets stuck at $\underline{u}_2^{s}=1$, i.e. $h_{2fp}^f(t)= 0.7h_{2fp}(t) $ and $h_{2fs}^f(t)\equiv 1$ for $t\geq25 \ s$.\par
 Figure \ref{followers trajectory S1} clearly shows that  if a reconfiguration control strategy is not invoked, the agents become unstable and their states grow exponentially unbounded. Therefore, it is necessary to reconfigure the agent's control law after the occurrence of this fault. \par 
\begin{figure}[thpb]
      \centering
            \includegraphics[scale=0.4]{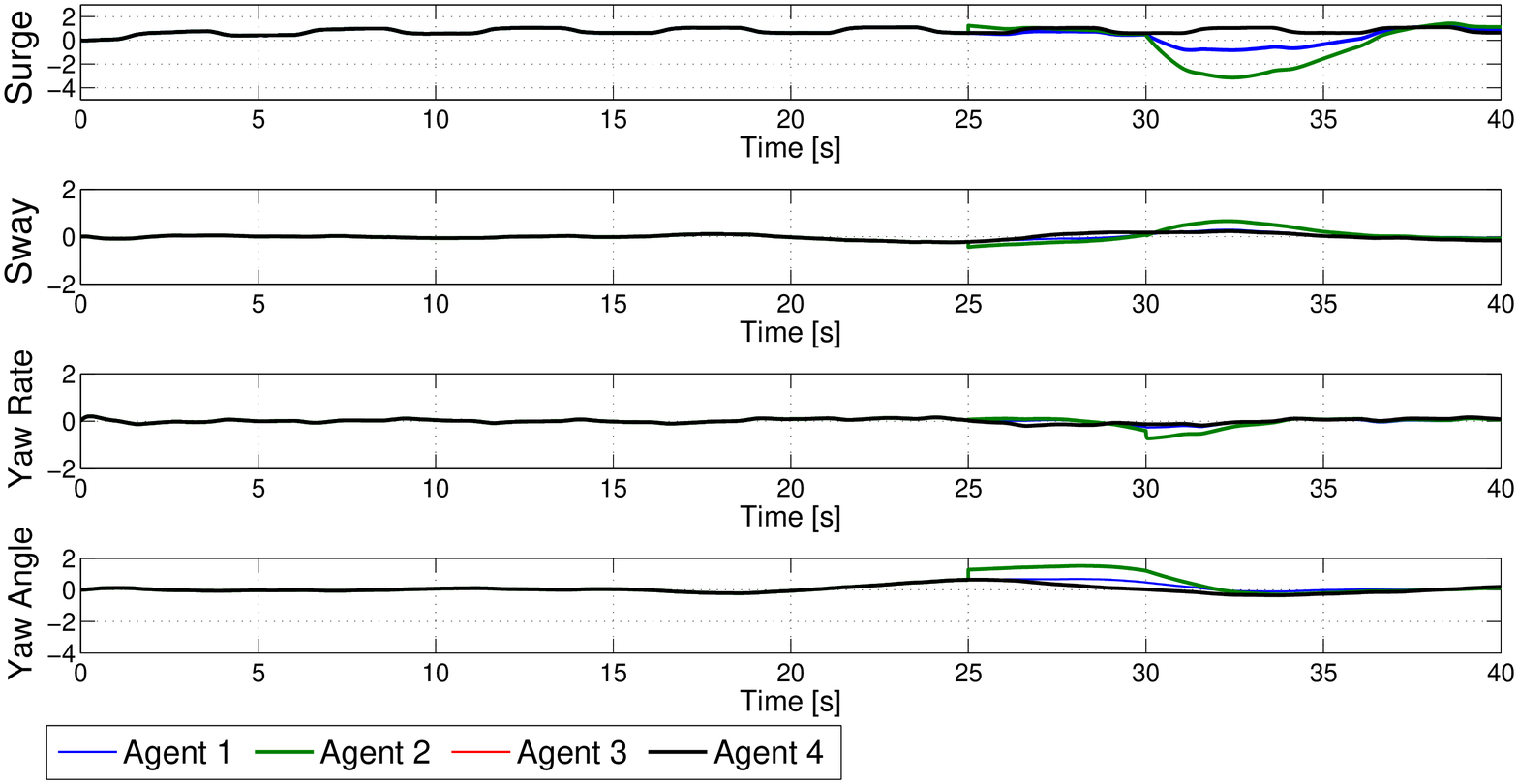}
      \caption{ The followers trajectories corresponding to the Scenario 2.}     
      \label{followers trajectory S2}
   \end{figure}

  \underline{\textbf{ Scenario 2}: Control reconfiguration subject to delays in invoking the reconfigured control law}: Unlike the previous scenario, in this scenario  control reconfiguration laws are invoked to the faulty agents. However, 
   it is assumed that there are delays in the time that the FDI module communicates this information to the faulty agents and the agents reconfigured controls are invoked.  
  The specifics for the execution of the mission  are as follows where the followers state trajectories are depicted in Figure \ref{followers trajectory S2}. \\
\textbf{A}) All the agents are healthy and the agent control law is similar to the Scenario 1.  \\
\textbf{B}) At time $t=t_f=25 \ s$, the agents $1$ and $2$ become faulty. The fault scenario that is considered is the same as that of  Step \textbf{B)} in Scenario 1. \\
\textbf{C}) The control laws for both faulty agents are reconfigured according to  Theorem \ref{theorem 2} at $t=t_r=30\ s$ and are set as 
$K_{11}^r=10e03  [k_{111};k_{112};k_{113}]$, 
$k_{111}=[ 2.627  , 0.090 ,  -1.895  , -4.328]$, $k_{112}=[-2.058 ,   0.09  ,  0.293   , 3.034]$, 
   $k_{113}=[-0.7549 ,  -0.1799  , 0.8115  , 0.4458]$,  
   
   $K_{21}^r=[k_{211},k_{212},k_{213}]$, 
   $k_{211}=[ 0.50,0.00,$ $0.50 ,0.00]$, $k_{212}=[ 0.50,0.00,0.499,-0.00]$, $k_{213}=[ 0.00,1.00, -0.00,1.00]$ and 
   $K_{12}^r=10e03[k_{121};k_{122};k_{123}]$, 
   $k_{121}=[  0.113  ,  0.130  , -0.485  , -2.89]$, 
    $k_{122}=[0.881 , -0.089 ,  -1.263  , 0.664]$, 
    $k_{123}=[-0.969  ,  0.18,$ $ 0.697  ,  0.4]$, 
  $K_{22}^r=[k_{211},k_{212},k_{213}]$, 
   $k_{211}=[ 0.7140,0.00,$ $0.7143 ,0.00]$, $k_{212}=[ 0.50,0.00,0.499,-0.00]$, $k_{213}=[ 0.00,1.00, -0.00,1.00]$ and 
 
   \par
   Figure \ref{followers trajectory S2}, depicts  that by invoking the reconfigured control laws one can now stabilize all the agents. 
   The delay in invoking the control reconfiguration causes a transient period in which the agent states diverge and will not follow the leader (refer to discussion in Subsection \ref{subsection 3}, \textbf{Case \RNum{5}}). However, after the transients have died out, the agent reach a consensus with the leader state.   \par
\begin{figure}[thpb]
      \centering
            \includegraphics[scale=0.4]{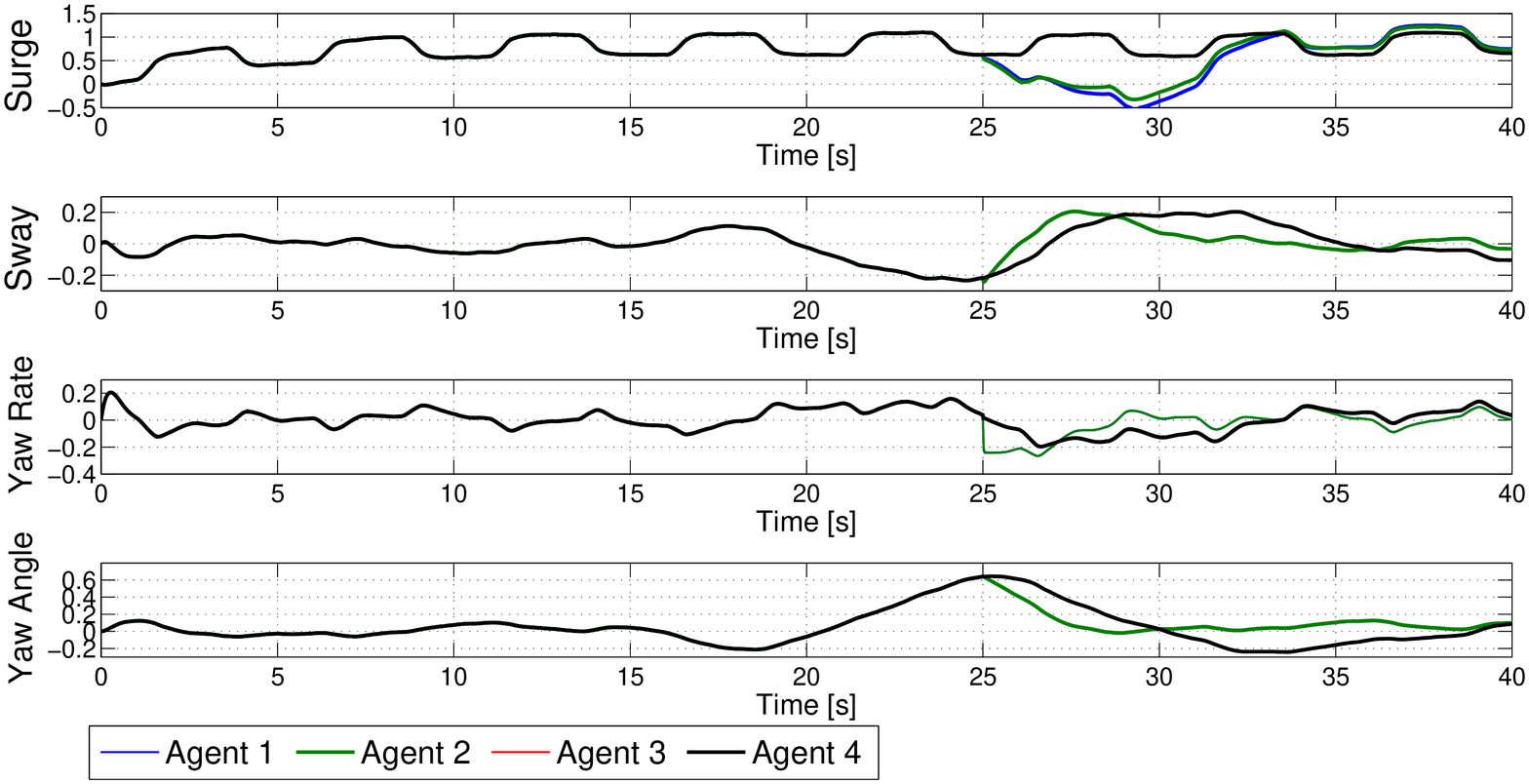}
      \caption{ The followers trajectories corresponding to the Scenario 3.}     
      \label{followers trajectory S3}
   \end{figure}

  \underline{  \textbf{Scenario 3}: Control reconfiguration subject to fault estimation uncertainties}: In this scenario, we  consider a similar  fault scenario as in the previous scenarios. However, it is assumed that the estimated fault severities are subject to unreliabilities, errors and uncertainties. Using the inequality (\ref{uncertainty bound}) the upper bound on  uncertainties is obtained as $ \bar{\epsilon}_{2max}^1=0.146$, implying that  the reconfigured control law stabilizes the errors provided that it  is designed based on $0.554\leq\hat{\Gamma}_2^1\leq0.846$. To investigate  how accurate this range is, 
   various levels of uncertainties and mismatches are considered and it is observed that the control gains that are designed for   {$\hat{\Gamma}_2^1< 0.86$} stabilize the errors whereas for {$\hat{\Gamma}_2^1\geq 0.87$} the state consensus errors become \underline{unstable}. This indicates that the bound  provided by the inequality (\ref{uncertainty bound}) provides  an acceptable approximation to the maximum allowable fault severities estimation errors and uncertainties. The  agents state simulation responses correspond to 
 $\hat\Gamma_2^1=0.6$ and $\hat{\underline{u}}_i^s=0.9$, and are depicted in Figure \ref{followers trajectory S3}.\par
 Figure \ref{followers trajectory S3} shows that by invoking the reconfigured control law, the agent states will no longer diverge and the recovery control strategy stabilizes the agent states. In fact, in this scenario the agents do follow the changes in the leader speed  trajectory, although the error between the faulty agent speed trajectory and the leader speed trajectory will not vanish but converges asymptotically to a small constant value. 
 
  \underline{  \textbf{Scenario 4}: Control reconfiguration subject to uncertainties in the fault isolation}: In this scenario, the effects of uncertainties in the fault isolation decision made by the FDI module are studied. It is assumed that the FDI module of agents $1$ and $2$ are subject to fault isolation uncertainties. Two cases are considered as follows: \\
 \underline{  \textbf{Scenario 4.1}}:\\
  \textbf{A}) Similar to Step \textbf{A)} in the Scenario 1.\\
  \textbf{B}) At  time $t=20$s the FDI module  launches a  false fault alarm for  the agent $2$, that the second actuator gets stuck at $\underline{u}_2^s=1$ and its first actuator loses $30\%$ of its efficiency.  
  Then, a reconfigured control is invoked to this agent.\\
  \underline{   \textbf{Scenario 4.2}}:\\
   \textbf{A}) Similar to Step \textbf{A)} in the Scenario 1.\\
  \textbf{B}) At  time $t=20$ s agent 2 becomes faulty and a fault scenario similar to Step \textbf{B)} in the Scenario $1$ occurs. However, the FDI module  does not detect and isolate this fault in the agent $2$ and instead the FDI module  wrongly initiates a fault alarm and a reconfigured control that is applied to the agent $1$. \par

\section{Conclusions }\label{conclusion}
In this work, cooperative and distributed reconfigurable control law strategies are developed and designed to control and reconfigure  faulty agents from three types of actuator faults, namely loss of effectiveness, outage, and stuck faults that guarantee boundedness  of the state consensus errors for a network of multi-agent systems.
  It is shown that the proposed control strategies can ensure an $H_{\infty}$ performance bound attenuation for the  team agents  when they are  subjected to environmental disturbances and actuator faults. 
 Our proposed reconfigured control laws ensure that the output of the faulty agent matches that of the healthy agent   in absence of disturbances.  Moreover, the control laws also guarantee that 
 the state consensus errors either remain bounded. Furthermore, in presence of environmental disturbances the $H_\infty$ disturbance attenuation bound is ensured to be minimized. The effectiveness of our proposed cooperative control and reconfigurable approaches are evaluated by applying them to a network of five autonomous underwater vehicles. Extensive simulation case studies are also considered to demonstrate  the capabilities and advantages of our proposed strategies subject to FDI module uncertainties, erroneous decisions, and imperfections.
\bibliographystyle{plain}
\bibliography{References.bib}

\begin{thebibliography}{10}

\bibitem{Azizi11}
S.~M. Azizi and K.~Khorasani.
\newblock Cooperative actuator fault accommodation in formation flight of
  unmanned vehicles using relative measurements.
\newblock {\em International Journal of Control}, 84(5):876--894, 2011.

\bibitem{Bacciotti99}
A.~Bacciotti and F.~Ceragioli.
\newblock Stability and stabilization of discontinuous systems and nonsmooth
  lyapunov functions.
\newblock {\em Control, Optimization and Calculus of Variations}, 4, 1999.

\bibitem{Basile92}
G.~Basile and G.~Marro.
\newblock {\em Controlled and Conditioned Invariants in Linear System Theory}.
\newblock Prentice Hall, Englewood Cliffs, NJ, 1992.

\bibitem{breger2008}
L.~Breger and J.~How.
\newblock Safe trajectories for autonomous rendezvous of spacecraft.
\newblock {\em Journal of Guidance, Control, and Dynamics}, 31(5):1478--1489,
  2008.

\bibitem{Chen2014}
S.~Chen, D.~Ho, L.~Li, and M.~Liu.
\newblock Fault-tolerant consensus of multi-agent system with distributed
  adaptive protocol.
\newblock {\em IEEE Transactions on Cybernetics}, 2015.

\bibitem{croomes2006}
S.~Croomes.
\newblock Overview of the $\text{DART}$ mishap investigation results.
\newblock {\em NASA Report}, pages 1--10, 2006.

\bibitem{fossen94}
T.~Fossen.
\newblock {\em Guidance and Control of Ocean Vehicles}.
\newblock Wiley, 1994.

\bibitem{fossen2002}
T.~Fossen.
\newblock {\em Marine control systems: Guidance, navigation and control of
  ships, rigs and underwater vehicles}.
\newblock Marine Cybernetics Trondheim, 2002.

\bibitem{ACC2014}
Z.~Gallehdari, N.~Meskin, and K.~Khorasani.
\newblock Cost performance based control reconfiguration in multi-agent
  systems.
\newblock In {\em 2014 American Control Conference}, pages 509--516, 2014.

\bibitem{ECC2014}
Z.~Gallehdari, N.~Meskin, and K.~Khorasani.
\newblock Robust cooperative control reconfiguration/recovery in multi-agent
  systems.
\newblock In {\em 2014 European Control Conference (ECC)}, pages 1554--1561,
  2014.

\bibitem{elec2:2015:MISC}
Z.~Gallehdari, N.~Meskin, and K.~Khorasani.
\newblock An {$H_\infty$} cooperative control fault recovery of multi-agent
  systems.
\newblock \url{http://arxiv.org/abs/1508.07076}, 2015.

\bibitem{jakuba2003}
M.~Jakuba.
\newblock Modeling and control of an autonomous underwater vehicle with
  combined foil/thruster actuators.
\newblock Master's thesis, Massachusetts Institute of Technology and Woods Hole
  Oceanographic Institution, 2003.

\bibitem{Li2012}
L.~Junquan and K.~Kumar.
\newblock Decentralized fault-tolerant control for satellite attitude
  synchronization.
\newblock {\em IEEE Transactions on Fuzzy Systems}, 20(3):572--586, 2012.

\bibitem{Khosrowjerdi2004}
M.~J. Khosrowjerdi, R.~Nikoukhah, and N.~Safari-Shad.
\newblock A mixed {H$_2$/H$_\infty$} approach to simultaneous fault detection
  and control.
\newblock {\em Automatica}, 40(2):261 -- 267, 2004.

\bibitem{Li13}
Z.~Li, X.~Liu, W.~Ren, and L.~Xie.
\newblock Distributed tracking control for linear multi-agent systems with a
  leader of bounded unknown input.
\newblock {\em IEEE Transactions on Automatic Control}, 58(2):518--523, 2013.

\bibitem{Liu2012}
L.~Liu, Y.~Shen, E.~H. Dowell, and C.~Zhu.
\newblock A general fault tolerant control and management for a linear system
  with actuator faults.
\newblock {\em Automatica}, 48(8):1676 -- 1682, 2012.

\bibitem{Lunze2006}
J.~Lunze and T.~Steffen.
\newblock Control reconfiguration after actuator failures using disturbance
  decoupling methods.
\newblock {\em IEEE Transactions on Automatic Control}, 51(10):1590--1601,
  2006.

\bibitem{Basil:2010:MISC}
G.~Marro.
\newblock Geometric approach toolbox.
\newblock
  \url{http://www3.deis.unibo.it/Staff/FullProf/GiovanniMarro/geometric.htm},
  2010.

\bibitem{Mehrabian2011-2}
A.R. Mehrabian, S.~Tafazoli, and K.~Khorasani.
\newblock Reconfigurable control of networked nonlinear {Euler-Lagrange}
  systems subject to fault diagnostic imperfections.
\newblock In {\em 50th IEEE Conference on Decision and Control and European
  Control Conference (CDC-ECC)}, pages 6380--6387, 2011.

\bibitem{Mehrabian2011}
A.R. Mehrabian, S.~Tafazoli, and K.~Khorasani.
\newblock State synchronization of networked {Euler-Lagrange} systems with
  switching communication topologies subject to actuator faults.
\newblock In {\em 18th IFAC World Congress Milano (Italy)}, 2011.

\bibitem{Movric2014}
K.~H. Movric and F.~L. Lewis.
\newblock Cooperative optimal control for multi-agent systems on directed graph
  topologies.
\newblock {\em IEEE Transactions on Automatic Control}, 59(3):769--774, 2014.

\bibitem{Xavier2012}
Xavier O.
\newblock $\text{FDI(R)}$ for satellites: How to deal with high availability
  and robustness in the space domain?
\newblock {\em International Journal of Applied Mathematics and Computer
  Science}, 22(1):99Ð107, 2012.

\bibitem{Semsar2009}
E.~Semsar-Kazerooni and K.~Khorasani.
\newblock Multi-agent team cooperation: A game theory approach.
\newblock {\em Automatica}, 45(10):2205--2213, 2009.

\bibitem{Shevitz94}
D.~Shevitz and B.~Paden.
\newblock Lyapunov stability theory of nonsmooth systems.
\newblock {\em IEEE Transactions on Automatic Control}, 39(9):1910--1914, 1994.

\bibitem{Tousi2012}
M.~M. Tousi and K.~Khorasani.
\newblock Optimal hybrid fault recovery in a team of unmanned aerial vehicles.
\newblock {\em Automatica}, 48(2):410 -- 418, 2012.

\bibitem{Wang2015-3}
X.~Wang and G.~H. Yang.
\newblock Cooperative adaptive fault-tolerant tracking control for a class of
  multi-agent systems with actuator failures and mismatched parameter
  uncertainties.
\newblock {\em IET Control Theory Applications}, 9(8):1274--1284, 2015.

\bibitem{Wang2015-2}
Y.~Wang, Y.~Song, and F.L. Lewis.
\newblock Robust adaptive fault-tolerant control of multiagent systems with
  uncertain nonidentical dynamics and undetectable actuation failures.
\newblock {\em IEEE Transactions on Industrial Electronics}, 62(6):3978--3988,
  2015.

\bibitem{Ren2010a}
R.~Wei.
\newblock Consensus tracking under directed interaction topologies: Algorithms
  and experiments.
\newblock {\em IEEE Transactions on Control Systems Technology},
  18(1):230--237, 2010.

\bibitem{Xiao2013}
B.~Xiao, Q.~Hu, and P.~Shi.
\newblock Attitude stabilization of spacecrafts under actuator saturation and
  partial loss of control effectiveness.
\newblock {\em IEEE Transactions on Control Systems Technology},
  21(6):2251--2263, 2013.

\bibitem{Wang2010}
G.~H. Yang and D.~Ye.
\newblock Adaptive actuator failure compensation control of uncertain nonlinear
  systems with guaranteed transient performance.
\newblock {\em Automatica}, 46(12):2082 -- 2091, 2010.

\bibitem{Yedavalli14}
R.~K. Yedavalli.
\newblock {\em Robust Control of Uncertain Dynamic Systems:A Linear State Space
  Approach}.
\newblock Springer New York, 2014.

\bibitem{Zhao2014}
L.~Zhao and Y.~Jia.
\newblock Neural network-based adaptive consensus tracking control for
  multi-agent systems under actuator faults.
\newblock {\em International Journal of Systems Science}, 21(6):2251--2263,
  2014.

\bibitem{Zhou2014}
B.~Zhou, W.~Wang, and H.~Ye.
\newblock Cooperative control for consensus of multi-agent systems with
  actuator faults.
\newblock {\em Computers $\&$ Electrical Engineering}, 40(7):2154 -- 2166,
  2014.

\end{thebibliography}







@book{rao1998,
  title={Matrix algebra and its applications to statistics and econometrics},
  author={Rao, Calyampudi Radhakrishna and Rao, M Bhaskara},
  year={1998},
  publisher={World Scientific}
}

@article{Marschak1955,
	title = {Elements for a Theory of Teams},
	author = {Marschak, J.},
	journal = {Management Science},
	volume = {1},
	number = {2},
	pages = {pp. 127-137},
	year = {1955},
	publisher = {INFORMS},
}
@article{ho1972,
	title={Team decision theory and information structures in optimal control problems--Part I},
	author={Ho, Yu-Chi and Chu, K.C.},
	journal={ IEEE Transactions on Automatic Control},
	volume={17},
	number={1},
	pages={15--22},
	year={1972},
	publisher={IEEE}
}
@ARTICLE{Chu72, 
	author={Chu, K.C.}, 
	journal={ IEEE Transactions on Automatic Control}, 
	title={Team decision theory and information structures in optimal control problems--Part II}, 
	year={1972}, 
	volume={17}, 
	number={1}, 
	pages={22-28}, 
	,}

@ARTICLE{Ho80, 
	author={Ho, Y.-C.}, 
	journal={Proceedings of the IEEE}, 
	title={Team decision theory and information structures}, 
	year={1980}, 
	volume={68}, 
	number={6}, 
	pages={644-654}, 
	,}
@inproceedings{guo2010,
	title={Adaptive leader-follower formation control for autonomous mobile robots},
	author={Guo, Jing and Lin, Zhiyun and Cao, Ming and Yan, Gangfeng},
	booktitle={Proceedings of the 2010 American Control Conference},
	pages={6822--6827},
	year={2010},
	organization={IEEE}
}
@inproceedings{ren2004formation,
	title={Formation feedback control for multiple spacecraft via virtual structures},
	author={Ren, Wei and Beard, Randal W},
	booktitle={IEE Proceedings Control Theory and Applications},
	volume={151},
	number={3},
	pages={357--368},
	year={2004},
	organization={IET}
}
@article{beard2001coordination,
	title={A coordination architecture for spacecraft formation control},
	author={Beard, Randal W and Lawton, Jonathan and Hadaegh, Fred Y},
	journal={  IEEE Transactions on Control Systems Technology},
	volume={9},
	number={6},
	pages={777--790},
	year={2001},
	publisher={IEEE}
}

@INPROCEEDINGS{Broek2009, 
	author={Van den Broek, T. H A and Van de Wouw, N. and Nijmeijer, H.}, 
	booktitle={ Proceedings of the 48th IEEE Conference on Decision and Control held jointly with the 28th Chinese Control Conference. CDC/CCC 2009 }, 
	title={Formation control of unicycle mobile robots: a virtual structure approach}, 
	year={2009}, 
	pages={8328-8333}, 
	,}
@phdthesis{fax2001,
	title={Optimal and cooperative control of vehicle formations},
	author={Fax, J Alexander},
	year={2001},
	school={California Institute of Technology}
}
@INPROCEEDINGS{Saber03, 
	author={Saber, R.O. and Murray, R.M.}, 
	booktitle={Proceedings of the 2003 American Control Conference. }, 
	title={Consensus protocols for networks of dynamic agents}, 
	year={2003}, 
	volume={2}, 
	pages={951-956}, 
	,}
@ARTICLE{Saber04, 
	author={Olfati-Saber, R. and Murray, R.M.}, 
	journal={ IEEE Transactions on Automatic Control}, 
	title={Consensus problems in networks of agents with switching topology and time-delays}, 
	year={2004}, 
	volume={49}, 
	number={9}, 
	pages={1520-1533}, 
	,}
@ARTICLE{Jadbabaie2003, 
	author={Jadbabaie, A. and Jie Lin and Morse, A.S.}, 
	journal={IEEE Transactions on Automatic Control}, 
	title={Coordination of groups of mobile autonomous agents using nearest neighbor rules}, 
	year={2003}, 
	volume={48}, 
	number={6}, 
	pages={988-1001}, 
	,}
@ARTICLE{Ren2008, 
	author={Wei Ren}, 
	journal={IEEE Transactions on Automatic Control}, 
	title={On Consensus Algorithms for Double-Integrator Dynamics}, 
	year={2008}, 
	volume={53}, 
	number={6}, 
	pages={1503-1509}, 
	,}
@article{yang2011,
	title={Consensus of second-order multi-agent systems with exogenous disturbances},
	author={Yang, Hongyong and Zhang, Zhenxing and Zhang, Siying},
	journal={International Journal of Robust and Nonlinear Control},
	volume={21},
	number={9},
	pages={945--956},
	year={2011},
	publisher={Wiley Online Library}
}

@ARTICLE{Qin2012, 
	author={Jiahu Qin and Wei-Xing Zheng and Huijun Gao}, 
	journal={IEEE Transactions on Systems, Man, and Cybernetics, Part B: Cybernetics }, 
	title={Coordination of Multiple Agents With Double-Integrator Dynamics Under Generalized Interaction Topologies}, 
	year={2012}, 
	volume={42}, 
	number={1}, 
	pages={44-57}, 
	,}
@article{Qin2011,
	title = "Second-order consensus for multi-agent systems with switching topology and communication delay ",
	journal = "Systems and Control Letters ",
	volume = "60",
	number = "6",
	pages = "390 - 397",
	year = "2011",
	author = "Jiahu Qin and Huijun Gao and Wei Xing Zheng",
}
@INPROCEEDINGS{Gupta2004, 
	author={V. Gupta, B. Hassibi, and R. Murray}, 
	booktitle={Proceedings of the 2004 American Control Conference}, 
	title={On the Synthesis of Control Laws for a Network of Autonomous  Agents}, 
	year={2004}, 
	volume={6}, 
	pages={4927-4932}, 
	,}
@INPROCEEDINGS{Bauso2006, 
	author={Bauso, D. and Giarre, L. and Pesenti, R.}, 
	booktitle={45th IEEE Conference on Decision and Control}, 
	title={Mechanism Design for Optimal Consensus Problems}, 
	year={2006}, 
	pages={3381-3386}, 
	,}
@article{Delvenne2009,
	title = "Optimal strategies in the average consensus problem ",
	journal = "Systems and Control Letters ",
	volume = "58",
	number = "10-11",
	pages = "759 - 765",
	year = "2009",
	author = "Jean-Charles Delvenne and Ruggero Carli and Sandro Zampieri",
}
@ARTICLE{Semsar2010, 
	author={Semsar-Kazerooni, E. and Khorasani, K.}, 
	journal={IEEE Transactions on Systems, Man, and Cybernetics, Part B: Cybernetics}, 
	title={Optimal Consensus Seeking in a Network of Multi-Agent Systems: An LMI Approach}, 
	year={2010}, 
	volume={40}, 
	number={2}, 
	pages={540-547}, 
	,}

@ARTICLE{Semsar2010-2,
	author={Semsar-Kazerooni, E. and Khorasani, K.},
	journal={Control Systems Technology, IEEE Transactions on},
	title={Team Consensus for a Network of Unmanned Vehicles in Presence of Actuator Faults},
	year={2010},
	volume={18},
	number={5},
	pages={1155-1161},
	,}
@ARTICLE{Ma2010,
author={ Ma, C. Q. and  Zhang, J. F.},
journal={ IEEE Transactions on Automatic Control},
title={Necessary and Sufficient Conditions for Consensusability of Linear Multi-Agent Systems},
year={2010},
volume={55},
number={5},
pages={1263-1268},}

@INPROCEEDINGS{Scharf2004, 
	author={Scharf, D.P. and Hadaegh, F.Y. and Ploen, S.R.}, 
	booktitle={Proceedings of the 2004 American Control Conference}, 
	title={A survey of spacecraft formation flying guidance and control. Part II: control}, 
	year={2004}, 
	volume={4}, 
	pages={2976-2985 vol.4}}

@article{SemsarKazerooni2008,
title = "Optimal consensus algorithms for cooperative team of agents subject to partial information ",
journal = "Automatica ",
volume = "44",
number = "11",
pages = "2766 - 2777",
year = "2008",
author = "Semsar-Kazerooni, E.  and  Khorasani, K.",
}


@ARTICLE{Cui2010,
author={ Ma, C. Q. and  Zhang, J. F.},
journal={ IEEE Transactions on Automatic Control},
title={Necessary and Sufficient Conditions for Consensusability of Linear Multi-Agent Systems},
year={2010},
volume={55},
number={5},
pages={1263-1268},
}

@article{beard2001,
  title={A coordination architecture for spacecraft formation control},
  author={Beard, R. W. and Lawton, J. and Hadaegh, F. Y.},
  journal={  IEEE Transactions on Control Systems Technology},
  volume={9},
  number={6},
  pages={777-790},
  year={2001},
  publisher={IEEE}
}

@ARTICLE{Ren2008, 
author={Wei, R.}, 
journal={IEEE Transactions on Automatic Control}, 
title={On Consensus Algorithms for Double-Integrator Dynamics}, 
year={2008}, 
volume={53}, 
number={6}, 
pages={1503-1509}, 
,}

@article{jiang2005,
  title={Fault-tolerant control systems-an introductory overview},
  author={Jiang, J.},
  journal={Acta Automatica Sinica},
  volume={31},
  number={1},
  pages={161--174},
  year={2005}
}

@article{Wang2010,
author={ Yang, G. H. and  Ye, D.}, 
title = "Adaptive actuator failure compensation control of uncertain nonlinear systems with guaranteed transient performance",
journal = "Automatica",
volume = "46",
number = "12",
pages = "2082 - 2091",
year = "2010",
}

@article{zhang2008,
  title={Bibliographical review on reconfigurable fault-tolerant control systems},
  author={Zhang, Y. and Jiang, J.},
  journal={Annual Reviews in Control},
  volume={32},
  number={2},
  pages={229--252},
  year={2008},
  publisher={Elsevier}
}

@article{Jiang2012,
title = "Fault-tolerant control systems: A comparative study between active and passive approaches",
journal = "Annual Reviews in Control",
volume = "36",
number = "1",
pages = "60 - 72",
year = "2012",
author = " Jiang, J. and  Yu, X.",
}
@article{Liao02,
author={ Liao, F. and  Liang Wang, J. and  Yang, G. H.},
journal="IEEE Transactions on Control Systems Technology",
 title="Reliable robust flight tracking control: an {L}{M}{I} approach",
year="2002",
volume="10",
number="1",
pages="76 -89",
ISSN="1063-6536",}

@article{Khosrowjerdi2004,
 author = {Khosrowjerdi, M. J. and Nikoukhah, R. and Safari-Shad, N.},
title = "A mixed {H$_2$/H$_\infty$} approach to simultaneous fault detection and control",
journal = "Automatica",
volume = "40",
number = "2",
pages = "261 - 267",
year = "2004",
note = "",
issn = "0005-1098",
}


@ARTICLE{Benosman10,
author={Benosman, M. and Lum, K. Y.},
journal={ IEEE Transactions on Control Systems Technology},
 title={Passive Actuators' Fault-Tolerant Control for Affine Nonlinear Systems},
year={2010},
volume={18},
number={1},
pages={152 -163},
ISSN={1063-6536},}

@article{Edwards2006,
title = "Sensor fault tolerant control using sliding mode observers",
journal = "Control Engineering Practice",
volume = "14",
number = "8",
pages = "897 - 908",
year = "2006",
author = " Edwards, C. and  Tan, C. P.",
}
@article{Alwi2008,
title = "Fault tolerant control using sliding modes with on-line control allocation",
journal = "Automatica",
volume = "44",
number = "7",
pages = "1859 - 1866",
year = "2008",
author = "Alwi, H. and  Edwards, C.",
}

@ARTICLE{Wen12,
author={ Liang, Y. W. and  Ting, L. W. and  Lin, L. G.},
journal={ IEEE Transactions on Industrial Electronics},
 title={Study of Reliable Control Via an Integral-Type Sliding Mode Control Scheme},
year={2012},
volume={59},
number={8},
pages={3062 -3068},}

@ARTICLE{Li12,
author={Li, X. J. and Yang, G. H.},
journal={ IET Control Theory Applications},
 title={Robust adaptive fault-tolerant control for uncertain linear systems with actuator failures},
year={2012},
volume={6},
number={10},
pages={1544 -1551},
ISSN={1751-8644},
}

@article{Wang2010,
author={ Yang, G. H. and  Ye, D.}, 
title = "Adaptive actuator failure compensation control of uncertain nonlinear systems with guaranteed transient performance",
journal = "Automatica",
volume = "46",
number = "12",
pages = "2082 - 2091",
year = "2010",
}

@ARTICLE{Yang10,
author={ Yang, G. H. and  Ye, D.},
journal={ IEEE Transactions on Automatic Control}, 
title={Reliable ${H_\infty}$ Control of Linear Systems With Adaptive Mechanism},
year={2010},
volume={55},
number={1},
pages={242 -247}, 
}

@article{Mao2011,
title = "Observer-based fault-tolerant control for a class of networked control systems with transfer delays",
journal = "Journal of the Franklin Institute",
volume = "348",
number = "4",
pages = "763 - 776",
year = "2011",
author = "Mao, Z. and Jiang, B. and  Shi, P."
}

@article{ZHAO2012,
title = "Fault Tolerant Control for Networked Control Systems with Access Constraints",
journal = "Acta Automatica Sinica",
volume = "38",
number = "7",
pages = "1119 - 1126",
year = "2012",
author = " Zhao, M. Y. and  Liu, H. P. and  Li, Z. J. and  Sun, D. H. and  Liu, K. P.",
}

@article{MAO2007,
title = "Fault Estimation and Accommodation for Networked Control Systems with Transfer Delay",
journal = "Acta Automatica Sinica",
volume = "33",
number = "7",
pages = "738 - 743",
year = "2007",
author = " Hui Mao, Z. and  Jiang, B. ",
}

@ARTICLE{Lavretsky09,
author={Lavretsky, E.},
journal={ IEEE Transactions on Automatic Control},
 title={Combined/Composite Model Reference Adaptive Control},
year={2009},
volume={54},
number={11},
pages={2692 -2697},
}
@article{Zhang2009,
title = "Adaptive actuator fault compensation for linear systems with matching and unmatching uncertainties",
journal = "Journal of Process Control",
volume = "19",
number = "6",
pages = "985 - 990",
year = "2009",
author = " Zhang, Y. and  Joe Qin, S.",
}


@article{Liu2012,
title = "A general fault tolerant control and management for a linear system with actuator faults",
journal = "Automatica",
volume = "48",
number = "8",
pages = "1676 - 1682",
year = "2012",
author = " Liu, L. and  Shen, Y. and  Dowell, E. H. and  Zhu, C.",
}

@ARTICLE{HaoYang10,
author={ Yang, H. and Cocquempot, V. and  Jiang, B.},
journal={ IEEE Transactions on Intelligent Transportation Systems}, 
title={Optimal Fault-Tolerant Path-Tracking Control for 4WS4WD Electric Vehicles},
year={2010},
volume={11},
number={1},
pages={237 -243},
}

@ARTICLE{CLiu11,
author={Liu, C. S. and Jiang, B. and Zhang, S. J.},
journal={ IET Control Theory Applications}, 
title={Fault-tolerant synthesis controller design for a flight-tracking system},
year={2011},
volume={5},
number={11},
pages={1243 -1254},}

@article{Juan12,
year={2012},
journal={International Journal of Control, Automation and Systems},
volume={10},
issue={1},
title={Fault diagnosis and optimal fault-tolerant control for systems with delayed measurements and states},
publisher={Institute of Control, Robotics and Systems and The Korean Institute of Electrical Engineers},
author={Li, J. and Gao, H. W. and Zhang, P. and Huang, D. R.},
pages={150-157},
}

@article{abel2011,
  title={Coordinated Fault Tolerant Optimal Control of Autonomous Agents: Geometry and Communications Architecture},
  author={Abel, R. O. and Dasgupta, S. and Kuhl, J. G.},
  journal={Communications in Information and Systems},
  volume={11},
  number={2},
  pages={173},
  year={2011}
}
@article{izadi2011,
  title={Decentralized model predictive control for cooperative multiple vehicles subject to communication loss},
  author={Izadi, H. A. and Gordon, B. W. and Zhang, Y.},
  journal={International Journal of Aerospace Engineering},
  volume={2011},
  year={2011},
  publisher={Hindawi Publishing Corporation}
}

@article{pasqualetti2012,
  title={Consensus computation in unreliable networks: A system theoretic approach},
  author={Pasqualetti, F. and Bicchi, A. and Bullo, F.},
  journal={IEEE Transactions on Automatic Control},
  volume={57},
  number={1},
  pages={90--104},
  year={2012},
  publisher={IEEE}
}

@article{Zhu2013,
title = "On distributed constrained formation control in operator⠀"vehicle adversarial networks ",
journal = "Automatica ",
volume = "49",
number = "12",
pages = "3571 - 3582",
year = "2013",
author="Zhu, M. and Martinez, S."
}


@INPROCEEDINGS{Tousi09,
author={Tousi, M. M. and Khorasani, K.},
booktitle={ 3rd Annual IEEE Systems Conference}, 
title={An optimal hybrid fault recovery for a team of unmanned vehicles},
year={2009},
volume={},
number={},
pages={241 -246},
}

@article{Tousi2012,
title = "Optimal hybrid fault recovery in a team of unmanned aerial vehicles",
journal = "Automatica",
volume = "48",
number = "2",
pages = "410 - 418",
year = "2012",
author = " Tousi, M. M. and K. Khorasani",
}

@article{Azizi11,
author = {Azizi, S. M. and Khorasani, K.},
title = {Cooperative actuator fault accommodation in formation flight of unmanned vehicles using relative measurements},
journal = {International Journal of Control},
volume = {84},
number = {5},
pages = {876-894},
year = {2011},
}

@INPROCEEDINGS{Azizi10,
author={Azizi, S. M. and Khorasani, K.},
booktitle={ 49th IEEE Conference on Decision and Control },
 title={Cooperative actuator fault accommodation of formation flying vehicles with absolute measurements},
year={2010},
volume={},
number={},
pages={6299 -6304},}

@INPROCEEDINGS{mehrabian2010,
	author={Mehrabian, A. R.  and S. Tafazoli, and K. Khorasani},
	booktitle={2010 Conference on Control and Fault Tolerant Systems },
	title={Velocity Synchronization of Networked {Euler-Lagrange} Systems with Switching Network Topologies Subject to Actuator Faults},
	year={2010},
	volume={},
	number={},
	pages={6299 -6304}
	}
@INPROCEEDINGS{Mehrabian2011-2,
	author={Mehrabian, A.R. and Tafazoli, S. and Khorasani, K.},
	booktitle={ 50th IEEE Conference on Decision and Control and European Control Conference (CDC-ECC)},
	title={Reconfigurable control of networked nonlinear {Euler-Lagrange} systems subject to fault diagnostic imperfections},
	year={2011},
	pages={6380-6387},
	}

@INPROCEEDINGS{mehrabian2011,
	author={Mehrabian, A.R.  and S. Tafazoli, and K. Khorasani},
	booktitle={18th IFAC World Congress Milano (Italy) },
	title={State Synchronization of Networked {Euler-Lagrange} Systems with Switching Communication Topologies Subject to Actuator Faults},
	year={2011},
	volume={},
	number={},
	pages={6299 -6304}
	}
@ARTICLE{mehrabian2012, 
	author={Mehrabian, A.R. and Tafazoli, S. and Khorasani, K.}, 
	journal={ IET Control Theory Applications}, 
	title={Reconfigurable cooperative control of networked {Lagrangian} systems under actuator saturation constraints}, 
	year={2012}, 
	volume={6}, 
	number={4}, 
	pages={578-587}, 
}
@INPROCEEDINGS{Mehrabian2011, 
author={Mehrabian, A.R. and Tafazoli, S. and Khorasani, K.}, 
booktitle={50th IEEE Conference on Decision and Control and European Control Conference (CDC-ECC)}, 
title={Distributed $H_\infty$ optimal control of networked uncertain nonlinear {Euler-Lagrange} systems with switching communication network topologies}, 
year={2011}, 
month={Dec}, 
pages={1379-1386},}


@ARTICLE{Boskovic10,
author={Boskovic, J. D. and Mehra, R. K.},
journal={ IEEE Transactions on Control Systems Technology},
 title={A Decentralized Fault-Tolerant Control System for Accommodation of Failures in Higher-Order Flight Control Actuators},
year={2010},
volume={18},
number={5},
pages={1103 -1115},
}

@article{SaludesRodil2010,
title = "Fault tolerance in the framework of support vector machines based model predictive control",
journal = "Engineering Applications of Artificial Intelligence",
volume = "23",
number = "7",
pages = "1127 - 1139",
year = "2010",
author = " Saludes Rodil, S. and  Fuente, M. J.",
}

@article{Ye2010,
title = "Fault-Tolerant Control for a Class of Uncertain Systems with Actuator Faults",
journal = " Tsinghua Science and Technology",
volume = "15",
number = "2",
pages = "174 - 183",
year = "2010",
author = "Ye, S.  and  Zhang, Y. and Wang,  X. and  Jiang, B.",
}

@article{Simandl2009,
title = "Active fault detection and control: Unified formulation and optimal design",
journal = "Automatica",
volume = "45",
number = "9",
pages = "2052 - 2059",
year = "2009",
author = "Simandl, M.  and  Puncochar, I. ",
}
@article{Chilin2010,
title = "Detection, isolation and handling of actuator faults in distributed model predictive control systems",
journal = "Journal of Process Control",
volume = "20",
number = "9",
pages = "1059 - 1075",
year = "2010",
author =" Chilin, D. and  Liu, J. and  de la Pena, D. M. and  Christofides, P. D. and  Davis, J. F.",
}

@incollection{Ciubotaru2007,
title = "Fault tolerant control of the {Boeing} 747 short-period mode using the admissible model matching technique",
editor = "Hong-Yue Zhang",
booktitle = "Fault Detection, Supervision and Safety of Technical Processes",
publisher = "Elsevier Science Ltd",
year = "2007",
pages = "819 - 824",
author = "Ciubotaru, B. and Staroswiecki,  M. and  Christophe, C."
}

@article{Richter2011,
title = "Reconfigurable control of piecewise affine systems with actuator and sensor faults: Stability and tracking",
journal = "Automatica",
volume = "47",
number = "4",
pages = "678 - 691",
year = "2011",
author = " Richter, J.H. and  Heemels, W. P. M. H. and van de Wouw ,  N. and  Lunze, J.",
}
@article{Zou2011,
title = "Adaptive fuzzy fault-tolerant attitude control of spacecraft",
journal = "Control Engineering Practice",
volume = "19",
number = "1",
pages = "10 - 21",
year = "2011",
author = " Zou, A. M. and  Dev Kumar, K.",
}

@INPROCEEDINGS{Boskovic05,
author={Boskovic, J. D. and Bergstrom, S. E. and Mehra, R. K.},
booktitle={ Proceedings of the 2005 American Control Conference, ACC2005},
 title={Retrofit reconfigurable flight control in the presence of control effector damage},
year={2005},
volume={4},
number={},
pages={ 2652 - 2657},
}


@article{yetendje2012,
  title={Robust multisensor fault tolerant model-following MPC design for constrained systems},
  author={Yetendje, A. and Seron, M. and De Don{\'a}, J.},
   journal={Int. J. Appl. Math. Comput. Sci.},
   volume={22},
  pages={211 - 223},
    year={2012}
}

@article{chung1998,
  title={A game theoretic fault detection filter},
  author={Chung, W. H. and Speyer, J. L.},
  journal={ IEEE Transactions on Automatic Control},
  volume={43},
  number={2},
  pages={143--161},
  year={1998},
  publisher={IEEE}
}

@article{Niemann05,
author = {Niemann  , H. and Stoustrup, J.},
title = {An architecture for fault tolerant controllers},
journal = {International Journal of Control},
volume = {78},
number = {14},
pages = {1091-1110},
year = {2005},
}

@article{massoumnia1989,
  title={Failure detection and identification},
  author={Massoumnia, M. and Verghese, G. and Willsky, A. S.},
  journal={IEEE Transactions on Automatic Control},
  volume={34},
  number={3},
  pages={316--321},
  year={1989},
  publisher={IEEE}
}

@article{Ran1988,
title = "Existence and comparison theorems for algebraic Riccati equations for continuous- and discrete-time systems ",
journal = "Linear Algebra and its Applications ",
volume = "99",
number = "0",
pages = "63 - 83",
year = "1988",
author = " Ran, A. C. M. and  Vreugdenhil, R."
}

@book{bang2008,
  title={Digraphs: theory, algorithms and applications},
  author={Bang-Jensen, J. and Gutin, G. Z.},
  year={2008},
  publisher={Springer}
}
@article{olfati2004,
  title={Consensus problems in networks of agents with switching topology and time-delays},
  author={Olfati-Saber, R. and Murray, R. M.},
  journal={IEEE Transactions on Automatic Control},
  volume={49},
  number={9},
  pages={1520--1533},
  year={2004},
  publisher={IEEE}
}
@article{olfati2007,
  title={Consensus and cooperation in networked multi-agent systems},
  author={Olfati-Saber, R. and Fax, J. A. and Murray, R. M.},
  journal={Proceedings of the IEEE},
  volume={95},
  number={1},
  pages={215--233},
  year={2007},
  publisher={IEEE}
  }
@incollection{Ren05,
year={2005},
isbn={978-3-540-22861-5},
booktitle={Cooperative Control},
volume={309},
series={Lecture Notes in Control and Information Science},
doi={10.1007/978-3-540-31595-7_10},
title={Coordination Variables and Consensus Building in Multiple Vehicle Systems},
publisher={Springer Berlin Heidelberg},
author={Ren, W. and Beard, R. and McLain, T.},
pages={171-188}
}

@ARTICLE{Ma10,
author={ Ma, C. Q. and  Zhang, J. F.},
journal={IEEE Transactions on Automatic Control},
 title={Necessary and Sufficient Conditions for Consensusability of Linear Multi-Agent Systems},
year={2010},
volume={55},
number={5},
pages={1263-1268},
}

@book{ren2007,
  title={Distributed consensus in multi-vehicle cooperative control: theory and applications},
  author={Ren, W. and Beard, R.},
  year={2007},
  publisher={Springer}
}

@book{boyd1987,
  title={Linear matrix inequalities in system and control theory},
  author={Boyd, S. and El Ghaoul, L. and Feron, E. and Balakrishnan, V.},
  volume={15},
  year={1987},
  publisher={Society for Industrial Mathematics}
}

@article{Semsar2009,
 author = {Semsar-Kazerooni, E. and Khorasani, K.},
 title = {Multi-agent team cooperation: A game theory approach},
 journal = {Automatica},
 issue_date = {October, 2009},
 volume = {45},
 number = {10},
 year = {2009},
 pages = {2205--2213},
 numpages = {9},
 publisher = {Pergamon Press, Inc.},
} 
@ONLINE{penbmi,
 title ={PENOPT}
url ={http://www.penopt.com/}
}


@article{CAO1998,
title = "Static Output Feedback Stabilization: An {ILMI} Approach",
journal = "Automatica",
volume = "34",
number = "12",
pages = "1641 - 1645",
year = "1998",
author = " Cao, Y. and  Lam , J. and  Sun, Y. X.",
}
@ARTICLE{Zhongkui10,
author={ Li, Z. and  Duan, Z. and  Chen, G. and Huang,  L.},
journal={ IEEE Transactions on Circuits and Systems I: Regular Papers},
 title={Consensus of Multiagent Systems and Synchronization of Complex Networks: A Unified Viewpoint},
year={2010},
volume={57},
number={1},
pages={213 -224},}




@ARTICLE{Kim11,
author={ Kim, H. and  Shim, H. and  Seo, J. H.},
journal={IEEE Transactions on Automatic Control},
title={Output Consensus of Heterogeneous Uncertain Linear Multi-Agent Systems},
year={2011},
volume={56},
number={1},
pages={200-206},
}
@ARTICLE{Xu13,
author={ Xu, J,  and  Xie, L. and  Li,  T. and  Yew Lum, K.},
journal={Control Theory Applications, IET}, 
title={Consensus of multi-agent systems with general linear dynamics via dynamic output feedback control},
year={2013},
volume={7},
number={1},
pages={108-115},}

@book{Anderson89,
  title={Optimal Control Linear Quadratic Methods},
  author={ Anderson, B.  and  Moore, J. B.},
  volume={},
  year={1989},
  publisher={Prentice-Hall, Inc}
}

@book{Naidu03,
  title={Optimal Control Systems},
  author={Subbaram Naidu, D. },
  volume={},
  year={2003},
  publisher={CRC PRESS}
}
@inproceedings{yu2004,
  title={Optimal guaranteed cost control of linear uncertain systems with input constraints},
  author={Yu, L. and Xu, J. M. and Han, Q. L.},
  booktitle={Intelligent Control and Automation, 2004. WCICA 2004. Fifth World Congress on},
  volume={1},
  pages={553--557},
  year={2004},
  organization={IEEE}
}

@INPROCEEDINGS{Yang11,
author={Yang, T. and Saberi, A. and Stoorvogel, A. A. and Grip, H. F.},
booktitle={50th IEEE Conference on Decision and Control and European Control Conference (CDC-ECC 2011)}
, title={Output consensus for networks of non-identical introspective agents},
year={2011},
pages={1286-1292},}

@book{Johnson1985,
  title={Matrix Analysis},
  author={Horn, R. A. and Johnson, C. R. },
  volume={},
  year={1985},
  publisher={Cambridge University Press}
}

@mastersthesis{jakuba2003,
  title={Modeling and control of an autonomous underwater vehicle with combined foil/thruster actuators},
  author={Jakuba, M.},
  year={2003},
  school={Massachusetts Institute of Technology and Woods Hole Oceanographic Institution}
}

@ARTICLE{Healey93, 
author={ Healey,  A.J. and  Lienard, D. }, 
journal={ IEEE Journal of Oceanic Engineering}, 
title={Multivariable sliding mode control for autonomous diving and steering of unmanned underwater vehicles}, 
year={1993}, 
volume={18}, 
number={3}, 
pages={327-339}}

@INPROCEEDINGS{Staroswiecki09, 
author={ Staroswiecki M. and Berdjag,  D.   and Jiang, B.  and  Zhang, K. }, 
booktitle={ Proceedings of the 48th IEEE Conference on Decision and Control, 2009 held jointly with the 2009 28th Chinese Control Conference. CDC/CCC 2009 }, 
title={PACT : a passive / active approach to fault tolerant stability under actuator outages}, 
year={2009}, 

pages={7819-7824}, 
}
@article{Staroswiecki10,
author = { Staroswiecki, M. and Berdjag,  D. },
title = {A general fault tolerant linear quadratic control strategy under actuator outages},
journal = {International Journal of Systems Science},
volume = {41},
number = {8},
pages = {971-985},
year = {2010
}
@ARTICLE{Staroswiecki12, 
author={ Staroswiecki, M. and  Zhang, K. and  Berdjag, D. and Abbas-Turki,  M.  }, 
journal={ IEEE Transactions on Automatic Control,}, 
title={Reducing the Reliability Over-Cost in Reconfiguration-Based Fault Tolerant Control Under Actuator Faults}, 
year={2012}, 
volume={57}, 
number={12}, 
pages={3181-3186} }

@inproceedings{moreau2004,
  title={Stability of continuous-time distributed consensus algorithms},
  author={Moreau, L.},
  booktitle={43rd IEEE Conference on Decision and Control, 2004. CDC. 2004},
  volume={4},
  pages={3998--4003},
  year={2004},
}
@inproceedings{yang2011,
  title={Output consensus for networks of non-identical introspective agents},
  author={Yang, T. and Saberi, A. and Stoorvogel, A. A. and Grip, H. F.},
  booktitle={50th IEEE Conference on Decision and Control and European Control Conference (CDC-ECC) },
  pages={1286--1292},
  year={2011},
  organization={IEEE}
}
@article{grip2012,
  title={Output synchronization for heterogeneous networks of non-introspective agents},
  author={Grip, H. F. and Yang, T. and Saberi, A. and Stoorvogel, A. A.},
  journal={Automatica},
    volume={48},
  issue={10},
 pages={2444 - 2453},
  year={2012},
  publisher={Elsevier}
}

@article{lunze2012,
  title={Synchronization of heterogeneous agents},
  author={Lunze, J.},
   journal={IEEE Transactions on Automatic Control, },
  volume={57},
  issue={11},
 pages={2885 - 2890},
  year={2012},
  publisher={IEEE}
}

@article{liu2010,
  title={Consensus of multi-agent systems with unbounded time-varying delays},
  author={Liu, X. and Lu, W. and Chen, T.},
  journal={IEEE Transactions on Automatic Control, },
  volume={55},
  number={10},
  pages={2396-2401},
  year={2010},
  publisher={IEEE}
}

@book{Boyd2004,
 author = {Boyd, S. and Vandenberghe, L.},
 title = {Convex Optimization},
 year = {2004},
 isbn = {0521833787},
 publisher = {Cambridge University Press},
 address = {New York, NY, USA},
} 

@ARTICLE{Rami00, 
author={Ait Rami, M. and Xun Y. Z.}, 
journal={IEEE Transactions on Automatic Control}, 
title={Linear matrix inequalities, $\text{Riccati}$ equations, and indefinite stochastic linear quadratic controls}, 
year={2000}, 
volume={45}, 
number={6}, 
pages={1131-1143}}


 @conference{ACC2014,
author={Gallehdari, Z. and Meskin, N. and Khorasani, K.},
  booktitle={ 2014 American Control Conference }, 
title={Cost Performance Based Control Reconfiguration in Multi-Agent Systems},
 year={ 2014},
 pages={509-516}
 organization={IEEE}
}

  

@ARTICLE{Loan1993,
author={Loan, C. F. and Pitsianis, N.},
year={1993},
journal={Linear Algebra for Large Scale and Real-Time Applications},
volume={232},
title={Approximation with Kronecker Products},
publisher={Springer Netherlands},

pages={293-314},
}



  @ARTICLE{Meskin2009-2, 
author={Meskin, N.  and  Khorasani, K.}, 
journal={ IEEE Transactions on Automatic Control}, 
title={Actuator Fault Detection and Isolation for a Network of Unmanned Vehicles}, 
volume = {54},
number = {4},
pages = {835-840},
year = {2009},
}

@inproceedings{guo2012,
  title={Distributed real-time fault detection and isolation for cooperative multi-agent systems},
  author={Guo, M. and Dimarogonas, D. and Johansson, K.  H.},
  booktitle={2012 American Control Conference (ACC)},
  pages={5270--5275},
  year={2012},
}

@article {Germund1961,
author = { Dahlquist, G.},
title = {Stability and error bounds in the numerical integration of ordinary differential equations},
journal = {ZAMM - Journal of Applied Mathematics and Mechanics},
volume = {41},
number = {6},
publisher = {WILEY-VCH Verlag},
pages = {267--268},
year = {1961},
}

@article{Kagstrom,
year={1977},
journal={BIT Numerical Mathematics},
volume={17},
number={1},
title={Bounds and perturbation bounds for the matrix exponential},
publisher={Kluwer Academic Publishers},
author={ Kagstrom, B.},
pages={39-57},
}
@INPROCEEDINGS{Qiu1988,
author={Qiu, L. and Davison, E. J.},
booktitle={ Proceedings of the 27th IEEE Conference on Decision and Control},
title={A new method for the stability robustness determination of state space models with real perturbations},
year={1988},
pages={538-543},
volume={1},
}

@article{Strom1975,
author = {Strom, T. },
title = {On Logarithmic Norms},
journal = {SIAM Journal on Numerical Analysis},
volume = {12},
number = {5},
pages = {741-753},
year = {1975},

}

@article{Hu2004,
title = "The weighted logarithmic matrix norm and bounds of the matrix exponential ",
journal = "Linear Algebra and its Applications ",
volume = "390",
number = "0",
pages = "145 - 154",
year = "2004",
author = " Hu, G. and  Liu,  M.",
}


@article{zhang2011a,
  title={Optimal design for synchronization of cooperative systems: state feedback, observer and output feedback},
  author={Zhang, H. and Lewis, F. L. and Das, A.},
  journal={IEEE Transactions on Automatic Control},
  volume={56},
  number={8},
  pages={1948--1952},
  year={2011},
  publisher={IEEE}
}
@ARTICLE{Semsar2010a,
author={Semsar-Kazerooni, E. and Khorasani, K.},
journal={IEEE Transactions on Systems, Man, and Cybernetics, Part B: Cybernetics},
title={Optimal Consensus Seeking in a Network of Multi-agent Systems: An \text{LMI} Approach},
year={2010},
month={April},
volume={40},
number={2},
pages={540-547}}

@ARTICLE{Movric2014, 
author={Movric, K. H. and Lewis, F. L.}, 
journal={ IEEE Transactions on Automatic Control}, 
title={Cooperative Optimal Control for Multi-Agent Systems on Directed Graph Topologies}, 
year={2014}, 
volume={59}, 
number={3}, 
pages={769-774}, }

@ARTICLE{Ren2010, 
author={Cao, Y. and Wei, R.}, 
journal={IEEE Transactions on Systems, Man, and Cybernetics, Part B: Cybernetics, }, 
title={Optimal Linear-Consensus Algorithms: An LQR Perspective}, 
year={2010}, 
volume={40}, 
number={3}, 
pages={819-830}, }

@article{sheng2013,
  title={Optimal consensus control of linear multi-agent systems with communication time delay},
  author={Sheng, J. and Ding, Z.},
  journal={ IET Control Theory $\&$ Applications,},
  volume={7},
  number={15},
  pages={1899--1905},
  year={2013},
  publisher={IET}
}
@ARTICLE{Cui2012, 
author={Cui, Y. and Jia, Y.}, 
journal={IET Control Theory Applications}, 
title={Brief Paper : ${L}_2-{L}_{\infty}$ consensus control for high-order multi-agent systems with switching topologies and time-varying delays}, 
year={2012}, 
volume={6}, 
number={12}, 
pages={1933-1940}, 
,}
@ARTICLE{Liu2012adaptive, 
author={Liu, Y. and Jia, Y.}, 
journal={IET Control Theory Applications}, 
title={Adaptive leader-following consensus control of multi-agent systems using model reference adaptive control approach}, 
year={2012}, 
volume={6}, 
number={13}, 
pages={2002-2008}, 
,}
@article{liu2012h,
  title={${H}_{\infty}$ consensus control for multi-agent systems with linear coupling dynamics and communication delays},
  author={Liu, Yang and Jia, Yingmin},
  journal={International Journal of Systems Science},
  volume={43},
  number={1},
  pages={50--62},
  year={2012},
  publisher={Taylor and Francis}
}



@ARTICLE{Yu2012,
author={Yu, X. and Jiang, J.},
journal={IEEE Transactions on Control Systems Technology},
title={Hybrid Fault-Tolerant Flight Control System Design Against Partial Actuator Failures},
year={2012},
volume={20},
number={4},
pages={871-886},
}

@ARTICLE{mes2009, 
author={ Meskin, N.  and  Khorasani, K.}, 
journal="Automatica", 
title={Fault Detection and Isolation of discrete-time {Markovian} Jump systems with Application to a Network of Multi-Agent System Having Imperfect Communication Channels}, 
volume = {45},
number = {9},
pages = { 2032-2040},
year = {2009},
}

@ARTICLE{Liao2002,
author={Fang, L. and Jian Liang, W. and Yang, G. H.},
journal={ IEEE Transactions on Control Systems Technology},
title={Reliable robust flight tracking control: an {LMI} approach},
year={2002},
volume={10},
number={1},
pages={76-89},
}

@ARTICLE{Ren2010a,
author={Wei, R.},
journal={ IEEE Transactions on Control Systems Technology},
title={Consensus Tracking Under Directed Interaction Topologies: Algorithms and Experiments},
year={2010},
volume={18},
number={1},
pages={230-237},
}
@ARTICLE{Dan2006,
author={ Ye, D. and  Yang, G. H.},
journal={IEEE Transactions on Control Systems Technology},
title={Adaptive Fault-Tolerant Tracking Control Against Actuator Faults With Application to Flight Control},
year={2006},
volume={14},
number={6},
pages={1088-1096},
}

@ARTICLE{Li2012,
author={Junquan, L. and Kumar, K.},
journal={IEEE Transactions on Fuzzy Systems},
title={Decentralized Fault-Tolerant Control for Satellite Attitude Synchronization},
year={2012},
volume={20},
number={3},
pages={572-586},}


@inproceedings{schenker2000,
  title={Reconfigurable robots for all-terrain exploration},
  author={Schenker, P. and Pirjanian, P. and Balaram, J. and Ali, K. and Trebi-Ollennu, A. and Huntsberger, T. and Aghazarian, H. and Kennedy, B. A. and Baumgartner, E. and Iagnemma, K. and others},
  booktitle={Intelligent Systems and Smart Manufacturing},
  pages={454--468},
  year={2000},
}

@article{Xavier2012,
title = "$\text{FDI(R)}$ for satellites: How to deal with high availability and robustness in the space domain?",
journal = " International Journal of Applied Mathematics and Computer Science ",
volume = "22",
number = "1",
pages = " 99Ð107 ",
year = "2012",
author = "Xavier O."
}

@ARTICLE{Huntsberger2003,
author={Huntsberger, T. and Pirjanian, P. and Trebi-Ollennu, A. and Das Nayar, H. and Aghazarian, H. and Ganino, A. and Garrett, M. and Joshi, S. and Schenker, P. },
journal={IEEE Transactions on Systems, Man and Cybernetics, Part A: Systems and Humans},
title={CAMPOUT: a control architecture for tightly coupled coordination of multi-robot systems for planetary surface exploration},
year={2003},
volume={33},
number={5},
pages={550-559},}

@ARTICLE{Weisbin2000,
author={Weisbin, C. and Rodriguez, G.},
journal={IEEE Robotics Automation Magazine},
title={$\text{NASA}$ robotics research for planetary surface exploration},
year={2000},
volume={7},
number={4},
pages={25-34},}

@ARTICLE{Yuh2000,
year={2000},
issn={0929-5593},
journal={Autonomous Robots},
volume={8},
number={1},
title={Design and Control of Autonomous Underwater Robots: A Survey},
author={Yuh, J.},
pages={7-24},}

@INPROCEEDINGS{Kunz2008,
author={Kunz, C. and Murphy, C. and Camilli, R. and Singh, H. and Bailey, J. and Eustice, R. and Jakuba, M. and Nakamura, K. and Roman, C. and Sato, T. and Sohn, R. and Willis, C.},
booktitle={2008 IEEE/RSJ International Conference on Intelligent Robots and Systems},
title={Deep sea underwater robotic exploration in the ice-covered Arctic ocean with AUVs},
year={2008},
pages={3654-3660},}

@article{Yoerger2007,
author = {Yoerger, D. and Jakuba, M. and Bradley, A. and Bingham, B.}, 
title = {Techniques for Deep Sea Near Bottom Survey Using an Autonomous Underwater Vehicle},
volume = {26}, 
number = {1}, 
pages = {41-54}, 
year = {2007}, 
journal = {The International Journal of Robotics Research} 
}

@phdthesis{braman2009,
  title={Safety verification and failure analysis of goal-based hybrid control systems},
  author={Braman, J.},
  year={2009},
  school={California Institute of Technology}
}

@book{wander2013,
  title={Innovative Fault Detection, Isolation and Recovery Strategies on-board Spacecraft: State of the Art and Research Challenges},
  author={Wander, A. and F{\"o}rstner, R.},
  year={2013},
  publisher={Deutsche Gesellschaft f{\"u}r Luft-und Raumfahrt-Lilienthal-Oberth eV}
}
@article{Jin2008,
title = "Fault tolerant control for satellites with four reaction wheels ",
journal = "Control Engineering Practice ",
volume = "16",
number = "10",
pages = "1250 - 1258",
year = "2008",
author = "Jaehyun. J. and Sangho. K. and Chang-Kyung. R.",
}

@article{xiao2012,
  title={Adaptive sliding mode fault tolerant attitude tracking control for flexible spacecraft under actuator saturation},
  author={Xiao, B. and Hu, Q. and Zhang, Y.},
  journal={ IEEE Transactions on Control Systems Technology,},
  volume={20},
  number={6},
  pages={1605--1612},
  year={2012},
  publisher={IEEE}
}

@article{yim2007,
  title={Modular self-reconfigurable robot systems: grand challenges of robotics},
  author={Yim, M. and Shen, W. and Salemi, B. and Rus, D. and Moll, M. and Lipson, H. and Klavins, E. and Chirikjian, G.},
  journal={IEEE Robotics \& Automation Magazine},
  volume={14},
  number={1},
  pages={43--52},
  year={2007},
  publisher={IEEE}
}

@article{croomes2006,
  title={Overview of the $\text{DART}$ mishap investigation results},
  author={Croomes, S.},
  journal={NASA Report},
  pages={1--10},
  year={2006}
}
@article{breger2008,
  title={Safe trajectories for autonomous rendezvous of spacecraft},
  author={Breger, L. and How, J. },
  journal={Journal of Guidance, Control, and Dynamics},
  volume={31},
  number={5},
  pages={1478--1489},
  year={2008}
}
@article{sabol2001,
  title={Satellite formation flying design and evolution},
  author={Sabol, C. and Burns, R. and McLaughlin, C.},
  journal={Journal of Spacecraft and Rockets},
  volume={38},
  number={2},
  pages={270--278},
  year={2001}
}

@ARTICLE{Lee2010,
author={Lee, H. and Kim, Y.},
journal={IET Control Theory Applications},
title={Fault-tolerant control scheme for satellite attitude control system},
year={2010},
volume={4},
number={8},
pages={1436-1450},}

@book{edwards2004,
  title={A leader-follower algorithm for multiple AUV formations},
  author={Edwards, D. B. and Bean, T. and Odell, D. L. and Anderson, M.},
  year={2004},
  publisher={IEEE}
}

@ARTICLE{Semsar-Kazerooni2010,
author={Semsar-Kazerooni, E. and Khorasani, K.},
journal={IEEE Transactions on Control Systems Technology},
title={Team Consensus for a Network of Unmanned Vehicles in Presence of Actuator Faults},
year={2010},
volume={18},
number={5},
pages={1155-1161},}

@ARTICLE{Mori88,
author={Mori, T.},
journal={Automatic Control, IEEE Transactions on},
title={Comments on "A matrix inequality associated with bounds on solutions of algebraic Riccati and Lyapunov equation" by J.M. Saniuk and I.B. Rhodes},
year={1988},
month={Nov},
volume={33},
number={11},
pages={1088},}

@ARTICLE{Lunze2006,
author={Lunze, J. and Steffen, T.},
journal={IEEE Transactions on Automatic Control},
title={Control Reconfiguration After Actuator Failures Using Disturbance Decoupling Methods},
year={2006},
volume={51},
number={10},
pages={1590-1601}}

@ARTICLE{Drea1997,
author={Drea, C.E.T. and Milani, B.E.A.},
journal={ IEEE Transactions on Automatic Control},
title={Disturbance decoupling in a class of linear systems},
year={1997},
volume={42},
number={10},
pages={1427-1431}}

@book{wohnam,
  title={linear multivariable control- a geometric approach},
  author={Wohnam, W. M.},
  year={1979},
  publisher={Springer-Verlag}
}


@book{Basile92,
title={Controlled and Conditioned Invariants in Linear System Theory},
author={ Basile, G. and  Marro, G.},
year={1992},
publisher={Prentice Hall, Englewood Cliffs, NJ}
}
 @MISC{Basil:2010:MISC,
   author = "Marro, G.",
   title = "Geometric Approach  Toolbox",
   year = "2010",
   howpublished= {\url{http://www3.deis.unibo.it/Staff/FullProf/GiovanniMarro/geometric.htm}},
 }
 
  @MISC{elec2:2015:MISC,
   author = "Gallehdari, Z. and Meskin, N. and Khorasani, K.",
   title = "An {$H_\infty$} Cooperative Control Fault Recovery  of Multi-Agent Systems",
   year = "2015",
   howpublished= {\url{http://arxiv.org/abs/1508.07076}},
 }


 

@book{Yedavalli14,
title={Robust Control of Uncertain Dynamic Systems:A Linear State Space Approach},
author={Yedavalli, R. K.},
year={2014},
publisher={Springer New York}
}

@book{huang2004,
title={Nonlinear output regulation: Theory and applications},
author={Huang, J.},
volume={8},
year={2004},
publisher={SIAM}
}

@ARTICLE{Karanam83,
author={Karanam, V.},
journal={ IEEE Transactions on Automatic Control},
title={A note on eigenvalue bounds in algebraic Riccati equation},
year={1983},
volume={28},
number={1},
pages={109-111}}

@ARTICLE{Wenk1980,
author={Wenk, C.J. and Knapp, C.},
journal={IEEE Transactions on Automatic Control},
title={Parameter optimization in linear systems with arbitrarily constrained controller structure},
year={1980},
volume={25},
number={3},
pages={496-500}}

@book{wright1999,
  title={Numerical optimization},
  author={Wright, SJ and Nocedal, J},
  volume={2},
  year={1999},
  publisher={Springer New York}
}

@book{dennis1996,
  title={Numerical methods for unconstrained optimization and nonlinear equations},
  author={Dennis  J. E. and Schnabel, R. B.},
  volume={16},
  year={1996},
  publisher={Siam}
}

@book{garren1968,
  title={Bounds for the eigenvalues of a matrix},
  author={Garren, K.R.},
  series={NASA technical note},
  year={1968},
  publisher={National Aeronautics and Space Administration; for sale by the Clearinghouse for the Federal Scientific and Technical Information, Springfield, Va.}
}

@ARTICLE{Yang20102,
author={Yang, G.  and  Ye, D.},
journal={ IEEE Transactions on Automatic Control},
title={Reliable ${H}_{\infty}$ Control of Linear Systems With Adaptive Mechanism},
year={2010},
volume={55},
number={1},
pages={242-247}}

@ARTICLE{Xiao2013,
author={ Xiao, B. and  Hu, Q. and Shi,  P.},
journal={IEEE Transactions on Control Systems Technology},
title={Attitude Stabilization of Spacecrafts Under Actuator Saturation and Partial Loss of Control Effectiveness},
year={2013},
volume={21},
number={6},
pages={2251-2263},}

@ARTICLE{Zhao2014,
author={ Zhao, L. and Jia, Y.},
journal={International Journal of Systems Science},
title={Neural network-based adaptive consensus tracking control for multi-agent systems under actuator faults},
year={2014},
volume={21},
number={6},
pages={2251-2263},}

@ARTICLE{Chen2014,
author={Chen, S. and Ho, D. and Li, L. and Liu, M.},
journal={IEEE Transactions on Cybernetics},
title={Fault-Tolerant Consensus of Multi-Agent System With Distributed Adaptive Protocol},
year={2015},
volume={PP},
number={99},
pages={1},
}

@ARTICLE{Wang2015-3,
author={Wang, X. and  Yang, G. H.},
journal={IET Control Theory Applications},
title={Cooperative adaptive fault-tolerant tracking control for a class of multi-agent systems with actuator failures and mismatched parameter uncertainties},
year={2015},
volume={9},
number={8},
pages={1274-1284},}

@ARTICLE{Wang2015-2,
author={ Wang, Y. and  Song, Y. and Lewis, F.L.},
journal={IEEE Transactions on Industrial Electronics},
title={Robust Adaptive Fault-Tolerant Control of Multiagent Systems With Uncertain Nonidentical Dynamics and Undetectable Actuation Failures},
year={2015},
volume={62},
number={6},
pages={3978-3988},}

@inproceedings{ECC2014,
  title={Robust cooperative control reconfiguration/recovery in multi-agent systems},
  author={Gallehdari, Z. and Meskin, N. and Khorasani, K.},
  booktitle={2014 European Control Conference (ECC) },
  pages={1554--1561},
  year={2014},
}


@ARTICLE{Wang13,
author={ Wang, X. and  Hong, Y. and  Huang, J. and  Jiang, Zh.},
journal={ IEEE Transactions on Automatic Control},
title={A Distributed Control Approach to A Robust Output Regulation Problem for Multi-Agent Linear Systems},
year={2010},
volume={55},
number={12},
pages={2891-2895}}

@article{Wieland2011,
title = "An internal model principle is necessary and sufficient for linear output synchronization ",
journal = "Automatica ",
volume = "47",
number = "5",
pages = "1068 - 1074",
year = "2011"}

@INPROCEEDINGS{Grip2013,
author={Grip, H.F. and Saberi, A. and Stoorvogel, A.A.},
booktitle={2013 European Control Conference (ECC) },
title={Regulated output synchronization for heterogeneous networks of non-introspective, minimum-phase SISO agents without exchange of controller states},
year={2013},
pages={3827-3832}}

@ARTICLE{Huang2014,
author={ Huang, C. and  Ye, X.},
journal={ IEEE Transactions on Automatic Control},
title={Cooperative Output Regulation of Heterogeneous Multi-Agent Systems: An $H_{\infty }$ Criterion},
year={2014},
volume={59},
number={1},
pages={267-273},}

@ARTICLE{Su2013,
author={Su, Y. and Hong, Y. and Huang, J.},
journal={ IEEE Transactions on Automatic Control},
title={A General Result on the Robust Cooperative Output Regulation for Linear Uncertain Multi-Agent Systems},
year={2013},
volume={58},
number={5},
pages={1275-1279}}

@ARTICLE{Zhang2011,
author={Zhang, H. and Lewis, F.L. and Das, A.},
journal={IEEE Transactions on Automatic Control},
title={Optimal Design for Synchronization of Cooperative Systems: State Feedback, Observer and Output Feedback},
year={2011},
volume={56},
number={8},
pages={1948-1952},}

@INPROCEEDINGS{Boyd1998,
author={Hindi, H. and Boyd, S.},
booktitle={Proceedings of the 37th IEEE Conference on Decision and Control, 1998},
title={Analysis of linear systems with saturation using convex optimization},
year={1998},
volume={1},
pages={903-908 vol.1}}
@book{Slotine1991,
  title={Applied Nonlinear Control},
  author={Slotine, J. J. and Li, W.},
  year={1991},
  publisher={Prentice Hall, Englewood Cliffs, NJ},}   

@ARTICLE{Li13,
author={Li, Z. and  Liu, X. and Ren, W.  and Xie, L.},
journal={ IEEE Transactions on Automatic Control},
title={Distributed Tracking Control for Linear Multi-agent Systems With a Leader of Bounded Unknown Input},
year={2013},
volume={58},
number={2},
pages={518-523},}

@article{bacciotti2005,
  title={Generalized solutions of differential inclusions and stability},
  author={Bacciotti, A.},
  journal={Ital. J. Pure Appl. Math},
  volume={17},
  number={183-192},
  pages={7},
  year={2005},
  publisher={Citeseer},}
  
  @ARTICLE{Paden87,
author={Paden, B.E. and Sastry, S.S.},
journal={ IEEE Transactions on Circuits and Systems},
title={A calculus for computing Filippov's differential inclusion with application to the variable structure control of robot manipulators},
year={1987},
volume={34},
number={1},
pages={73-82},}
@book{fossen2002,
  title={Marine control systems: Guidance, navigation and control of ships, rigs and underwater vehicles},
  author={Fossen, T.},
  year={2002},
  publisher={Marine Cybernetics Trondheim},}
  @book{fossen94,
  title={Guidance and Control of Ocean Vehicles},
  author={Fossen, T.},
  year={1994},
  publisher={ Wiley},}
  %
  @article{Zhou2014,
title = "Cooperative control for consensus of multi-agent systems with actuator faults ",
journal = "Computers $\&$ Electrical Engineering ",
volume = "40",
number = "7",
pages = "2154 - 2166",
year = "2014",
author = "Zhou, B. and  Wang, W. and  Ye, H."
}
@inproceedings{Hespanha1999,
  title={Stability of switched systems with average dwell-time},
  author={Hespanha, J. P.  and Morse, A. S.},
  booktitle={Proceedings of the 38th IEEE Conference on Decision and Control, 1999},
  volume={3},
  pages={2655--2660},
  year={1999},
  organization={IEEE}
}

@ARTICLE{Hu2003,
author={ Hu, T.  and  Lin, Z. },
journal={ IEEE Transactions on Automatic Control},
title={Composite quadratic Lyapunov functions for constrained control systems},
year={2003},
volume={48},
number={3},
pages={440-450},
}

@article{Hu2005,
title = "Conjugate Lyapunov functions for saturated linear systems ",
journal = "Automatica ",
volume = "41",
number = "11",
pages = "1949 - 1956",
year = "2005",
author = " Hu, T. and  Goebel, R. and  Teel, A. R. and  Lin, Z.",
}





@INPROCEEDINGS{ZhaiG2000,
author={Zhai, G. and Bo H. and Yasuda, K. and Michel, A.N.},
booktitle={American Control Conference, 2000. Proceedings of the 2000},
title={Stability analysis of switched systems with stable and unstable subsystems: an average dwell time approach},
year={2000},
volume={1},
number={6},
pages={200-204 vol.1},}

@ARTICLE{Su12,
author={ Su, Y. and  Huang, J.},
journal={ IEEE Transactions on Automatic Control},
title={Stability of a Class of Linear Switching Systems with Applications to Two Consensus Problems},
year={2012},
volume={57},
number={6},
pages={1420-1430},}

@ARTICLE{Hespanha2005,
author={Hespanha, J.P. and Liberzon, D. and Angeli, D. and Sontag, E.D.},
journal={ IEEE Transactions on Automatic Control},
title={Nonlinear norm-observability notions and stability of switched systems},
year={2005},
volume={50},
number={2},
pages={154-168},}

@ARTICLE{Hu2006,
author={Hu, T. and Teel, A.R. and Zaccarian, L.},
journal={ IEEE Transactions on Automatic Control},
title={Stability and Performance for Saturated Systems via Quadratic and Nonquadratic Lyapunov Functions},
year={2006},
volume={51},
number={11},
pages={1770-1786},
}

@ARTICLE{Shevitz94,
author={Shevitz, D. and Paden, B.},
journal={IEEE Transactions on Automatic Control},
title={Lyapunov stability theory of nonsmooth systems},
year={1994},
volume={39},
number={9},
pages={1910-1914}}

@ARTICLE{Bacciotti99,
author={Bacciotti, A. and Ceragioli, F.},
journal={Control, Optimization and Calculus of Variations},
title={Stability and Stabilization of Discontinuous Systems and Nonsmooth Lyapunov Functions},
year={1999},
volume={4},
pages={361{376}
}
@article{Molchanov1989,
title = {Criteria of asymptotic stability of differential and difference inclusions encountered in control theory },
author = { Molchanov, A.P. and  Pyatnitskiy, Ye.S.},
journal = {Systems \& Control Letters },
volume = {13},
number = {1},
pages = {59 - 64},
year = {1989},}
@INPROCEEDINGS{Hassibi99,
author={Hassibi, A. and How, J. and Boyd, S.},
booktitle={Proceedings of the 1999 American Control Conference.},
title={A path-following method for solving BMI problems in control},
year={1999},
volume={2},
pages={1385-1389},}

@ARTICLE{Tran2012,
author={ Tran Dinh, Q. and Gumussoy, S. and Michiels, W. and Diehl, M.},
journal={ IEEE Transactions on Automatic Control},
title={Combining Convex-Concave Decompositions and Linearization Approaches for Solving BMIs, With Application to Static Output Feedback},
year={2012},
volume={57},
number={6},
pages={1377-1390},}

@ARTICLE{Tuan2000,
author={Tuan, H.D. and Apkarian, P.},
journal={IEEE Transactions on Automatic Control},
title={Low nonconvexity-rank bilinear matrix inequalities: algorithms and applications in robust controller and structure designs},
year={2000},
volume={45},
number={11},
pages={2111-2117},}

@ARTICLE{Yamada98,
author={Yamada, Y. and Hara, S.},
journal={IEEE Transactions on Automatic Control},
title={Global optimization for H infin; control with constant diagonal scaling},
year={1998},
volume={43},
number={2},
pages={191-203},}

@article{Goh95,
year={1995},
journal={Journal of Global Optimization},
volume={7},
number={4},
title={Global optimization for the Biaffine Matrix Inequality problem},
author={Goh, K. and Safonov, M. G. and Papavassilopoulos, G. P.},
pages={365-380},}

@ARTICLE{Henrion06,
author={Henrion, D. and Lasserre, J.-B.},
journal={ IEEE Transactions on Automatic Control},
title={Convergent relaxations of polynomial matrix inequalities and static output feedback},
year={2006},
volume={51},
number={2},
pages={192-202},}

@INPROCEEDINGS{Zhu93,
author={Zhu, G. and Rotea, M.A. and Skelton, R.E.},
booktitle={Proceedings of the 1993 American Control Conference},
title={A Convergent Feasible Algorithm for the Output Covariance Constraint Problem},
year={1993},
pages={1675-1679},
}

@INPROCEEDINGS{Ghaoui94,
author={El Ghaoui, L. and Balakrishnan, V.},
booktitle={Proceedings of the 33rd IEEE Conference on Decision and Control},
title={Synthesis of fixed-structure controllers via numerical optimization},
year={1994},
volume={3},
pages={2678-2683}
,}

@ARTICLE{Iwasaki99,
author={Iwasaki, T.},
journal={IEEE Transactions on Automatic Control},
title={The dual iteration for fixed-order control},
year={1999},
volume={44},
number={4},
pages={783-788},}

@article{kanev2004,
  title={Robust output-feedback controller design via local BMI optimization},
  author={Kanev, S. and Scherer, C. and Verhaegen, M. and De Schutter, B.},
  journal={Automatica},
  volume={40},
  number={7},
  pages={1115--1127},
  year={2004},
  publisher={Elsevier}
}

@inproceedings{simon2011,
  title={LMIs-based coordinate descent method for solving BMIs in control design},
  author={Simon, E. and R-Ayerbe, P. and Stoica, C. and Dumur, D.and Wertz, V.},
  booktitle={Proceedings of the 18th IFAC World Congr.},
  pages={10180-10186},
  year={2011}
}


@book{khalil96,
  title={Nonlinear systems},
  author={Khalil, H. K and Grizzle, J.},
  volume={3},
  year={1996},
  publisher={Prentice hall New Jersey}
}
\end{document}